\documentstyle[12pt,a41,epsfig,cite,amssymb,amstex]{article}
\newcommand\SHU{\,\mbox{$\sqcup \! \sqcup$}\,}
\newcommand{\bra}[1]{\langle#1\mid}
 \newcommand{\ket}[1]{\mid#1\rangle}
 \newcommand{\MOM}{\sf MOM}
 \newcommand{\MS}{\overline{\sf MS}}
 \newcommand{\Ahathat}{\hat{\hspace*{-1mm}\hat{A}}}
 \newcommand{\Atiltil}{\tilde{\hspace*{-1mm}\tilde{A}}}
 \newcommand{\Atil}{\tilde{A}}
 \newcommand{\Ctil}{\tilde{C}}
 
 \newcommand{\ep}{\varepsilon}

 \newcommand{\N}{\nonumber}
 \newcommand{\adag}{/\!\!\! }

\newcommand{\bea}{\begin{eqnarray}}
\newcommand{\bq}{\begin{equation}}
\newcommand{\eea}{\end{eqnarray}}
\newcommand{\eq}{\end{equation}}

\newcommand{\gsim}{\raisebox{-0.07cm   }
{$\, \stackrel{>}{{\scriptstyle\sim}}\, $}}

\newcommand\Mvec{\mbox{\boldmath $M$}}

\newcommand\be{\begin{eqnarray}}
\newcommand\ee{\end{eqnarray}}

\begin{document}
\noindent
\sloppy
\thispagestyle{empty}

\begin{flushleft}
DESY 09--057 \hfill {\tt arXiv:0904.3563}
\\
SFB/CPP--09--033\\
IFIC/09--16\\
April 2009
\end{flushleft}

\vspace*{\fill}
\begin{center}
{\Large \bf Mellin Moments of the \boldmath{$O(\alpha_s^3$)} Heavy Flavor}

\vspace*{2mm}
{\Large \bf Contributions to Unpolarized Deep-Inelastic Scattering}

\vspace*{2mm}
{\Large \bf \boldmath at $Q^2 \gg m^2$ and Anomalous 
Dimensions}

\vspace{2cm}
\large
Isabella Bierenbaum~\footnote{Present address: Instituto de Fisica
Corpuscular, CSIC-Universitat de Val\`{e}ncia, Apartado de Correros 22085,
E-46071 Valencia, Spain.},
Johannes Bl\"umlein  and Sebastian Klein
\\
\vspace{2em}
\normalsize
{\it Deutsches Elektronen--Synchrotron, DESY,\\
Platanenallee 6, D--15738 Zeuthen, Germany}
\\
\vspace{2em}
\end{center}
\vspace*{\fill}
%
\begin{abstract}
\noindent
We calculate the $O(\alpha_s^3)$ heavy flavor contributions to the Wilson coefficients 
of the structure function $F_2(x,Q^2)$ and the massive operator matrix elements (OMEs) 
for the twist--2 operators of unpolarized deeply inelastic scattering in the region 
$Q^2 \gg m^2$. The massive Wilson coefficients are obtained as convolutions of massive 
OMEs and the known light flavor Wilson coefficients. We also compute the  massive OMEs 
which are needed to evaluate heavy flavor parton distributions in the variable flavor 
number scheme (VFNS) to 3--loop order. All contributions to the Wilson coefficients and 
operator matrix elements but the genuine constant terms at $O(\alpha_s^3)$ of the OMEs 
are derived in terms of quantities, which are known for general values in the Mellin 
variable $N$. For the operator matrix elements $A_{Qg}^{(3)}, A_{qg,Q}^{(3)}$ and 
$A_{gg,Q}^{(3)}$ the moments $N = 2$ to 10, for $A_{Qq}^{(3), \rm PS}$ to $N = 12$, 
and for $A_{qq,Q}^{(3), \rm NS}$, $A_{qq,Q}^{(3),\rm PS}$, $A_{gq,Q}^{(3)}$ to $N=14$ 
are computed. These terms contribute to the light flavor $+$-combinations. For the 
flavor non-singlet terms, we calculate as well the odd moments $N=1$ to $13$, 
corresponding to the light flavor $-$-combinations. We also obtain the moments of 
the 3--loop anomalous dimensions, their color projections for the present processes 
respectively, in an independent calculation, which agree with the results given in the 
literature.  
\end{abstract} 
\vspace*{\fill} \newpage

\section{Introduction}
\renewcommand{\theequation}{\thesection.\arabic{equation}}
\setcounter{equation}{0}
\label{sec-1}

\vspace{1mm}
\noindent
Deep-inelastic scattering processes of charged or neutral leptons off proton
and deuteron targets, in the region of large enough values of the gauge boson
virtuality $Q^2 = -q^2$ and hadronic mass $W^2 = (q+p)^2$, allow to measure
the leading twist parton densities of the nucleon, the QCD-scale
$\Lambda_{\rm QCD}$ and the strong coupling constant $a_s(Q^2) =
\alpha_s(Q^2)/(4\pi)$, to high precision. The precise value of
$\Lambda_{\rm QCD}$, a fundamental parameter of the Standard Model, is of
central importance for the quantitative understanding of all strongly
interacting processes. Moreover, the possible unification of the gauge forces
\cite{UNIV} depends crucially on its value. Of similar importance is the
detailed knowledge of the parton densities for all hadron-induced processes
\cite{HERALHC}, notably for the interpretation of all scattering cross
sections measured at the Tevatron and the LHC. For example, the process of Higgs-boson
production at the LHC \cite{HIGGS} depends on the gluon density
and its accuracy is widely determined by this distribution.

Let us consider the kinematic region in deeply inelastic scattering, where 
processes of higher twist can be safely disregarded and the hard scales $Q^2$ 
and $W^2$ are large enough to allow the application of the light-cone
expansion, saturated by the twist--2 contributions. The scattering processes are then 
described by 
structure functions $F_i(x,Q^2)$, which decompose into {\sf non-perturbative} 
{\it massless} parton densities $f_j(x,\mu^2)$ and {\sf perturbative} 
coefficient functions $C_{i}^{j}(x,Q^2/\mu^2)$ by   
\begin{eqnarray}
\label{eqFi}
F_i(x,Q^2) = \sum_{j = q, \overline{q}, g} C_{i}^{j}\left(x,\frac{Q^2}{\mu^2}\right) \otimes 
f_j(x,\mu^2)~.
\end{eqnarray}
The scale $\mu$
denotes the factorization scale, which is arbitrary and cancels between the
coefficient functions and parton distribution functions in the respective
orders in perturbation theory. The symbol $\otimes$ denotes the Mellin
convolution
\begin{eqnarray}
\label{MELC}
[A \otimes B](x) = \int_0^1 dx_1 \int_0^1 dx_2 \delta(x - x_1 x_2)
A(x_1) B(x_2)~.
\end{eqnarray}
The Mellin transformation 
\begin{eqnarray}
\label{MELTRA}
\Mvec\left[f(x)\right](N) = \int_0^1 dx x^{N-1} f(x)~,
\end{eqnarray}
if applied to (\ref{MELC}), resolves the convolution into a product.

Since we strictly consider twist-2 parton densities in the Bjorken limit, no
transverse momentum effects in the initial distributions will be allowed, which
otherwise is related in the kinematic sense to higher twist operators. As 
is well known,
the leading--twist approximation and the QCD improved parton model are equivalent
descriptions for the dominant contributions to the deep-inelastic structure functions
at sufficiently large scales $Q^2$. The condition for the validity of the parton model 
\cite{DRELL} demands that
\begin{eqnarray}
\label{eqtim}
\frac{\tau_{\rm int}}{\tau_{\rm life}} \ll 1~,
\end{eqnarray}
with $\tau_{\rm int}$ being the interaction time of the virtual gauge boson
with a hadronic quantum-fluctuation, the life--time of which is given by $\tau_{\rm
  life}$. The latter can be interpreted as a partonic state, provided
(\ref{eqtim}) holds. 
Both times are measured in an infinite momentum frame and
they are given by
\begin{eqnarray}
\tau_{\rm int} &\sim& \frac{1}{q_0} = \frac{4 P x}{Q^2 (1-x)} \\
\tau_{\rm life}&\sim& \frac{1}{\displaystyle \sum_i E_i - E} 
                    = \frac{2 P}{\displaystyle  \sum_i (k_{\perp,i}^2 + 
m_i^2)/x_i - M_N^2}, \hspace*{7mm} \sum_i x_i = 1~,   
\end{eqnarray}
with $P$ the large momentum of the hadron, $q_0$ the energy component of the
virtual gauge boson in the infinite momentum frame, $E_i$ the energy of the
$i$th fluctuating parton, $k_{\perp, i}$, $m_i$, $x_i$ its transverse
momentum, mass, and momentum fraction, $E$ the total energy, and $M_N$ the
nucleon mass. In the region of not too small values, nor values near the
elastic region $x \simeq 1$, of the Bjorken variable $x$, the partonic
description holds for massless partons. Evidently, iff $Q^2 (1-x)^2/m^2_i
\gg \hspace*{-5mm}/ \hspace*{2mm}1$ {\sf no} partonic description for a
potential heavy quark distribution can be obtained. In the general kinematic 
region the parton densities in Eq.~(\ref{eqFi}) are enforced to be 
massless and the heavy quark mass effects are contained in the Wilson 
coefficients  $C_i^j$, which are perturbatively calculable. Due to this, one 
may identify the massless flavor contributions and separate the Wilson 
coefficients into a purely light part $C_i^{j, \rm light}$ and ${\sf H}_i^j$, 
which 
accounts for the heavy quark contributions,  
\begin{eqnarray}
\label{eqLH}
C_i^j\left(x,\frac{Q^2}{\mu^2}\right) = 
  C_i^{j, \rm light}\left(x,\frac{Q^2}{\mu^2}\right)  
+ {\sf H}_i^j\left(x,\frac{Q^2}{\mu^2},  
\frac{m_k^2}{\mu^2}\right)~, k = c,b~. 
\end{eqnarray}
The question, under which circumstances one may introduce a heavy flavor 
parton density, will be discussed later. 
Both, the measurements of the heavy 
flavor part of the deep-inelastic structure functions, cf.~\cite{HEXP}, and 
numerical studies \cite{BR} based on the leading \cite{LO} and next-to-leading 
order (NLO) heavy flavor Wilson coefficients \cite{NLO}, show that the scaling 
violations of the light and the heavy contributions to (\ref{eqLH}) exhibit 
a different behaviour over a wide range of $Q^2$. This is both due to the 
logarithmic contributions $\ln^k(Q^2/m^2)$ and power corrections $\propto 
(m^2/Q^2)^k,~k \geq 1$. Moreover, in the region of smaller values of $x$ the 
heavy flavor contributions amount to 20--40\%. Therefore, the precision measurement of 
the QCD parameter $\Lambda_{\rm QCD}$ \cite{LAMB} and the parton distribution 
functions in deeply inelastic scattering require the description of the light 
and heavy flavor contributions at the same accuracy. 
The separation (\ref{eqLH}) 
allows the definition of the light flavor contributions and the related heavy 
flavor contributions to $F_i(x,Q^2)$ applying the factorization Eq.~(\ref{eqFi}). 

The perturbative accuracy reached for $F_i^{\rm light}(x,Q^2)$ is of 3--loop order
\cite{Floratos:1977au,Floratos:1978ny,GonzalezArroyo:1979df,GonzalezArroyo:1979he,
  Curci:1980uw,Furmanski:1980cm,Hamberg:1991qt,Ellis:1996nn,Vogt:2004mw,Moch:2004pa,
  WIL1,WIL2,Larin:1993vu,Larin:1996wd,Retey:2000nq,Blumlein:2004xt,
  Vermaseren:2005qc}, which requires to calculate the 3--loop heavy flavor Wilson
coefficients as well.  The NLO heavy flavor corrections in the complete kinematic
range are available only in semi-analytic form \cite{NLO} due to the
complexity of the contributing phase space integrals.~\footnote{A precise
numerical implementation in Mellin space was given in \cite{AB}.} Heavy 
flavor corrections to different sum rules for deep-inelastic structure 
functions were calculated in \cite{Blumlein:1998sh}. An
important part of the kinematic region is that of larger values of $Q^2$. As
has been shown in Ref.~\cite{Buza:1995ie}, the heavy flavor Wilson
coefficients ${\sf H}_2^j(x,{Q^2}/{\mu^2},m_i^2/\mu^2)$ can be calculated
analytically at NLO for $Q^2/m^2 \gsim 10$.~\footnote{In case of
${\sf H}_L^j(x,{Q^2}/{\mu^2},m_i^2/\mu^2)$ this approximation is only valid 
for $Q^2/m^2 \gsim 800$, \cite{Buza:1995ie}.  The 3--loop corrections were
calculated in Ref.~\cite{BFNK}.}  This is due to a factorization of the
heavy quark Wilson coefficients into massive OMEs, $A_{jk}$,
and massless Wilson coefficients, $C_i^{j, \rm light}$ in case one heavy quark
flavor of mass $m$ and $n_f$ light flavors are considered.  This restriction
to only one heavy quark flavor is required beginning with the 3--loop
corrections and will be adopted in the following.
In the present paper, we calculate the massive operator matrix elements
$A_{jk}$ contributing to the heavy flavor Wilson coefficients for the
structure function $F_2(x,Q^2)$ in the region $Q^2/m^2 \gsim 10$ to 3--loop
order for fixed moments of the Mellin variable $N$. In case of the flavor
non-singlet ({\sf NS}) contributions, we also present the odd moments of the $-$-projection. 
We further calculate the operator matrix elements, which are required to define
heavy quark densities in the VFNS~\cite{Buza:1996wv}.
Due to renormalization, higher order contributions in
$\varepsilon$ to corrections of lower order in $a_s$,
cf.~\cite{Buza:1995ie,Buza:1996wv,
  Bierenbaum:2007dm,Bierenbaum:2007qe,Bierenbaum:2008yu,Bierenbaum:2009zt},
and other renormalization terms, such as the anomalous dimensions and the
expansion coefficients of the QCD $\beta$--function and mass anomalous
dimensions, contribute. For these reasons, the present calculation yields also
the moments of the complete 2--loop anomalous dimensions and the terms $\propto T_F$ of the
3--loop anomalous dimensions $\gamma_{ij}(N)$. In the pure singlet ({\sf PS}) case,
$\gamma_{qq}^{\sf+, PS}(N)$, and for $\gamma_{qg}(N)$, these are the complete anomalous 
dimensions
given in \cite{Moch:2004pa,Vogt:2004mw}, to which we agree. Since the present
calculation is completely independent by method, formalism, and codes, it
provides a check on the previous results. Except for the constant part of the
unrenormalized heavy flavor operator matrix elements, we obtain the heavy
quark Wilson coefficients in the asymptotic region for all values of the
Mellin variable $N$. The analytic continuation of these expressions to complex
values of $N$ can be performed with the help of the representations in
\cite{ANCONT} and those given for the anomalous dimensions and massless
Wilson coefficients in \cite{Moch:2004pa,Vogt:2004mw,Vermaseren:2005qc}.

The paper is organized as follows. In Section~2, a brief outline of the 
basic formalism is given. The renormalization of the different massive operator
matrix elements is described in Section~3. In Section~4, we present details on
the unrenormalized and renormalized operator matrix elements. Technical
details of the calculation and the main results are discussed in
Section~5. Depending on the  CPU time and storage
size required, the moments up to $N=10, 12$, and $14$ of the different operator matrix
elements could be calculated.  In Section~6, representations for heavy quark
parton densities in the region $\mu^2 \gg m^2$ are given and Section~7
contains the conclusions. In the Appendices, we give a consistent set of
Feynman rules for the composite operators up to 3--loop order,
present the moments of the 3--loop anomalous dimensions, and of the constants part of the
different 3--loop massive operator matrix elements.
\section{The Formalism}
\renewcommand{\theequation}{\thesection.\arabic{equation}}
\setcounter{equation}{0}
\label{sec-forma}

\vspace{1mm}
\noindent
The heavy quark contribution to the structure function $F_2(x,Q^2)$ for one 
heavy flavor of mass $m$ and $n_f$ light flavors is given by, \cite{Buza:1996wv},
\begin{eqnarray}
\label{eqF2}
F_{2,Q}(x,Q^2,n_f,m) &=& \sum_{k=1}^{n_f} e_k^2 \Biggl\{
L_{2,q}^{\sf NS}\left(n_f,\frac{Q^2}{m^2},\frac{m^2}{\mu^2} \right)
\otimes \Bigl[f_k(x,\mu^2,n_f) + f_{\overline{k}}(x,\mu^2,n_f)\Bigr]
\nonumber\\ 
&&
+{\tilde{L}}_{2,q}^{\sf PS}\left(n_f,\frac{Q^2}{m^2},\frac{m^2}{\mu^2} 
\right) \otimes \Sigma(x,\mu^2,n_f)
\nonumber\\
&&
+
{\tilde{L}}_{2,g}^{\sf S}\left(n_f,\frac{Q^2}{m^2},\frac{m^2}{\mu^2} 
\right) \otimes G(x,\mu^2,n_f) \Biggr\}
\nonumber\\ &&
+ e_Q^2 \Biggl[H_{2,q}^{\sf PS} 
\left(n_f,\frac{Q^2}{m^2},\frac{m^2}{\mu^2}\right) \otimes \Sigma(x,\mu^2,n_f)
\nonumber\\ &&
+ H_{2,g}^{\sf S} 
\left(n_f,\frac{Q^2}{m^2},\frac{m^2}{\mu^2}\right) \otimes G(x,\mu^2,n_f)
\Biggr]~,
\end{eqnarray}
with {\sf (S)} the singlet contributions. 
Here, we denote the heavy flavor Wilson coefficients ${\sf H}_i^j$ by
$L_i^j$, $H_i^j$ respectively, depending on whether the photon couples to a light 
$(L)$ or the heavy $(H)$ quark line. $f_k(x,\mu^2)$ and $f_{\overline{k}}(x,\mu^2)$ 
denote the quark-
and antiquark distribution functions, $G(x,\mu^2)$ is the gluon distribution and
\begin{eqnarray}
\Sigma(x,\mu^2) = \sum_{k=1}^{n_f} \left[ f_k(x,\mu^2) + f_{\overline{k}}(x,\mu^2)\right]
\end{eqnarray}
denotes the flavor singlet distribution. $e_Q$ is the electric charge of the heavy quark.
Due to the difference 
  of quantities taken at $n_f+1$ and $n_f$ flavors, it is useful 
  to adopt the following notation for a function $f(n_f)$,
  \begin{eqnarray}
    \hat{f}(n_f)&\equiv&f(n_f+1)-f(n_f)~, \label{gammapres1} \\
    {\tilde{f}}(n_f)&\equiv&\frac{f(n_f)}{n_f}~,
\label{gammapres2}
  \end{eqnarray}
and $\hat{\hspace*{-1mm}{\tilde{f}}}(n_f) \equiv 
\widehat{[{\tilde{f}}(n_f)]}$.~\footnote{Later on, the 
symbol $~\hat{\empty}~$ will also be used for the bare coupling $\hat{a}_s$, the mass 
$\hat{m}$, and the bare OMEs, where (\ref{gammapres1}) is not applied.}  As has been 
shown in
Ref.~\cite{Buza:1995ie}, the heavy quark Wilson coefficients in
deeply--inelastic scattering, ${\sf H}_i^j$, factorize in the region $Q^2 \gg
m^2$, in which power corrections can be disregarded, into massive operator
matrix elements $A_{kl}^{\sf NS,S}$ and the light flavor Wilson coefficients
$C_{i,k}^{\sf NS,S}$,
\begin{eqnarray}
\label{FAC1}
{\sf H}_{i,l}^{\sf NS, S} = {A}_{kl}^{\sf NS, S} \otimes C_{i,k}^{\sf NS, S}~, 
\end{eqnarray}
where $i = 2, L$ specifies the structure function
considered. 

The operator matrix elements $A_{k,l}^{\sf NS,S}$ are the partonic expectation
values
\begin{eqnarray}
\label{eqAmain}
A_{kl}^{\sf NS,S}\left(N, \frac{m^2}{\mu^2}\right) = \langle l| O_k^{\sf 
NS,S} | l \rangle~, \hspace*{5mm} l = q,g~,
\end{eqnarray}
with the local twist--2 operators given by
\begin{eqnarray}
\label{COMP1}
O^{\sf NS}_{F,a;\mu_1, \ldots, \mu_n} &=& i^{n-1} {\bf S} [\overline{\psi}
\gamma_{\mu_1} D_{\mu_2} \ldots D_{\mu_n} \frac{\lambda_a}{2}
\psi] - {\rm trace~terms}~, \\
\label{COMP2}
O^{\sf S}_{F;\mu_1, \ldots, \mu_n} &=& i^{n-1} {\bf S} [\overline{\psi}
\gamma_{\mu_1} D_{\mu_2} \ldots D_{\mu_n} \psi] - {\rm trace~terms}~, \\
\label{COMP3}
O^{\sf S}_{V;\mu_1, \ldots, \mu_n} &=& 2 i^{n-2} {\bf S} {\rm 
\bf Sp}[F_{\mu_1 \alpha}^a
D_{\mu_2} \ldots D_{\mu_{n-1}} F_{\mu_n}^{\alpha,a}] - {\rm trace~terms}~,
\label{eqglO}
\end{eqnarray}
for the fermionic non--singlet, singlet, and gluonic case,~\cite{Geyer:1977gv}.
Here, $\bf S$ denotes the symmetrization operator of the Lorentz indices
$\mu_1, \ldots, \mu_n$; $\lambda_a$ is the flavor matrix of $SU(n_f)$ with
$n_f$ light flavors, $\psi$ denotes the quark field, $F_{\mu\nu}^a$ the gluon
field--strength tensor, and $D_{\mu}$ the covariant derivative. ${\bf Sp}$ in
(\ref{eqglO}) is the color--trace. The quarkonic operator matrix element 
can be represented by
\begin{eqnarray}
A_{qq}^{\sf S} = A_{qq}^{\sf NS} + A_{qq}^{\sf PS}~.
\end{eqnarray}

The different contributions to (\ref{FAC1}) were given in~\cite{Buza:1996wv}, 
Eqs.~(2.31--2.35). To
$O(a_s^3)$, the Wilson coefficients ${\sf H}_i^j$ in Mellin space are~:
\begin{eqnarray}
\label{eqWIL1}
L_{2,q}^{\sf NS}(n_f) &=& 
a_s^2 \left[A_{qq,Q}^{{\sf NS}, (2)}(n_f) +
\hat{C}^{{\sf NS}, (2)}_{2,q}(n_f)\right]
\nonumber
\end{eqnarray}
\begin{eqnarray}
&+&
a_s^3 \left[A_{qq,Q}^{{\sf NS}, (3)}(n_f) 
+  A_{qq,Q}^{{\sf NS}, (2)}(n_f) C_{2,q}^{{\sf NS}, (1)}(n_f)
+ \hat{C}^{{\sf NS}, (3)}_{2,q}(n_f)\right]
\\
\label{eqWIL2}
{\tilde{L}}_{2,q}^{\sf PS}(n_f) &=& 
a_s^3 \left[~\Atil_{qq,Q}^{{\sf PS}, (3)}(n_f) 
+  A_{gq,Q}^{(2)}(n_f)~~ \Ctil_{2,g}^{(1)}(n_f+1)
+ \hat{\Ctil}^{{\sf PS}, (3)}_{2,q}(n_f)\right]
\\
\label{eqWIL3}
{\tilde{L}}_{2,g}^{\sf S}(n_f) &=& 
a_s^2 A_{gg,Q}^{(1)}(n_f) \Ctil_{2,g}^{(1)}(n_f+1)
\nonumber\\ &+&
a_s^3 \Bigl[~\Atil_{qg,Q}^{(3)}(n_f) 
+  A_{gg,Q}^{(1)}(n_f)~~\Ctil_{2,g}^{(2)}(n_f+1)
+  A_{gg,Q}^{(2)}(n_f)~~\Ctil_{2,g}^{(1)}(n_f+1)
\nonumber\\ && \hspace*{5mm}
+  ~A_{Qg}^{(1)}(n_f)~~\Ctil_{2,q}^{{\sf PS}, (2)}(n_f+1)
+ \hat{\Ctil}^{(3)}_{2,g}(n_f)\Bigr]
\\
\label{eqWIL4}
H_{2,q}^{\sf PS}(n_f)
&=& a_s^2 \left[~A_{Qq}^{{\sf PS}, (2)}(n_f) 
+~\Ctil_{2,q}^{{\sf PS}, (2)}(n_f+1)\right]
\nonumber\\
&+& a_s^3 \Bigl[~A_{Qq}^{{\sf PS}, (3)}(n_f)
+~\Ctil_{2,q}^{{\sf PS}, (3)}(n_f+1) 
+ A_{gq,Q}^{(2)}(n_f)~\Ctil_{2,g}^{(1)}(n_f+1) 
\nonumber\\ && \hspace*{5mm}
+ A_{Qq}^{{\sf PS},(2)}(n_f)~C_{2,q}^{{\sf NS}, (1)}(n_f+1) 
\Bigr]
\\
\label{eqWIL5}
H_{2,g}^{\sf S}(n_f) &=& a_s \left[~A_{Qg}^{(1)}(n_f)
+~\Ctil^{(1)}_{2,g}(n_f+1) \right] \nonumber\\ 
&+& a_s^2 \Bigl[~A_{Qg}^{(2)}(n_f) 
+~A_{Qg}^{(1)}(n_f)~C^{{\sf NS}, (1)}_{2,q}(n_f+1)
+~A_{gg,Q}^{(1)}(n_f)~\Ctil^{(1)}_{2,g}(n_f+1)
\nonumber\\ && 
\hspace*{5mm}
+~\Ctil^{(2)}_{2,g}(n_f+1) \Bigr]
\nonumber\\ &+&
a_s^3 \Bigl[~A_{Qg}^{(3)}(n_f) 
+~A_{Qg}^{(2)}(n_f)~C^{{\sf NS}, (1)}_{2,q}(n_f+1)
+~A_{gg,Q}^{(2)}(n_f)~\Ctil^{(1)}_{2,g}(n_f+1)
\nonumber\\ &&
\hspace*{5mm}
+~A_{Qg}^{(1)}(n_f)\left[
C^{{\sf NS}, (2)}_{2,q}(n_f+1)
+~\Ctil^{{\sf PS}, (2)}_{2,q}(n_f+1)\right]
+~A_{gg,Q}^{(1)}(n_f)~\Ctil^{(2)}_{2,g}(n_f+1)
\nonumber\\ && \hspace*{5mm}
+~\Ctil^{(3)}_{2,g}(n_f+1) \Bigr]~.
\end{eqnarray}
For brevity, we have dropped here part of the arguments of the Wilson
coefficients and operator matrix elements by identifying ${\sf H}_i^j = {\sf
  H}_i^j(N,Q^2/\mu^2,\mu^2/m^2,n_f), {C}_i^j = {C}_i^j(N,Q^2/\mu^2,n_f)$ and
${A}_{ij} = {A}_{ij}(N,m^2/\mu^2,n_f)$.  These representations were verified in
the LO and NLO case comparing with the results in \cite{LO,NLO} for $Q^2 \gg
m^2$.

The massive operator matrix elements are calculated keeping the external
massless parton lines on--shell, while the heavy quark mass $m$ sets the
scale.  The massless Wilson coefficients $C_i^j$ in (\ref{eqWIL1}--\ref{eqWIL5}) were 
calculated 
in Refs.~\cite{WIL1,WIL2,Larin:1996wd,Retey:2000nq,Blumlein:2004xt,Vermaseren:2005qc}.
\section{\bf\boldmath Renormalization of the Massive Operator Matrix Elements}
\renewcommand{\theequation}{\thesection.\arabic{equation}}
\setcounter{equation}{0}
\label{Sec-REN}

\vspace{1mm}
\noindent
We perform the calculation of the massive operator matrix elements in $D=4+\ep$ 
dimensions and apply dimensional regularization. For each loop integral a 
factor $S_\ep$
\begin{equation}
S_\ep = \exp \left[\frac{\ep}{2} \left(\gamma_E - \ln(4\pi)\right)\right],
\end{equation}
with $\gamma_E$ the Euler--Mascheroni constant, is obtained which collects
universal terms, and  $S_\ep := 1$ in the $\MS$--scheme. 
The following equation shows the perturbative expansion of
the unrenormalized OMEs, denoted by a double--hat, in the bare coupling
constant $\hat{a}_s$ in Mellin space
\begin{eqnarray}
\hspace*{-6mm}
    \Ahathat_{ij}\Bigl(\frac{\hat{m}^2}{\mu^2},\ep,N\Bigr)&=&
                 \delta_{ij} +         \sum_{l=1}^{\infty}
                         \hat{a}_s^l
                              ~\Ahathat_{ij}^{(l)}
                               \Bigl(\frac{\hat{m}^2}{\mu^2},\ep,N\Bigr) 
= \delta_{ij} +
                        \sum_{l=1}^{\infty}
                        \hat{a}_s^l\Bigl(\frac{\hat{m}^2}{\mu^2}\Bigr)^{l\ep/2}
                              \Ahathat_{ij}^{(l)}
                              \Bigl(\hat{m}^2=\mu^2,\ep,N\Bigr),
                         \label{pertome1}
\end{eqnarray}
with
\begin{eqnarray}
\label{epAhat}
\Ahathat_{ij}^{(l)}\Bigl(\hat{m}^2=\mu^2,\ep,N\Bigr) 
= \sum_{k=0}^\infty \frac{1}{\ep^{l-k}} 
\hspace*{2mm} a_{ij}^{(l,k)}(N)~.
\end{eqnarray}
Here, $N$ is the Mellin--parameter, (\ref{MELTRA}), $\hat{m}$ the bare mass, and $\mu =
\mu_R$ is the renormalization scale. Also the factorization scale $\mu_F$ will
be identified with $\mu$ in the following.

The factorization between the massive OMEs and the massless Wilson coefficients 
(\ref{FAC1}) requires the external legs of the operator matrix 
elements to be on--shell, 
\begin{eqnarray}
\label{OS}
                      p^2 = 0~,
\end{eqnarray}
where $p$ denotes the external momentum. Unlike in the massless case, where
the scale of the OMEs is set by an off--shell momentum
$-p^2 < 0$, in our framework the internal heavy quark mass sets the scale. In
the former case, one observes a mixing of the physical OMEs with non--gauge
invariant (NGI) operators,
cf. \cite{Hamberg:1991qt,Harris:1994tp,Collins:1994ee}, and contributions
originating in the violation of the equations of motion (EOM). Terms of this
kind do not contribute in the present case.

The renormalization of the massive OMEs is performed in four steps. First mass
renormalization is carried out, for which we use the on--mass--shell scheme
and later also compare to the results in the $\MS$--scheme. Afterwards, charge
renormalization is performed in the $\MS$--scheme. To maintain condition
(\ref{OS}), which is of physical importance, we will, however, 
first introduce a
$\MOM$--scheme for the strong coupling constant and then perform a finite
renormalization changing to the $\MS$--scheme. The former scheme is implied by
keeping the external massless parton lines on shell. Note, that there are
other, differing $\MOM$--schemes in the literature, cf.
e.g.~\cite{Chetyrkin:2008jk}. After mass and coupling constant
renormalization, the OMEs are denoted by a single hat, $\hat{A}_{ij}$. The
ultraviolet singularities of the composite operators are canceled via the
corresponding $Z_{ij}$--factors and the UV--finite OMEs are denoted by a 
double tilde, $\tilde{\hspace*{-1mm}\tilde{A}}_{ij}$. Finally, the collinear 
divergences are removed via mass factorization.
\subsection{\bf\boldmath Mass Renormalization}
\label{SubSec-RENMa}

\vspace{1mm}\noindent
There are two main schemes to perform mass renormalization: $i)$ the 
on--shell scheme and $ii)$ the $\MS$--scheme. We will apply the 
on--shell scheme in the 
following, defining the heavy quark mass as the pole mass, and compare to 
the $\MS$--scheme later. The bare mass in  
(\ref{pertome1}) is replaced by the on--shell mass $m$ through
\begin{eqnarray}
\hat{m}&=&Z_m m 
           = m \Bigl[ 1 
                       + \hat{a}_s \Bigl(\frac{m^2}{\mu^2}\Bigr)^{\ep/2}
                                   \delta m_1 
                       + \hat{a}_s^2 \Bigl(\frac{m^2}{\mu^2}\Bigr)^{\ep}
                                     \delta m_2
                 \Bigr] + O(\hat{a}_s^3)~.
            \label{mren1}
\end{eqnarray}
The constants in the above equation are~\footnote{Note that there 
is a misprint in the double--pole term of Eq. (28) in Ref. \cite{Bierenbaum:2008yu}.}
\begin{eqnarray}
    \delta m_1 &=&C_F
                  \left[\frac{6}{\ep}-4+\left(4+\frac{3}{4}\zeta_2\right)\ep
                  \right] \label{delm1}  \\
               &\equiv&  \frac{\delta m_1^{(-1)}}{\ep}
                        +\delta m_1^{(0)}
                        +\delta m_1^{(1)}\ep~, \label{delm1exp} \\
    \delta m_2 &=& C_F
                   \Biggl[\frac{1}{\ep^2}\left(18 C_F-22 C_A+8T_F(n_f+N_h)
                    \right)
                  +\frac{1}{\ep}\left(-\frac{45}{2}C_F+\frac{91}{2}C_A-14T_F
                   (n_f+N_h)\right)
\nonumber\end{eqnarray}\begin{eqnarray} 
&&
                  +C_F\left(\frac{199}{8}-\frac{51}{2}\zeta_2+48\ln(2)\zeta_2
                   -12\zeta_3\right)+C_A\left(-\frac{605}{8}
                  +\frac{5}{2}\zeta_2-24\ln(2)\zeta_2+6\zeta_3\right)
 \N\\ &&
                  +T_F\left[n_f\left(\frac{45}{2}+10\zeta_2\right)+N_h
                  \left(\frac{69}{2}-14\zeta_2\right)\right]\Biggr]
                  \label{delm2}  \\
               &\equiv&  \frac{\delta m_2^{(-2)}}{\ep^2}
                        +\frac{\delta m_2^{(-1)}}{\ep}
                        +\delta m_2^{(0)}~, \label{delm2exp}
\end{eqnarray}
   with $C_F = (N_c^2-1)/(2 N_c), C_A = N_c, T_F =1/2$ for $SU(N_c)$ and
   $N_c = 3$ in case of QCD. $\zeta_k$ denotes the Riemann $\zeta$--function.
   In (\ref{delm2}), $n_f$ denotes the number of light flavors and $N_h$ the
   number of heavy flavors, which we will set equal to one from now on. The
   pole terms were given in \cite{Tarrach:1980up,Nachtmann:1981zg}, and the
   constant term in \cite{Gray:1990yh,Broadhurst:1991fy}, see also
   \cite{Fleischer:1998dw}.  In Eqs. (\ref{delm1exp}, \ref{delm2exp}), we have
   defined the expansion coefficients in $\ep$ of the corresponding
   quantities. The following equation shows the general structure of the OMEs
   up to $O(\hat{a}_s^3)$ after mass renormalization
\begin{eqnarray}
    \Ahathat_{ij}\Bigl(\frac{m^2}{\mu^2},\ep,N\Bigr) 
                 &=& \delta_{ij}
                 +\hat{a}_s~ 
                   \Ahathat_{ij}^{(1)}\Bigl(\frac{m^2}{\mu^2},\ep,N\Bigr) 
\N\\ &&
                         + \hat{a}_s^2 \left[
                                        \Ahathat^{(2)}_{ij}
                                        \Bigl(\frac{m^2}{\mu^2},\ep,N\Bigr) 
                                      + {\delta m_1} 
                                        \Bigl(\frac{m^2}{\mu^2}\Bigr)^{\ep/2}
                                        m \frac{d}{dm} 
                                                   \Ahathat_{ij}^{(1)}
                                           \Bigl(\frac{m^2}{\mu^2},\ep,N\Bigr) 
                                \right]
\N\\ &&
                         + \hat{a}_s^3 \Biggl[ 
                                         \Ahathat^{(3)}_{ij}
                                           \Bigl(\frac{m^2}{\mu^2},\ep,N\Bigr) 
                                        +{\delta m_1} 
                                         \Bigl(\frac{m^2}{\mu^2}\Bigr)^{\ep/2}
                                         m \frac{d}{dm} 
                                                    \Ahathat_{ij}^{(2)}
                                           \Bigl(\frac{m^2}{\mu^2},\ep,N\Bigr)
\N\\ &&
                                        + {\delta m_2} 
                                          \Bigl(\frac{m^2}{\mu^2}\Bigr)^{\ep}
                                          m \frac{d}{dm} 
                                                    \Ahathat_{ij}^{(1)}
                                           \Bigl(\frac{m^2}{\mu^2},\ep,N\Bigr) 
                                        + \frac{\delta m_1^2}{2}
                                          \Bigl(\frac{m^2}{\mu^2}\Bigr)^{\ep}
                                                   m^2  \frac{d^2}{{dm}~^2}
                                                     \Ahathat_{ij}^{(1)}
                                           \Bigl(\frac{m^2}{\mu^2},\ep,N\Bigr) 
    \Biggr]~. 
\nonumber\\
\label{maren}
\end{eqnarray}
  \subsection{\bf\boldmath Renormalization of the Coupling}
   \label{SubSec-RENCo}
  As the next step, we consider charge renormalization. We briefly summarize
  first the main steps in the massless case in the $\MS$--scheme. Afterwards,
  we extend the description to the massive case in the {\sf MOM}-scheme which
  we use, before we transform back to the $\MS$--scheme.

  The bare coupling constant $\hat{a}_s$ is expressed by the renormalized
  coupling $a_s^{\MS}$ via
  \begin{eqnarray}
   \hat{a}_s             &=& {Z_g^{\MS}}^2(\ep,n_f) 
                             a^{\MS}_s(\mu^2) \N\\
                         &=& a^{\MS}_s(\mu^2)\left[
                                   1 
                                 + \delta a^{\MS}_{s, 1}(n_f) 
                                      a^{\MS}_s(\mu^2)
                                 + \delta a^{\MS}_{s, 2}(n_f) 
                                      {a^{\MS}_s}^2(\mu)    
                                     \right] + O({a^{\MS}_s}^3)~. 
                            \label{asrenMSb}
  \end{eqnarray}
The coefficients in Eq. (\ref{asrenMSb}) are, 
\cite{Khriplovich:1969aa,tHooft:unpub,Politzer:1973fx,Gross:1973id} and 
\cite{Caswell:1974gg,Jones:1974mm}, 
  \begin{eqnarray}
    \delta a^{\MS}_{s, 1}(n_f) &=& \frac{2}{\ep} \beta_0(n_f)~,
                             \label{deltasMSb1} \\
    \delta a^{\MS}_{s, 2}(n_f) &=& \frac{4}{\ep^2} \beta_0^2(n_f)
                           + \frac{1}{\ep} \beta_1(n_f)~,
                             \label{deltasMSb2}
  \end{eqnarray}
with 
  \begin{eqnarray}
   \beta_0(n_f)
                 &=& \frac{11}{3} C_A - \frac{4}{3} T_F n_f \label{beta0}~, \\
   \beta_1(n_f)
                 &=& \frac{34}{3} C_A^2 
               - 4 \left(\frac{5}{3} C_A + C_F\right) T_F n_f \label{beta1}~.
  \end{eqnarray}
The evolution equation for the renormalized coupling constant is then given by
  \begin{eqnarray}
   \frac{d a_s(\mu^2)}{d \ln(\mu^2)} 
     &=& \frac{1}{2} \ep a_s(\mu^2)-\sum_{k=0}^\infty \beta_k a_s^{k+2}(\mu^2)~.
         \label{runningas}
  \end{eqnarray}

The factorization relation (\ref{FAC1}) strictly requires that the 
external
massless particles are on shell. Massive loop corrections to the gluon-- and
ghost--propagators violate this condition, which has to be enforced subtracting
the corresponding corrections. They can be uniquely absorbed into the strong
coupling constant applying the background field
method~\cite{Abbott:1980hw,Rebhan:1985yf,Jegerlehner:1998zg}.  Here, $Z_g$ can
be obtained by only considering the gluon propagator. After mass
renormalization in the on--shell scheme via Eq. (\ref{mren1}), we obtain
for the heavy quark contributions to the gluon self--energy
  \begin{eqnarray}
   \hat{\Pi}^{\mu\nu}_{H,ab,\mbox{\tiny{BF}}}(p^2,m^2,\mu^2,\ep,\hat{a}_s)&=&
                                i (-p^2g^{\mu\nu}+p^{\mu}p^{\nu})\delta_{ab}
\hat{\Pi}_{H,\mbox{\tiny{BF}}}(p^2,m^2,\mu^2,\ep,\hat{a}_s)~, \N\\
   \hat{\Pi}_{H,\mbox{\tiny{BF}}}(0,m^2,\mu^2,\ep,\hat{a}_s)&=&
                    \hat{a}_s   \frac{2\beta_{0,Q}}{\ep}
                         \Bigl(\frac{m^2}{\mu^2}\Bigr)^{\ep/2}
                          \exp \Bigl(\sum_{i=2}^{\infty}\frac{\zeta_i}{i}
                          \Bigl(\frac{\ep}{2}\Bigr)^{i}\Bigr)
\N\\ &&
                   +\hat{a}_s^2 \Bigl(\frac{m^2}{\mu^2}\Bigr)^{\ep}
                        \Biggl[
                       \frac{1}{\ep}\Bigl(
                                          -\frac{20}{3}T_FC_A
                                          -4T_FC_F
                                    \Bigr)
                      -\frac{32}{9}T_FC_A
                      +15T_FC_F
\N\\ &&
                     +\ep            \Bigl(
                                          -\frac{86}{27}T_FC_A
                                          -\frac{31}{4}T_FC_F
                                          -\frac{5}{3}\zeta_2T_FC_A
                                          -\zeta_2T_FC_F
                                   \Bigr)
                         \Biggl] + O(\hat{a}_s^3)~. \label{GluSelfBack}
\nonumber\\  
\end{eqnarray}
  Note, that although the $O(\hat{a}_s)$--term in the above formula is an
  expression to all orders in $\ep$, the $O(\hat{a}_s^2)$--term and hence the
  formula in general only holds up to  $O(\ep)$. We have used the Feynman
  rules of the background field formalism as given in Ref. \cite{YND}.  In the
  following, we define
  \begin{eqnarray}
   f(\ep)&\equiv&
                 \Bigl(\frac{m^2}{\mu^2}\Bigr)^{\ep/2}
    \exp \Bigl(\sum_{i=2}^{\infty}\frac{\zeta_i}{i}
                       \Bigl(\frac{\ep}{2}\Bigr)^{i}\Bigr)~. \label{fep}
  \end{eqnarray}
  The renormalization constant of the background field $Z_A$ 
  is related to $Z_g$ via 
  \begin{eqnarray} 
   Z_A=Z_g^{-2}~. \label{ZAZg}
  \end{eqnarray}
  The light--flavor contributions to $Z_A$, $Z_{A,l}$, can thus be determined
  by combining Eqs. (\ref{asrenMSb}) and (\ref{ZAZg}). The heavy flavor part,
  $Z_{A,H}$, follows from the condition
  \begin{eqnarray}
   \Pi_{H,\mbox{\tiny{BF}}}(0,\mu^2, a_s, m^2)+Z_{A,H}\equiv 0~, 
\label{ZAcond}
  \end{eqnarray}
which ensures that the on--shell gluon remains strictly massless.  
Thus we define the renormalization constant 
  of the strong coupling with $n_f$ light and one heavy flavor as 
  \begin{eqnarray}
   Z^{\MOM}_g(\ep,n_f+1,\mu,m)
         \equiv \frac{1}{(Z_{A,l}+Z_{A,H})^{1/2}}~ \label{Zgnfp1}
  \end{eqnarray}
  and obtain
  \begin{eqnarray}
   {Z_g^{\MOM}}^2(\ep,m,\mu,n_f+1)&=&
                  1+a^{\MOM}_s(\mu^2) \Bigl[
                              \frac{2}{\ep} (\beta_0(n_f)+\beta_{0,Q}f(\ep))
                        \Bigr]
\N\\ &&
                  +{a^{\MOM}_s}^2(\mu^2) \Bigl[
                                \frac{\beta_1(n_f)}{\ep}
                         +\frac{4}{\ep^2} (\beta_0(n_f)+\beta_{0,Q}f(\ep))^2
\N\\ &&
                          +\frac{1}{\ep}\Bigl(\frac{m^2}{\mu^2}\Bigr)^{\ep}
                           \Bigl(\beta_{1,Q}+\ep\beta_{1,Q}^{(1)}
                                            +\ep^2\beta_{1,Q}^{(2)}
                           \Bigr)
                          \Bigr]+O(\ep^2, {a^{\MOM}_s}^3)~, \label{Zgheavy2}
  \end{eqnarray}
  with
  \begin{eqnarray}
   \beta_{0,Q} &=&-\frac{4}{3}T_F~, \label{b0Q} \\
   \beta_{1,Q} &=&- 4 \left(\frac{5}{3} C_A + C_F \right) T_F~, \label{b1Q} \\
   \beta_{1,Q}^{(1)}&=&
                           -\frac{32}{9}T_FC_A
                           +15T_FC_F~, \label{b1Q1} \\
   \beta_{1,Q}^{(2)}&=&
                               -\frac{86}{27}T_FC_A
                               -\frac{31}{4}T_FC_F
                               -\zeta_2\left(\frac{5}{3}T_FC_A
                                        +T_FC_F\right)~. \label{b1Q2}
  \end{eqnarray}
  The coefficients corresponding to Eq. (\ref{asrenMSb}) expressed in the
  {\sf MOM}--scheme read
  \begin{eqnarray}
   \delta a_{s,1}^{\MOM}&=&\Bigl[\frac{2\beta_0(n_f)}{\ep}
                           +\frac{2\beta_{0,Q}}{\ep}f(\ep)
                            \Bigr]~,\label{dela1} \\
   \delta a_{s,2}^{\MOM}&=&\Bigl[\frac{\beta_1(n_f)}{\ep}+
                            \Bigl\{\frac{2\beta_0(n_f)}{\ep}
                              +\frac{2\beta_{0,Q}}{\ep}f(\ep)\Bigr\}^2
                          +\frac{1}{\ep}\Bigl(\frac{m^2}{\mu^2}\Bigr)^{\ep}
                           \Bigl(\beta_{1,Q}+\ep\beta_{1,Q}^{(1)}
                                            +\ep^2\beta_{1,Q}^{(2)}
                           \Bigr)\Bigr] + O(\ep^2)~.\label{dela2}
\nonumber\\
  \end{eqnarray}

Since the $\MS$--scheme is commonly used, we transform our results back from
the {\sf MOM}--description into the $\MS$--scheme, in order to be able to
compare to other analyzes. This is achieved by observing that the bare
coupling does not change under this transformation and one thus obtains the
condition
  \begin{eqnarray}
      {Z_g^{\MS}}^2(\ep,n_f+1) a^{\MS}_s(\mu^2) = 
      {Z_g^{\MOM}}^2(\ep,m,\mu,n_f+1) a^{\MOM}_s(\mu^2) \label{condas1}~.
  \end{eqnarray}
The following relations hold~:  
  \begin{eqnarray}
   a_s^{\MOM}&=& a_s^{\MS}
                -\beta_{0,Q}\ln \Bigl(\frac{m^2}{\mu^2}\Bigr) {a_s^{\MS}}^2
                +\Biggl[ \beta^2_{0,Q}\ln^2 \Bigl(\frac{m^2}{\mu^2}\Bigr) 
                        -\beta_{1,Q}\ln \Bigl(\frac{m^2}{\mu^2}\Bigr) 
                        -\beta_{1,Q}^{(1)}
                 \Biggr] {a_s^{\MS}}^3~, \label{asmoma}
  \end{eqnarray}
  or, 
  \begin{eqnarray}
   a_s^{\MS}&=&
               a_s^{\MOM}
              +{a_s^{\MOM}}^2\Biggl(
                          \delta a^{\MOM}_{s, 1}
                         -\delta a^{\MS}_{s, 1}(n_f+1)
                             \Biggr)
              +{a_s^{\MOM}}^{3}\Biggl(
                          \delta a^{\MOM}_{s, 2}
                         -\delta a^{\MS}_{s, 2}(n_f+1)
     \N\\ &&
                        -2\delta a^{\MS}_{s, 1}(n_f+1)\Bigl[
                             \delta a^{\MOM}_{s, 1}
                            -\delta a^{\MS}_{s, 1}(n_f+1)
                                                      \Bigr]
                             \Biggr)+O({a_s^{\MOM}}^4)~, \label{asmsa}
  \end{eqnarray}
vice versa. 
Eq.~(\ref{asmsa}) is valid to all orders in $\ep$. Here, $a_s^{\sf \MS} = a_s^{\sf \MS}(n_f + 1)$. 
  Applying the on--shell scheme for mass renormalization 
  and the described {\sf MOM}--scheme for the renormalization of the 
  coupling, one obtains as general formula for mass and coupling constant
  renormalization up to $O({a^{\MOM}_s}^3)$
  \begin{eqnarray}
   {\hat{A}}_{ij} &=&  \delta_{ij} 
                     + a^{\MOM}_s \Ahathat_{ij}^{(1)}
                     + {a^{\MOM}_s}^2 \left[\Ahathat^{(2)}_{ij}
                     + \delta m_1 \Bigl(\frac{m^2}{\mu^2}\Bigr)^{\ep/2} 
                                 m  \frac{d}{dm} \Ahathat_{ij}^{(1)}
                     + \delta a^{\MOM}_{s,1} \Ahathat_{ij}^{(1)}\right]
\N\\ &&
   + {a^{\MOM}_s}^3 \Biggl[ \Ahathat^{(3)}_{ij}
   + \delta a^{\MOM}_{s,2} \Ahathat_{ij}^{(1)}
   + 2 \delta a^{\MOM}_{s,1} \left( \Ahathat^{(2)}_{ij} 
   +  \delta m_1 \Bigl(\frac{m^2}{\mu^2}\Bigr)^{\ep/2}
      m \frac{d}{dm} \Ahathat_{ij}^{(1)} \right)
\N\\ &&
                + \delta m_1 \Bigl(\frac{m^2}{\mu^2}\Bigr)^{\ep/2}
                           m  \frac{d}{dm} \Ahathat_{ij}^{(2)}
                + \delta m_2 \Bigl(\frac{m^2}{\mu^2}\Bigr)^{\ep}
                          m  \frac{d}{dm} \Ahathat_{ij}^{(1)}
                + \frac{\delta m_1^2}{2} \Bigl(\frac{m^2}{\mu^2}\Bigr)^{\ep}
                m^2  \frac{d^2}{{dm}~^2} \Ahathat_{ij}^{(1)}
    \Biggr]~,\label{macoren}
  \end{eqnarray}
  where we have suppressed the dependence on $m,~\ep$ and $N$ in the 
arguments.~\footnote{Here we corrected a typographical error in \cite{Bierenbaum:2008yu}, 
Eq.~(48).} 
\subsection{\bf\boldmath Operator Renormalization}
\label{SubSec-RENOp}

\vspace{1mm}\noindent
The renormalization of the ultra-violet (UV) singularities of the composite 
operators is done introducing the corresponding $Z_{ij}$-factors. 
We consider first the case of $n_f$ massless flavors, 
cf.~\cite{Matiounine:1998ky},   
   \begin{eqnarray}
    A_{qq}^{\sf NS}\Bigl(\frac{-p^2}{\mu^2},a_s^{\MS},n_f,N\Bigr)
           &=&Z^{-1,{\sf NS}}_{qq}(a_s^{\MS},n_f,\ep,N)
              \hat{A}_{qq}^{\sf NS}\Bigl(\frac{-p^2}{\mu^2},a_s^{\MS},n_f,\ep,N\Bigr)
              \label{renAqqnf}
\\
    A_{ij}\Bigl(\frac{-p^2}{\mu^2},a_s^{\MS},n_f,N\Bigr)
           &=&Z^{-1}_{il}(a_s^{\MS},n_f,\ep,N)
              \hat{A}_{lj}\Bigl(\frac{-p^2}{\mu^2},a_s^{\MS},n_f,\ep,N\Bigr)
                   ~,~i,j,l=q,g,
              \label{renAijnf}
   \end{eqnarray}
   for the non--singlet and singlet case, with $p$ a space-like momentum.
   As mentioned before, we neglected all terms being associated to EOM
   and NGI parts, since they do not contribute in the renormalization of the
   massive on-shell operator matrix elements. The ${\sf NS}$ and ${\sf PS}$
   contributions are separated via
   \begin{eqnarray}
    Z_{qq}^{-1}&=&Z_{qq}^{-1, {\sf PS}}+Z_{qq}^{-1, {\sf NS}} 
\label{ZPSNS1}~,\\
    A_{qq}     &=&A_{qq}^{\sf PS}+A_{qq}^{\sf NS} \label{ZPSNS2}~. 
   \end{eqnarray}
   The anomalous dimensions $\gamma_{ij}$ of the operators are then given by 
   \begin{eqnarray}
    \gamma_{qq}^{\sf NS}(a_s^{\MS},n_f,N)&=&
                           \mu \frac{d}{d\mu} \ln Z_{qq}^{\sf 
NS}(a_s^{\MS},n_f,\ep,N)~,
                                               \label{gammazetNS}\\
    \gamma_{ij}(a_s^{\MS},n_f,N)&=&
                            Z^{-1}_{il}(a_s^{\MS},n_f,\ep,N)~
                           \mu \frac{d}{d\mu} 
Z_{lj}(a_s^{\MS},n_f,\ep,N)~.
                                               \label{gammazetS}
   \end{eqnarray}
   They can be expanded into a perturbative series as follows 
   \begin{eqnarray}
    \gamma_{ij}^{{\sf S,~PS,~NS}}(a_s^{\MS},n_f,N)
        &=&\sum_{l=1}^{\infty}{a^{\MS}_s}^l 
        \gamma_{ij}^{(l), {\sf S,~PS,~NS}}(n_f,N)~, 
            \label{pertgamma}
   \end{eqnarray}
   where the ${\sf PS}$ contribution starts at $O(a_s^2)$. The 
   anomalous dimensions are known for all $N$ at 
   ${\sf LO}$, \cite{Gross:1973ju,Georgi:1951sr}, and 
   ${\sf NLO}$, \cite{Floratos:1977au,GonzalezArroyo:1979df,
   Curci:1980uw,GonzalezArroyo:1979he,Floratos:1978ny,
   Furmanski:1980cm,Hamberg:1991qt}. Fixed 
   moments at ${\sf NNLO}$ have been calculated in 
   Refs. \cite{Larin:1996wd,Retey:2000nq,Blumlein:2004xt} and 
   the complete result has been obtained in Refs. 
\cite{Vogt:2004mw,Moch:2004pa}. 
   At the level of twist--$2$, they are connected to the 
   splitting functions, \cite{Altarelli:1977zs}, by a 
Mellin--transform~\footnote{
Due to our convention, Eqs. 
(\ref{gammazetNS}, \ref{gammazetS}), there is
   a relative factor of $2$ between the anomalous dimensions 
   considered in this work and Refs. \cite{Moch:2004pa,Vogt:2004mw}.}
   \begin{eqnarray}
    \gamma_{ij}^{(k)}(n_f,N) = - \int_0^1 dz z^{N-1} P_{ij}^{(k)}(n_f,z)~. 
\label{splitan}
   \end{eqnarray}
   In the following, we do not write the dependence on the Mellin--variable
   $N$ for the OMEs, the operator $Z$--factors and the anomalous dimensions
   explicitly. Furthermore, we will suppress the dependence on $\ep$ for
   unrenormalized quantities and $Z$--factors.  From Eqs. (\ref{gammazetNS},
   \ref{gammazetS}), one can determine the relation between the anomalous
   dimensions and the $Z$--factors order by order in perturbation theory. In
   the general case, one finds
  \begin{eqnarray}
   Z_{ij}(a^{\MS}_s,n_f) &=&
                            \delta_{ij}
                           +a^{\MS}_s \frac{\gamma_{ij}^{(0)}}{\ep}
                           +{a^{\MS}_s}^2 \Biggl\{
                                 \frac{1}{\ep^2} \Bigl(
                                     \frac{1}{2} \gamma_{il}^{(0)}
                                                 \gamma_{lj}^{(0)}
                                   + \beta_0 \gamma_{ij}^{(0)}
                                                 \Bigr)
                               + \frac{1}{2 \ep} \gamma_{ij}^{(1)}
                                   \Biggr\}
\nonumber 
\end{eqnarray}\begin{eqnarray}
&&
                           + {a^{\MS}_s}^3 \Biggl\{
                                 \frac{1}{\ep^3} \Bigl(
                                     \frac{1}{6}\gamma_{il}^{(0)}
                                                \gamma_{lk}^{(0)}
                                                \gamma_{kj}^{(0)}
                                   + \beta_0 \gamma_{il}^{(0)} 
                                             \gamma_{lj}^{(0)}
                                   + \frac{4}{3} \beta_0^2 \gamma_{ij}^{(0)}
                                                  \Bigr)
\N\\ &&
                               + \frac{1}{\ep^2}  \Bigl(
                                     \frac{1}{6} \gamma_{il}^{(1)} 
                                                 \gamma_{lj}^{(0)}
                                   + \frac{1}{3} \gamma_{il}^{(0)} 
                                                 \gamma_{lj}^{(1)}
                                   + \frac{2}{3} \beta_0 \gamma_{ij}^{(1)} 
                                   + \frac{2}{3} \beta_1 \gamma_{ij}^{(0)}
                                                  \Bigr)
                              + \frac{\gamma_{ij}^{(2)}}{3 \ep}
                                   \Biggr\}~. \label{Zijnf}
  \end{eqnarray}
  The ${\sf NS}$ and ${\sf PS}$ $Z$--factors are given by~\footnote{
In Eq.~(\ref{ZqqPSnf})
we corrected typographical errors contained in Eq. (34), \cite{Bierenbaum:2008yu}.}
  \begin{eqnarray}
   Z_{qq}^{\sf NS}(a^{\MS}_s,n_f) &=& 
                             1 
                           +a^{\MS}_s \frac{\gamma_{qq}^{(0),{\sf NS}}}{\ep}
                           +{a^{\MS}_s}^2 \Biggl\{
                                 \frac{1}{\ep^2} \Bigl(
                                     \frac{1}{2}{\gamma_{qq}^{(0),{\sf NS}}}^2 
                                   + \beta_0 \gamma_{qq}^{(0),{\sf NS}}
                                                 \Bigr)
                              + \frac{1}{2 \ep} \gamma_{qq}^{(1),{\sf NS}} 
                                        \Biggr\}
\N\\ &&
                           +{a^{\MS}_s}^3 \Biggl\{
                                 \frac{1}{\ep^3} \Bigl(
                                     \frac{1}{6} {\gamma_{qq}^{(0),{\sf NS}}}^3
                                   + \beta_0 {\gamma_{qq}^{(0),{\sf NS}}}^2 
                                   + \frac{4}{3} \beta_0^2 
                                                 \gamma_{qq}^{(0),{\sf NS}}
                                                 \Bigr)
\\
   Z_{qq}^{\sf PS}(a^{\MS}_s,n_f) &=&
                            {a^{\MS}_s}^2 \Biggl\{
                                 \frac{1}{2\ep^2} \gamma_{qg}^{(0)}
                                                  \gamma_{gq}^{(0)}   
                               + \frac{1}{2\ep}   \gamma_{qq}^{(1), {\sf PS}}
                                        \Biggr\}
                           +{a^{\MS}_s}^3 \Biggl\{
                                 \frac{1}{\ep^3} \Bigl(
                                     \frac{1}{3}\gamma_{qq}^{(0)} 
                                      \gamma_{qg}^{(0)}
                                      \gamma_{gq}^{(0)}
\N\\ &&
                                    +\frac{1}{6}\gamma_{qg}^{(0)}
                                     \gamma_{gg}^{(0)} \gamma_{gq}^{(0)}
                                    +\beta_0 \gamma_{qg}^{(0)}
                                             \gamma_{gq}^{(0)}
                                                 \Bigr)
                               + \frac{1}{\ep^2} \Bigl(
                                     \frac{1}{3}\gamma_{qg}^{(0)} 
                                                \gamma_{gq}^{(1)}
\N\\ &&
                                    +\frac{1}{6}\gamma_{qg}^{(1)} 
                                                \gamma_{gq}^{(0)}
                                    +\frac{1}{2} \gamma_{qq}^{(0)}
                                                 \gamma_{qq}^{(1), {\sf PS}}
                                    +\frac{2}{3} \beta_0
                                                 \gamma_{qq}^{(1), {\sf PS}}
                                                 \Bigr)
                                    +\frac{\gamma_{qq}^{(2), {\sf PS}}}{3\ep} 
                                        \Biggr\}~. \label{ZqqPSnf}
  \end{eqnarray}
  All quantities in Eqs. (\ref{Zijnf}--\ref{ZqqPSnf}) refer to 
  $n_f$ light flavors and renormalize 
  the massless off--shell OMEs given in Eqs. 
  (\ref{renAqqnf}, \ref{renAijnf}). 
  
  In the next step, we consider an additional heavy quark with mass $m$. 
  We keep the external momentum artificially
  off--shell for the moment, in order to deal with the 
  UV--singularities only.
  For the additional massive quark, one has to account for the prescription of
  the renormalization of the coupling constant we used in 
  Eqs.~(\ref{dela1}, \ref{dela2}). The $Z$--factors including one massive 
  quark are then obtained
  by taking Eqs. (\ref{Zijnf}-\ref{ZqqPSnf}) at $n_f+1$ flavors and performing
  the scheme transformation given in (\ref{asmsa}). The
  emergence of $\delta a_{s,k}^{\sf MOM}$ in $Z_{ij}$ is due to the finite 
  mass effects and cancels singularities which emerge for real radiation and 
  virtual processes at $p^2 \rightarrow 0$.  Thus one obtains
  \begin{eqnarray}
   Z_{ij}^{-1}(a_s^{\MOM},n_f+1,\mu)&=&
      \delta_{ij}
     -a_s^{\MOM}\frac{\gamma_{ij}^{(0)}}{\ep}
     +{a^{\MOM}_s}^2\Biggl[
          \frac{1}{\ep}\Bigl(
                       -\frac{1}{2}\gamma_{ij}^{(1)}
                       -\delta a^{\MOM}_{s,1}\gamma_{ij}^{(0)}
                       \Bigr)
         +\frac{1}{\ep^2}\Bigl(
                        \frac{1}{2}\gamma_{il}^{(0)}\gamma_{lj}^{(0)}
\N\\ &&
                       +\beta_0\gamma_{ij}^{(0)}
                        \Bigr)
         \Biggr]
     +{a^{\MOM}_s}^3\Biggl[
          \frac{1}{\ep}\Bigl(
                       -\frac{1}{3}\gamma_{ij}^{(2)}
                       -\delta a^{\MOM}_{s,1}\gamma_{ij}^{(1)}
                       -\delta a^{\MOM}_{s,2}\gamma_{ij}^{(0)}
                       \Bigr)
\N\\ &&
         +\frac{1}{\ep^2}\Bigl(
                        \frac{4}{3}\beta_0\gamma_{ij}^{(1)}
                       +2\delta a^{\MOM}_{s,1}\beta_0\gamma_{ij}^{(0)}
                       +\frac{1}{3}\beta_1\gamma_{ij}^{(0)}
                       +\delta a^{\MOM}_{s,1}\gamma_{il}^{(0)}\gamma_{lj}^{(0)}
                       +\frac{1}{3}\gamma_{il}^{(1)}\gamma_{lj}^{(0)}
\N\\ &&
                       +\frac{1}{6}\gamma_{il}^{(0)}\gamma_{lj}^{(1)}
                        \Bigr)
         +\frac{1}{\ep^3}\Bigl(
                       -\frac{4}{3}\beta_0^{2}\gamma_{ij}^{(0)}
                       -\beta_0\gamma_{il}^{(0)}\gamma_{lj}^{(0)}
                       -\frac{1}{6}\gamma_{il}^{(0)}\gamma_{lk}^{(0)}
                                    \gamma_{kj}^{(0)} 
                     \Bigr)
       \Biggr]~, \label{ZijInfp1}
  \end{eqnarray}
and
  \begin{eqnarray}
   Z_{qq}^{-1,{\sf NS}}(a_s^{\MOM},n_f+1)&=&
      1
     -a^{\MOM}_s\frac{\gamma_{qq}^{(0),{\sf NS}}}{\ep}
     +{a^{\MOM}_s}^2\Biggl[
          \frac{1}{\ep}\Bigl(
                       -\frac{1}{2}\gamma_{qq}^{(1),{\sf NS}}
                       -\delta a^{\MOM}_{s,1}\gamma_{qq}^{(0),{\sf NS}}
                       \Bigr)
\nonumber
\end{eqnarray}\begin{eqnarray}
&&
         +\frac{1}{\ep^2}\Bigl(
                        \beta_0\gamma_{qq}^{(0),{\sf NS}}
                       +\frac{1}{2}{\gamma_{qq}^{(0),{\sf NS}}}^{2}
                        \Bigr)
         \Biggr]
     +{a^{\MOM}_s}^3\Biggl[
          \frac{1}{\ep}\Bigl(
                       -\frac{1}{3}\gamma_{qq}^{(2),{\sf NS}}
                       -\delta a^{\MOM}_{s,1}\gamma_{qq}^{(1),{\sf NS}}
\N\\  && 
                       -\delta a^{\MOM}_{s,2}\gamma_{qq}^{(0),{\sf NS}}
                       \Bigr)
         +\frac{1}{\ep^2}\Bigl(
                       \frac{4}{3}\beta_0\gamma_{qq}^{(1),{\sf NS}}
                      +2\delta a^{\MOM}_{s,1}\beta_0\gamma_{qq}^{(0),{\sf NS}}
                       +\frac{1}{3}\beta_1\gamma_{qq}^{(0),{\sf NS}}
\N\\  && 
                     +\frac{1}{2}\gamma_{qq}^{(0),{\sf NS}}
                                   \gamma_{qq}^{(1),{\sf NS}}
                       +\delta a^{\MOM}_{s,1}{\gamma_{qq}^{(0),{\sf NS}}}^{2}
                        \Bigr)
         +\frac{1}{\ep^3}\Bigl(
                       -\frac{4}{3}\beta_0^{2}\gamma_{qq}^{(0),{\sf NS}}
                       -\beta_0{\gamma_{qq}^{(0),{\sf NS}}}^{2} \N\\ &&
                       -\frac{1}{6}{\gamma_{qq}^{(0),{\sf NS}}}^{3}
                     \Bigr)
       \Biggr]~, \label{ZNSInfp1} %
\\
   Z_{qq}^{-1,{\sf PS}}(a_s^{\MOM},n_f+1)&=&
      {a^{\MOM}_s}^2\Biggl[
          \frac{1}{\ep}\Bigl(
                       -\frac{1}{2}\gamma_{qq}^{(1), {\sf PS}}
                       \Bigr)
         +\frac{1}{\ep^2}\Bigl(
                        \frac{1}{2}\gamma_{qg}^{(0)}\gamma_{gq}^{(0)}
                        \Bigr)
         \Biggr]
     +{a^{\MOM}_s}^3\Biggl[
          \frac{1}{\ep}\Bigl(
                       -\frac{1}{3}\gamma_{qq}^{(2), {\sf PS}}
 \N\\  &&
                       -\delta a^{\MOM}_{s,1}\gamma_{qq}^{(1), {\sf PS}}
                       \Bigr)
         +\frac{1}{\ep^2}\Bigl(
                        \frac{1}{6}\gamma_{qg}^{(0)}\gamma_{gq}^{(1)}
                       +\frac{1}{3}\gamma_{gq}^{(0)}\gamma_{qg}^{(1)}
                       +\frac{1}{2}\gamma_{qq}^{(0)}\gamma_{qq}^{(1), {\sf PS}}
 \N\\  &&
                       +\frac{4}{3}\beta_0\gamma_{qq}^{(1), {\sf PS}}
                       +\delta a^{\MOM}_{s,1}\gamma_{qg}^{(0)}\gamma_{gq}^{(0)}
                        \Bigr)
         +\frac{1}{\ep^3}\Bigl(
                       -\frac{1}{3}\gamma_{qg}^{(0)}\gamma_{gq}^{(0)}
                                   \gamma_{qq}^{(0)}
                       -\frac{1}{6}\gamma_{gq}^{(0)}\gamma_{qg}^{(0)}
                                   \gamma_{gg}^{(0)}
  \N\\  &&
                       -\beta_0\gamma_{qg}^{(0)}\gamma_{gq}^{(0)}
                     \Bigr)
       \Biggr]~. \label{ZPSInfp1}
  \end{eqnarray} 
  The above equations are given for $n_f+1$ flavors.
  One rederives the expressions for $n_f$ light flavors 
  by setting $(n_f+1) =: n_f$ and $\delta a^{\MOM}_s=\delta a^{\MS}_s$.
  As a next step, we split the OMEs into
  a part involving only light flavors and the heavy flavor part 
  \begin{eqnarray}
\hspace*{-4mm}
    {\hat{A}}_{ij}(p^2,m^2,\mu^2,a_s^{\MOM},n_f+1)&=&
                {\hat{A}}_{ij}\Bigl(\frac{-p^2}{\mu^2},a_s^{\MS},n_f\Bigr)+
                {\hat{A}}^Q_{ij}(p^2,m^2,\mu^2,a_s^{\MOM},n_f+1)~.
\label{splitNSHL1}
  \end{eqnarray}
  In (\ref{splitNSHL1}, \ref{eqXX}), the light--flavor part depends on 
  $a_s^{\MS}$, since
  the prescription adopted for coupling constant renormalization only applies
  to the massive part. ${\hat{A}}^Q_{ij}$ denotes any massive OME we consider.
  The correct UV--renormalization prescription for the massive contribution 
  is obtained
  by subtracting from Eq. (\ref{splitNSHL1}) the terms applying to the light
  part only ~:
  \begin{eqnarray}
   \Atiltil^Q_{ij}(p^2,m^2,\mu^2,a_s^{\MOM},n_f+1)&=&
               Z^{-1}_{il}(a_s^{\MOM},n_f+1,\mu) 
                  \hat{A}^Q_{ij}(p^2,m^2,\mu^2,a_s^{\MOM},n_f+1)
\N\\ &&
              +Z^{-1}_{il}(a_s^{\MOM},n_f+1,\mu) 
                   \hat{A}_{ij}\Bigl(\frac{-p^2}{\mu^2},a_s^{\MS},n_f\Bigr)
\N\\ &&
              -Z^{-1}_{il}(a_s^{\MS},n_f,\mu)
                   \hat{A}_{ij}\Bigl(\frac{-p^2}{\mu^2},a_s^{\MS},n_f\Bigr)~,
\label{eqXX} 
 \end{eqnarray}
where
  \begin{eqnarray}
Z_{ij}^{-1} = \delta_{ij} + \sum_{k=1}^\infty a_s^k Z_{ij}^{-1, (k)}~.
  \end{eqnarray}
  In the limit $p^2=0$, integrals without a scale vanish within dimensional
  regularization. Hence for the light--flavor OMEs only the term 
  $\delta_{ij}$ remains and
  one obtains after expanding in $a_s$
  \begin{eqnarray}
\Atiltil^Q_{ij}\Bigl(\frac{m^2}{\mu^2},a_s^{\MOM},n_f+1\Bigr) &=& 
            a_s^{\MOM}\Biggl( \hat{A}_{ij}^{(1),Q}
                              \Bigl(\frac{m^2}{\mu^2}\Bigr)
                     +Z^{-1,(1)}_{ij}(n_f+1,\mu)
                     -Z^{-1,(1)}_{ij}(n_f)
               \Biggr)\nonumber
\end{eqnarray}\begin{eqnarray}
&& + {a_s^{\MOM}}^2\Biggl( \hat{A}_{ij}^{(2),Q}
                                    \Bigl(\frac{m^2}{\mu^2}\Bigr)
                       +Z^{-1,(2)}_{ij}(n_f+1,\mu)
                       -Z^{-1,(2)}_{ij}(n_f)
\nonumber\\ &&                  
     +Z^{-1,(1)}_{ik}(n_f+1,\mu)
                        \hat{A}_{kj}^{(1),Q}\Bigl(\frac{m^2}{\mu^2}\Bigr)
               \Biggr)
\N\\ &&
           +{a_s^{\MOM}}^3\Biggl( \hat{A}_{ij}^{(3),Q}
                                   \Bigl(\frac{m^2}{\mu^2}\Bigr)
                       +Z^{-1,(3)}_{ij}(n_f+1,\mu)
                       -Z^{-1,(3)}_{ij}(n_f)
                       +Z^{-1,(1)}_{ik}(n_f+1,\mu)
                        \hat{A}_{kj}^{(2),Q}\Bigl(\frac{m^2}{\mu^2}\Bigr)
  \N\\ &&\phantom{{a_s^{\MOM}}^3\Biggl(}
                       +Z^{-1,(2)}_{ik}(n_f+1,\mu)
                        \hat{A}_{kj}^{(1),Q}\Bigl(\frac{m^2}{\mu^2}\Bigr)
                        \Biggr)~. \label{GenRen1}
  \end{eqnarray}
  The $Z$--factors at $n_f+1$ flavors refer
  to Eqs. (\ref{ZijInfp1}--\ref{ZPSInfp1}), whereas those 
  at $n_f$ flavors correspond to the massless case. 
\subsection{\bf\boldmath Mass Factorization}
\label{SubSec-RENOp1}

\vspace{1mm}\noindent
  Finally, we have to remove the collinear singularities contained in 
  \hspace*{1.5mm} $\Atiltil_{ij}$, which emerge in the limit $p^2 =  0$. They  
  are absorbed into the parton distribution functions.
  As a generic renormalization formula, generalizing Eqs. (\ref{renAqqnf}, 
  \ref{renAijnf}), one finds
  \begin{eqnarray}
   A_{ij}&=&Z^{-1}_{il} \hat{A}_{lk} \Gamma_{kj}^{-1}~. \label{genren}
  \end{eqnarray}
  The renormalized operator matrix elements are obtained by 
  \begin{eqnarray}
   A^Q_{ij}\Bigl(\frac{m^2}{\mu^2},a_s^{\MOM},n_f+1\Bigr)&=&
     \Atiltil^Q_{il}\Bigl(\frac{m^2}{\mu^2},a_s^{\MOM},n_f+1\Bigr)
     \Gamma_{lj}^{-1}~. \label{genren1}
  \end{eqnarray}
  If all quarks were  massless, the identity, \cite{Buza:1995ie},  
  \begin{eqnarray}
    \Gamma_{ij} = Z^{-1}_{ij}~. \label{GammaZ}
  \end{eqnarray}
  would hold.  However, due to the presence of a heavy quark $Q$, the
  transition functions $\Gamma(n_f)$ refer only to massless sub-graphs.  Hence
  the $\Gamma$--factors contribute up to $O(a_s^2)$ only and do not involve
  the special scheme adopted for the renormalization of the coupling. Due to
  Eq. (\ref{GammaZ}), they can be read off from
  Eqs. (\ref{Zijnf}--\ref{ZqqPSnf}).

The renormalized operator matrix elements are then given by:
  \begin{eqnarray}
   && A^Q_{ij}\Bigl(\frac{m^2}{\mu^2},a_s^{\MOM},n_f+1\Bigr)=
\N\\&&\phantom{+}
                a^{\MOM}_s~\Biggl(
                      \hat{A}_{ij}^{(1),Q}\Bigl(\frac{m^2}{\mu^2}\Bigr)
                     +Z^{-1,(1)}_{ij}(n_f+1)
                     -Z^{-1,(1)}_{ij}(n_f)
                           \Biggr)
\N\\&&
           +{a^{\MOM}_s}^2\Biggl( 
                        \hat{A}_{ij}^{(2),Q}\Bigl(\frac{m^2}{\mu^2}\Bigr)
                       +Z^{-1,(2)}_{ij}(n_f+1)
                       -Z^{-1,(2)}_{ij}(n_f)
                       +Z^{-1,(1)}_{ik}(n_f+1)\hat{A}_{kj}^{(1),Q}
                                              \Bigl(\frac{m^2}{\mu^2}\Bigr)
\N\\
&&\phantom{+{a^{\MOM}_s}^2\Biggl(}
                       +\Bigl[ \hat{A}_{il}^{(1),Q}
                               \Bigl(\frac{m^2}{\mu^2}\Bigr)
                              +Z^{-1,(1)}_{il}(n_f+1)
                              -Z^{-1,(1)}_{il}(n_f)
                        \Bigr] 
                             \Gamma^{-1,(1)}_{lj}(n_f)
                         \Biggr)
\N\\ 
&&
          +{a^{\MOM}_s}^3\Biggl( 
                        \hat{A}_{ij}^{(3),Q}\Bigl(\frac{m^2}{\mu^2}\Bigr)
                       +Z^{-1,(3)}_{ij}(n_f+1)
                       -Z^{-1,(3)}_{ij}(n_f)
                       +Z^{-1,(1)}_{ik}(n_f+1)\hat{A}_{kj}^{(2),Q}
                                              \Bigl(\frac{m^2}{\mu^2}\Bigr)
\nonumber\end{eqnarray}\begin{eqnarray}
&&\phantom{+{a^{\MOM}_s}^3\Biggl(}
                       +Z^{-1,(2)}_{ik}(n_f+1)\hat{A}_{kj}^{(1),Q}
                                              \Bigl(\frac{m^2}{\mu^2}\Bigr)
                       +\Bigl[ 
                               \hat{A}_{il}^{(1),Q}
                                 \Bigl(\frac{m^2}{\mu^2}\Bigr)
                              +Z^{-1,(1)}_{il}(n_f+1)
 \N\\ &&\phantom{+{a^{\MOM}_s}^3\Biggl(}
                              -Z^{-1,(1)}_{il}(n_f)
                        \Bigr]
                              \Gamma^{-1,(2)}_{lj}(n_f)
                       +\Bigl[ 
                               \hat{A}_{il}^{(2),Q}
                                 \Bigl(\frac{m^2}{\mu^2}\Bigr)
                              +Z^{-1,(2)}_{il}(n_f+1)
                              -Z^{-1,(2)}_{il}(n_f)
 \N\\ 
&&\phantom{+{a^{\MOM}_s}^3\Biggl(}
                              +Z^{-1,(1)}_{ik}(n_f+1)\hat{A}_{kl}^{(1),Q}
                                              \Bigl(\frac{m^2}{\mu^2}\Bigr)
                        \Bigr]
                              \Gamma^{-1,(1)}_{lj}(n_f)
                        \Biggr)~. \label{GenRen3}
  \end{eqnarray}
  From (\ref{GenRen3}) it is obvious that the renormalization of $A^Q_{ij}$ to
  $O(a_s^3)$ requires the $1$--loop terms up to $O(\ep^2)$ and the $2$--loop
  terms up to $O(\ep)$, cf.~\cite{Buza:1995ie,Buza:1996wv,Bierenbaum:2007qe,
    Bierenbaum:2008yu,Bierenbaum:2009zt}.  Finally, we transform the coupling
  constant back to the $\MS$--scheme by using Eq.~(\ref{asmoma}). We do not
  give the explicit formula here, but present the individual renormalized OMEs
  after this transformation in the next Section as perturbative series in
  $a_s^{\MS}$,
  \begin{eqnarray}
   A_{ij}^{Q}\Bigl(\frac{m^2}{\mu^2},a_s^{\MS},n_f+1\Bigr)&=&
     a_s^{\MS}     A_{ij}^{Q, (1)}\Bigl(\frac{m^2}{\mu^2},n_f+1\Bigr)
   +{a_s^{\MS}}^2  A_{ij}^{Q, (2)}\Bigl(\frac{m^2}{\mu^2},n_f+1\Bigr)
\N \\ \phantom{A_{ij}^{Q}\Bigl(\frac{m^2}{\mu^2},a_s^{\MS},n_f+1\Bigr)}&& \!\!
   +{a_s^{\MS}}^3  A_{ij}^{Q, (3)}\Bigl(\frac{m^2}{\mu^2},n_f+1\Bigr)~. 
    \label{PertOmeren} 
  \end{eqnarray}
\section{\bf\boldmath General Structure of the Massive Operator Matrix Elements}
\renewcommand{\theequation}{\thesection.\arabic{equation}}
\setcounter{equation}{0}
\label{SubSec-RENPred}

\vspace{1mm}\noindent In the following, we present the unrenormalized and
renormalized massive operator matrix elements for the specific flavor
channels. The pole terms can all be expressed in terms of known
renormalization constants, which provides us with a strong check on our
calculation. In particular, we obtain the moments of the 
complete anomalous dimensions up to
$O(a_s^2)$, as well as their $T_F$--terms at $O(a_s^3)$.  The
moments of the
$O(\ep^0)$--terms of the unrenormalized OMEs at the $3$--loop level,
$a_{ij}^{(3)}$, are a new result. Previously, the $O(\ep)$ terms at the
$2$--loop level, $\overline{a}_{ij}^{(2)}$, for general values of $N$ were
calculated by the present authors in
Refs.~\cite{Bierenbaum:2008yu,Bierenbaum:2009zt}. The pole terms and the
$O(\ep^0)$ terms, $a_{ij}^{(2)}$, at the 2--loop level have been calculated
for the first time in Refs. \cite{Buza:1995ie,Buza:1996wv}. They were
confirmed in \cite{Bierenbaum:2007qe,Bierenbaum:2009zt}, as well as by the 
present
calculation, in which they appear in the renormalization of the respective
moments of the 3--loop OMEs. In order to keep up with the notation used in
\cite{Buza:1995ie,Buza:1996wv}, we define the 2--loop terms
$a_{ij}^{(2)},~\overline{a}_{ij}^{(2)}$ {\sf after} performing mass
renormalization in the on--shell scheme. This we {\sf do not} apply
for the $3$--loop terms.  We choose
to calculate one--particle reducible diagrams and therefore have to include
external self--energies containing massive quarks into our calculation. Before
presenting the operator matrix elements up to three loops, we first summarize
the necessary self--energy contributions.
%
 \subsection{Self--energy contributions}
  \label{Sec-elf}

\vspace{1mm}
\noindent
The gluon and quark self-energy contributions due to heavy quark lines
are given by
   \begin{eqnarray}
    \hat{\Pi}_{\mu\nu}^{ab}(p^2,\hat{m}^2,\mu^2,\hat{a}_s) &=& i\delta^{ab}
                            \left[-g_{\mu\nu}p^2 +p_\mu p_\nu\right] 
                            \hat{\Pi}(p^2,\hat{m}^2,\mu^2,\hat{a}_s)~, 
  \end{eqnarray}
   with
  \begin{eqnarray}
   \hat{\Pi}(p^2,\hat{m}^2,\mu^2,\hat{a}_s)&=&
        \sum_{k=1}^{\infty}\hat{a}_s^k\hat{\Pi}^{(k)}(p^2,\hat{m}^2,\mu^2).
        \label{pertPiGlu}
  \end{eqnarray}
and
  \begin{eqnarray}
   \hat{\Sigma}_{ij}(p^2,\hat{m}^2,\mu^2,\hat{a}_s)&=&
                      i~~\delta_{ij}~\adag p~~ 
                      \hat{\Sigma}(p^2,\hat{m}^2,\mu^2,\hat{a_s})~,
  \end{eqnarray}
  where
  \begin{eqnarray}
   \hat{\Sigma}(p^2,\hat{m}^2,\mu^2,\hat{a}_s)&=&
\sum_{k=2}^{\infty}\hat{a}_s^k\hat{\Sigma}^{(k)}(p^2,\hat{m}^2,\mu^2)~.
        \label{pertSiQu}
  \end{eqnarray}
Note, that the quark self--energy contributions start at 2--loop order. 
These self--energies are easily calculated using {\sf MATAD}, 
\cite{Steinhauser:2000ry}, cf. Section~\ref{sec-calc}. 
  The expansion coefficients for $p^2=0$ of Eq.~(\ref{pertPiGlu}, 
  \ref{pertSiQu})
  are needed for the calculation of the gluonic and quarkonic  OMEs.
  The contributions to the gluon vacuum polarization for general gauge 
  parameter $\xi$ are
  \begin{eqnarray}
  \label{eqPI1}
   \hat{\Pi}^{(1)}\Bigl(0,\frac{\hat{m}^2}{\mu^2}\Bigr)&=&
            T_F\left(\frac{\hat{m}^2}{\mu^2}\right)^{\ep/2} 
            \Biggl(
             -\frac{8}{3\ep}
              \exp \Bigl(\sum_{i=2}^{\infty}\frac{\zeta_i}{i}
                       \Bigl(\frac{\ep}{2}\Bigr)^{i}\Bigr)
             \Biggr)~.
               ~\label{GluSelf1}
\\
  \label{eqPI2}
    \hat{\Pi}^{(2)}\Bigl(0,\frac{\hat{m}^2}{\mu^2}\Bigr)&=&
    T_F\Bigl(\frac{\hat{m}^2}{\mu^2}\Bigr)^{\ep}\Biggl(
    -\frac{4}{\ep^2} C_A + \frac{1}{\ep} \left\{-12 C_F + 5 C_A\right\} +
    C_A \left(\frac{13}{12} -\zeta_2\right) - \frac{13}{3} C_F
    \N\\ &&
    + \ep \left\{C_A \left(\frac{169}{144} + \frac{5}{4} \zeta_2 - 
    \frac{\zeta_3}{3} \right) + C_F \left( - \frac{35}{12} -3 \zeta_2 \right) 
\right\}\Biggr) + O(\ep^2)\\
\label{Pia}
    \hat{\Pi}^{(3)}\Bigl(0,\frac{\hat{m}^2}{\mu^2}\Bigr)&=&
       T_F\Bigl(\frac{\hat{m}^2}{\mu^2}\Bigr)^{\ep}\Biggl(
                        \frac{1}{\ep^3}\Biggl\{
                                 -\frac{32}{9}T_FC_A\Bigl(2n_f+1\Bigr)
                                 +C_A^2\Bigl( 
                                              \frac{164}{9}
                                             +\frac{4}{3}\xi
                                       \Bigr)
                                       \Biggr\}
\N\\ 
&&
                       +\frac{1}{\ep^2}\Biggl\{
                               \frac{80}{27}\Bigl(
                                                  C_A-6C_F
                                                \Bigr)n_fT_F
                              +\frac{8}{27}   \Bigl(
                                                  35C_A-48C_F
                                                \Bigr)T_F
                              +\frac{C_A^2}{27} \Bigl(
                                                  -781+63\xi
                                                \Bigr)
\N\\ &&
                              +\frac{712}{9}C_AC_F
                                       \Biggr\}
                       +\frac{1}{\ep}\Biggl\{
                                \frac{4}{27}\Bigl(
                                                       C_A(-101-18\zeta_2)
                                                      -62C_F
                                                \Bigr)n_fT_F
\N\\ &&
                              +\frac{2}{27}   \Bigl(
                                                       C_A(-37-18\zeta_2)
                                                       -80C_F
                                                \Bigr)T_F
                              +C_A^2            \Bigl(
                                                  -12\zeta_3
                                                  +\frac{41}{6}\zeta_2
                                                  +\frac{3181}{108}
                                                  +\frac{\zeta_2}{2}\xi
                                                  +\frac{137}{36}\xi
                                                \Bigr)
\N\\ &&
                              +C_AC_F           \Bigl(
                                                   16\zeta_3
                                                  -\frac{1570}{27}
                                                \Bigr)
                              +\frac{272}{3}C_F^2
                                       \Biggr\}
                       +n_fT_F    \Biggl\{
                                       C_A\Bigl(
                                             \frac{56}{9}\zeta_3
                                            +\frac{10}{9}\zeta_2
                                            -\frac{3203}{243}
                                          \Bigr)
\N\\ 
&&
                                      +C_F\Bigl(
                                            -\frac{20}{3}\zeta_2
                                            -\frac{1942}{81}
                                          \Bigr)
                                       \Biggr\}
                       +T_F      \Biggl\{
                                       C_A\Bigl(
                                            -\frac{295}{18}\zeta_3
                                            +\frac{35}{9}\zeta_2
                                            +\frac{6361}{486}
                                          \Bigr)
\N\\ &&
                                      +C_F\Bigl(
                                            -7\zeta_3
                                            -\frac{16}{3}\zeta_2
                                            -\frac{218}{81}
                                          \Bigr)
                                   \Biggr\}
                       +C_A^2      \Biggl\{
                                       4{\sf B_4}
                                      -27\zeta_4
                                      +\frac{1969}{72}\zeta_3
                                      -\frac{781}{72}\zeta_2
                                      +\frac{42799}{3888}
\N\\ &&
                                      -\frac{7}{6}\zeta_3\xi
                                      +\frac{7}{8}\zeta_2\xi
                                      +\frac{3577}{432}\xi
                                   \Biggr\}
                       +C_AC_F      \Biggl\{
                                      -8{\sf B_4}
                                      +36\zeta_4
                                      -\frac{1957}{12}\zeta_3
                                      +\frac{89}{3}\zeta_2
                                      +\frac{10633}{81}
                                   \Biggr\}
\nonumber\end{eqnarray}\begin{eqnarray}
&&
                       +C_F^2      \Biggl\{
                                      \frac{95}{3}\zeta_3
                                      +\frac{274}{9}
                                   \Biggr\}
                                                   \Biggr) + O(\ep)~, 
                                          \label{GluSelf3}
\nonumber\\
   \end{eqnarray}
   and for the quark self--energy,
   \begin{eqnarray}
    \hat{\Sigma}^{(2)}(0,\frac{\hat{m}^2}{\mu^2}) &=&
        T_F C_F \Bigl(\frac{\hat{m}^2}{\mu^2}\Bigr)^{\ep} \left\{\frac{2}{\ep} 
+\frac{5}{6} + \left[\frac{89}{72} + \frac{\zeta_2}{2} \right] \ep \right\}
+ O(\ep^2)\\
\label{Sig3}
    \hat{\Sigma}^{(3)}(0,\frac{\hat{m}^2}{\mu^2}) &=&
        T_F C_F \Bigl(\frac{\hat{m}^2}{\mu^2}\Bigr)^{3\ep/2}
        \Biggl(
                        \frac{8}{3\ep^3}C_A \{1-\xi\}
                       +\frac{1}{\ep^2} \Bigl\{
                                   +\frac{32}{9}T_F(n_f+2)
                                   -C_A\Bigl(\frac{40}{9}+4\xi\Bigr)
                                   -\frac{8}{3}C_F
                                        \Bigl\}
\N\\ &&
                       +\frac{1}{\ep} \Biggl\{
                                       \frac{40}{27}T_F(n_f+2)
                                      +C_A\Bigl\{
                                             \zeta_2
                                            +\frac{454}{27}
                                            -\zeta_2\xi
                                            -\frac{70}{9}\xi
                                          \Bigr\}
                                       -26C_F
                                        \Biggl\}
                       +n_fT_F\Bigl\{
                                \frac{4}{3}\zeta_2
                               +\frac{674}{81}
                              \Bigr\}
\N\\ &&
                       +  T_F\Bigl\{
                                \frac{8}{3}\zeta_2
                               +\frac{604}{81}
                              \Bigr\}
                       +  C_A\Bigl\{
                                \frac{17}{3}\zeta_3
                               -\frac{5}{3}\zeta_2
                               +\frac{1879}{162}
                               +\frac{7}{3}\zeta_3\xi
                               -\frac{3}{2}\zeta_2\xi
                               -\frac{407}{27}\xi
                              \Bigr\}
\N\\ &&
                       +  C_F\Bigl\{
                               -8\zeta_3
                               -\zeta_2
                               -\frac{335}{18}
                              \Bigr\}
        \Biggr) + O(\ep)~,  \label{QuSelf3}
   \end{eqnarray}
see also~\cite{Chetyrkin:2008jk,CS1}.
 In Eq.~(\ref{Pia}) the constant
   \begin{eqnarray}
\label{eqB4}
          {\sf B_4}&=&-4\zeta_2\ln^2(2) +\frac{2}{3}\ln^4(2) 
-\frac{13}{2}\zeta_4
                  +16 {\sf Li}_4\Bigl(\frac{1}{2}\Bigr) \label{B4} \\
                  &\approx&  -1.762800093...~. \nonumber
   \end{eqnarray}
appears. 
%
%
%
 \subsection{\boldmath $A_{qq,Q}^{\sf NS}$}
  \label{Sec-NS}

\vspace{1mm}\noindent
  The lowest ${\sf NS}$--contribution is of $O(a_s^2)$,
  \begin{eqnarray}
   A_{qq,Q}^{\sf NS}&=&
                       a_s^2A_{qq,Q}^{(2), {\sf NS}}
                       +a_s^3A_{qq,Q}^{(3), {\sf NS}}
                       +O(a_s^4)~. \label{NSpert}
  \end{eqnarray}
  The expansion coefficients are  obtained in the {\sf MOM}--scheme
  from the bare quantities, using Eqs.~(\ref{macoren},~\ref{GenRen3}). 
  After operator renormalization and mass factorization, the OMEs are given by
  \begin{eqnarray}
    A_{qq,Q}^{(2), \sf NS, \MOM}&=&
                    \hat{A}_{qq,Q}^{(2),{\sf NS},\MOM}
                   +Z^{-1,(2), {\sf NS}}_{qq}(n_f+1)
                   -Z^{-1,(2), {\sf NS}}_{qq}(n_f)
                ~, \label{2LNSRen1} \\
    A_{qq,Q}^{(3), \sf NS, \MOM}&=&
                     \hat{A}_{qq,Q}^{(3), {\sf NS},\MOM}
                    +Z^{-1,(3), {\sf NS}}_{qq}(n_f+1)
                    -Z^{-1,(3), {\sf NS}}_{qq}(n_f)
\N\\ &&
                    +Z^{-1,(1), {\sf NS}}_{qq}(n_f+1)
                     \hat{A}_{qq,Q}^{(2), {\sf NS},\MOM}
\N\\ &&
                    +\Bigl[ \hat{A}_{qq,Q}^{(2), {\sf NS},\MOM}
                           +Z^{-1,(2), {\sf NS}}_{qq}(n_f+1)
                           -Z^{-1,(2), {\sf NS}}_{qq}(n_f)
                     \Bigr]\Gamma^{-1,(1)}_{qq}(n_f)
                ~. \label{3LNSRen1}
   \end{eqnarray}
   From (\ref{macoren}, \ref{GenRen3}, \ref{2LNSRen1}, \ref{3LNSRen1}), one 
   predicts the pole terms of the unrenormalized OME. 
   At second and third order they read
   \begin{eqnarray}
   \Ahathat_{qq,Q}^{(2),\sf NS}&=&
           \Bigl(\frac{\hat{m}^2}{\mu^2}\Bigr)^{\ep}\Biggl(
                    \frac{\beta_{0,Q}\gamma_{qq}^{(0)}}{\ep^2}
                   +\frac{\hat{\gamma}_{qq}^{(1), {\sf NS}}}{2\ep}
                   +a_{qq,Q}^{(2),{\sf NS}}
                   +\overline{a}_{qq,Q}^{(2),{\sf NS}}\ep
             \Biggr)~, 
\label{Ahhhqq2NSQ}
 \end{eqnarray}\begin{eqnarray} 
 \Ahathat_{qq,Q}^{(3),{\sf NS}}&=&
     \Bigl(\frac{\hat{m}^2}{\mu^2}\Bigr)^{3\ep/2}\Biggl\{
            -\frac{4\gamma_{qq}^{(0)}\beta_{0,Q}}{3\ep^3}
                   \Bigl(\beta_0+2\beta_{0,Q}\Bigr)
            +\frac{1}{\ep^2}
              \Biggl(
                      \frac{2\gamma_{qq}^{(1),{\sf NS}}\beta_{0,Q}}{3}
                     -\frac{4\hat{\gamma}_{qq}^{(1),{\sf NS}}}{3}
                             \Bigl[\beta_0+\beta_{0,Q}\Bigr]
\N\\ &&
                     +\frac{2\beta_{1,Q}\gamma_{qq}^{(0)}}{3}
                     -2\delta m_1^{(-1)}\beta_{0,Q}\gamma_{qq}^{(0)}
              \Biggr)
            +\frac{1}{\ep} 
              \Biggl(
                      \frac{\hat{\gamma}_{qq}^{(2), {\sf NS}}}{3}
                     -4a_{qq,Q}^{(2),{\sf NS}}\Bigl[\beta_0+\beta_{0,Q}\Bigr]
                     +\beta_{1,Q}^{(1)}\gamma_{qq}^{(0)}
\N\\ &&
                     +\frac{\gamma_{qq}^{(0)}\beta_0\beta_{0,Q}\zeta_2}{2}
                     -2 \delta m_1^{(0)} \beta_{0,Q} \gamma_{qq}^{(0)} 
                     -\delta m_1^{(-1)}\hat{\gamma}_{qq}^{(1),{\sf NS}}
                  \Biggr)
         +a_{qq,Q}^{(3), {\sf NS}}
                              \Biggr\}~. \label{Ahhhqq3NSQ}
   \end{eqnarray}
   Note, that we have already used the general structure of the 
   unrenormalized lower order OME in the evaluation of the $O(\hat{a}_s^3)$
   term, as we will always do in the following. 
   Using Eqs. (\ref{2LNSRen1}, \ref{3LNSRen1}, \ref{macoren}), one can 
   renormalize the above expressions. In addition, we finally transform back 
   to the $\MS$--scheme using Eq. (\ref{asmoma}). 
   Thus one obtains the renormalized expansion 
   coefficients of Eq.~(\ref{NSpert}) 
   \begin{eqnarray}
     A_{qq,Q}^{(2),\sf NS, \MS}&=&
                  \frac{\beta_{0,Q}\gamma_{qq}^{(0)}}{4}
                    \ln^2 \Bigl(\frac{m^2}{\mu^2}\Bigr)
                 +\frac{\hat{\gamma}_{qq}^{(1), {\sf NS}}}{2}
                    \ln \Bigl(\frac{m^2}{\mu^2}\Bigr)
                 +a_{qq,Q}^{(2),{\sf NS}}
                 -\frac{\beta_{0,Q}\gamma_{qq}^{(0)}}{4}\zeta_2~,
                  \label{Aqq2NSQMSren} \\
    A_{qq,Q}^{(3),{\sf NS}, \MS}&=&
     -\frac{\gamma_{qq}^{(0)}\beta_{0,Q}}{6}
          \Bigl(
                 \beta_0
                +2\beta_{0,Q}
          \Bigr)
             \ln^3 \Bigl(\frac{m^2}{\mu^2}\Bigr)
         +\frac{1}{4}
          \Biggl\{
                   2\gamma_{qq}^{(1),{\sf NS}}\beta_{0,Q}
                  -2\hat{\gamma}_{qq}^{(1),{\sf NS}}
                             \Bigl(
                                    \beta_0
                                   +\beta_{0,Q}
                             \Bigr)
\N\\ &&
                  +\beta_{1,Q}\gamma_{qq}^{(0)}
          \Biggr\}
             \ln^2 \Bigl(\frac{m^2}{\mu^2}\Bigr)
         +\frac{1}{2}
          \Biggl\{
                   \hat{\gamma}_{qq}^{(2),{\sf NS}}
                  -\Bigl(
                           4a_{qq,Q}^{(2),{\sf NS}}
                          -\zeta_2\beta_{0,Q}\gamma_{qq}^{(0)}
                                    \Bigr)(\beta_0+\beta_{0,Q})
\N\\ &&
                  +\gamma_{qq}^{(0)}\beta_{1,Q}^{(1)}
          \Biggr\}
             \ln \Bigl(\frac{m^2}{\mu^2}\Bigr)
         +4\overline{a}_{qq,Q}^{(2),{\sf NS}}(\beta_0+\beta_{0,Q})
         -\gamma_{qq}^{(0)}\beta_{1,Q}^{(2)}
         -\frac{\gamma_{qq}^{(0)}\beta_0\beta_{0,Q}\zeta_3}{6}
\N\\ &&
         -\frac{\gamma_{qq}^{(1),{\sf NS}}\beta_{0,Q}\zeta_2}{4}
         +2 \delta m_1^{(1)} \beta_{0,Q} \gamma_{qq}^{(0)}
         +\delta m_1^{(0)} \hat{\gamma}_{qq}^{(1),{\sf NS}}
         +2 \delta m_1^{(-1)} a_{qq,Q}^{(2),{\sf NS}}
\N\\ &&
         +a_{qq,Q}^{(3),{\sf NS}}
        ~. \label{Aqq3NSQMSren}
   \end{eqnarray}
   Note that in the ${\sf NS}$--case, one is generically provided 
   with even and odd moments due to a Ward--identity relating the 
   results in the polarized 
   and unpolarized case. The former refer 
   to the anomalous dimensions $\gamma_{qq}^{{\sf NS},+}$ 
   and the latter to $\gamma_{qq}^{{\sf NS},-}$ 
   as given in Eqs. (3.5, 3.7) and Eqs. (3.6, 3.8) in Ref. 
\cite{Moch:2004pa}.
   The relations above also apply to other twist--2 non--singlet massive OMEs, as to 
   transversity, for which  the 2- and 3--loop heavy flavor corrections are given in 
\cite{TRANS}.
%
%
%
 \subsection{\boldmath $A_{Qq}^{\sf PS}$ and $A_{qq,Q}^{\sf PS}$}
  \label{SubSec-PS}
  There are two different ${\sf PS}$--contributions. The term referring to the
  case in which the operator couples to a heavy quark, $A_{Qq}^{\sf PS}$,
  starts at $O(a_s^2)$, whereas the term in which it couples to an internal
  light quark line, $A_{qq,Q}^{\sf PS}$, emerges for the first time at
  $O(a_s^3)$,
  \begin{eqnarray}
   A_{Qq}^{\sf PS}&=&
                       a_s^2A_{Qq}^{(2), {\sf PS}}
                       +a_s^3A_{Qq}^{(3), {\sf PS}}
                       +O(a_s^4)~, \label{PSQqpert}\\
   A_{qq,Q}^{\sf PS}&=&
                        a_s^3A_{qq,Q}^{(3), {\sf PS}}
                       +O(a_s^4)~. \label{PSqqQpert}
  \end{eqnarray}
  Separating these contributions is not straightforward, since 
  the generic renormalization formula for operator 
  renormalization and mass factorization, Eq. (\ref{GenRen3}),
  applies to the sum of these terms only.
  At $O(a_s^2)$, this problem does not occur and renormalization proceeds 
  in the {\sf MOM}--scheme via
  \begin{eqnarray}
    A_{Qq}^{(2), \sf PS, \MOM}&=&
                    \hat{A}_{Qq}^{(2),{\sf PS}, \MOM}
                   +Z^{-1,(2), {\sf PS}}_{qq}(n_f+1)
                   -Z^{-1,(2), {\sf PS}}_{qq}(n_f) \N\\ &&
                   +\Bigl[ \hat{A}_{Qg}^{(1), \MOM}
                          +Z_{qg}^{-1,(1)}(n_f+1)
                          -Z_{qg}^{-1,(1)}(n_f)
                    \Bigr]\Gamma^{-1,(1)}_{gq}(n_f)~.
  \end{eqnarray}
  Thus the unrenormalized expression is given by
  \begin{eqnarray}
   \Ahathat_{Qq}^{(2),\sf PS}&=&
          \Bigl(\frac{\hat{m}^2}{\mu^2}\Bigr)^{\ep}\Biggl(
                  -\frac{\hat{\gamma}_{qg}^{(0)}
                         \gamma_{gq}^{(0)}}{2\ep^2}
                  +\frac{\hat{\gamma}_{qq}^{(1), {\sf PS}}}{2\ep}
                  +a_{Qq}^{(2),{\sf PS}}
                  +\overline{a}_{Qq}^{(2),{\sf PS}}\ep
            \Biggr)~.\label{AhhhQq2PS}
  \end{eqnarray}
  The renormalized result in the ${\sf \MS}$--scheme reads
  \begin{eqnarray}
    A_{Qq}^{(2),\sf PS, \MS}&=&
                -\frac{\hat{\gamma}_{qg}^{(0)}
                             \gamma_{gq}^{(0)}}{8}
                   \ln^2 \Bigl(\frac{m^2}{\mu^2}\Bigr)
                +\frac{\hat{\gamma}_{qq}^{(1), {\sf PS}}}{2}
                   \ln \Bigl(\frac{m^2}{\mu^2}\Bigr)
                +a_{Qq}^{(2),{\sf PS}}
                +\frac{\hat{\gamma}_{qg}^{(0)}
                             \gamma_{gq}^{(0)}}{8}\zeta_2~.
  \end{eqnarray}
  The corresponding renormalization relation at third order is given by
  \begin{eqnarray}
   &&A_{Qq}^{(3), \sf PS, \MOM}+
     A_{qq,Q}^{(2), \sf PS, \MOM}=
                     \hat{A}_{Qq}^{(3), {\sf PS}, \MOM} 
                    +\hat{A}_{qq,Q}^{(3), {\sf PS}, \MOM}
                    +Z^{-1,(3), {\sf PS}}_{qq}(n_f+1)
                    -Z^{-1,(3), {\sf PS}}_{qq}(n_f)
\N\\ && \phantom{abcde}
                    +Z^{-1,(1)}_{qq}(n_f+1)\hat{A}_{Qq}^{(2), {\sf PS}, \MOM}
                    +Z^{-1,(1)}_{qg}(n_f+1)\hat{A}_{gq,Q}^{(2), \MOM}
                    +\Bigl[
                            \hat{A}_{Qg}^{(1), \MOM}
                           +Z^{-1,(1)}_{qg}(n_f+1)
\N\\ && \phantom{abcde}
                           -Z^{-1,(1)}_{qg}(n_f)
                     \Bigr]\Gamma^{-1,(2)}_{gq}(n_f)
                    +\Bigl[ \hat{A}_{Qq}^{(2), {\sf PS}, \MOM}
                           +Z^{-1,(2), {\sf PS}}_{qq}(n_f+1)
\N\\ && \phantom{abcde}
                           -Z^{-1,(2), {\sf PS}}_{qq}(n_f)
                     \Bigr]\Gamma^{-1,(1)}_{qq}(n_f)
                    +\Bigl[ \hat{A}_{Qg}^{(2), \MOM}
                           +Z^{-1,(2)}_{qg}(n_f+1)
                           -Z^{-1,(2)}_{qg}(n_f)
\N\\ && \phantom{abcde}
                           +Z^{-1,(1)}_{qq}(n_f+1)A_{Qg}^{(1), \MOM}
                           +Z^{-1,(1)}_{qg}(n_f+1)A_{gg,Q}^{(1), \MOM}
                     \Bigr]\Gamma^{-1,(1)}_{gq}(n_f)~.
                  \label{AQqq3PSRen}
  \end{eqnarray}
  Taking into account the kinematic and UV--structure 
  of the contributing Feynman diagrams, the two
  contributions can be separated.  
  For the bare quantities we obtain 
   \begin{eqnarray}
    \Ahathat_{Qq}^{(3),{\sf PS}}&=&
     \Bigl(\frac{\hat{m}^2}{\mu^2}\Bigr)^{3\ep/2}\Biggl[
     \frac{\hat{\gamma}_{qg}^{(0)}\gamma_{gq}^{(0)}}{6\ep^3}
                  \Biggl(
                         \gamma_{gg}^{(0)}
                        -\gamma_{qq}^{(0)}
                        +6\beta_0
                        +16\beta_{0,Q}
                  \Biggr)
   +\frac{1}{\ep^2}\Biggl(     
                        -\frac{4\hat{\gamma}_{qq}^{(1),{\sf PS}}}{3}
                                 \Bigl[
                                        \beta_0
                                       +\beta_{0,Q}
                                 \Bigr]
 \N\\ &&
                        -\frac{\gamma_{gq}^{(0)}\hat{\gamma}_{qg}^{(1)}}{3}
                        +\frac{\hat{\gamma}_{qg}^{(0)}}{6}
                                 \Bigl[
                                        2\hat{\gamma}_{gq}^{(1)}
                                       -\gamma_{gq}^{(1)}
                                 \Bigr]
                        +\delta m_1^{(-1)} \hat{\gamma}_{qg}^{(0)}
                                           \gamma_{gq}^{(0)}
                 \Biggr)
   +\frac{1}{\ep}\Biggl(     
                           \frac{\hat{\gamma}_{qq}^{(2),{\sf PS}}}{3}
                        -n_f\frac{\hat{\tilde{\gamma}}_{qq}^{(2),{\sf PS}}}{3}
         \N\\ &&
                          +\hat{\gamma}_{qg}^{(0)}a_{gq,Q}^{(2)}
                          -\gamma_{gq}^{(0)}a_{Qg}^{(2)}
                          -4(\beta_0+\beta_{0,Q})a_{Qq}^{(2),{\sf PS}}
                   -\frac{\hat{\gamma}_{qg}^{(0)}\gamma_{gq}^{(0)}\zeta_2}{16}
                            \Bigl[
                               \gamma_{gg}^{(0)}
                              -\gamma_{qq}^{(0)}
                              +6\beta_0
                            \Bigr]
         \N\\ &&
                   +\delta m_1^{(0)} \hat{\gamma}_{qg}^{(0)}
                                           \gamma_{gq}^{(0)}
                   -\delta m_1^{(-1)} \hat{\gamma}_{qq}^{(1),{\sf PS}}
                 \Biggr)
   +a_{Qq}^{(3),{\sf PS}}
                              \Biggr]~, \label{AhhhQq3PS} \\
    \Ahathat_{qq,Q}^{(3),{\sf PS}}&=&
           n_f\Bigl(\frac{\hat{m}^2}{\mu^2}\Bigr)^{3\ep/2}\Biggl[
            \frac{2\hat{\gamma}_{qg}^{(0)}\gamma_{gq}^{(0)}\beta_{0,Q}}{3\ep^3}
                +\frac{1}{3\ep^2} \Biggl(
                    2\hat{\gamma}_{qq}^{(1),{\sf PS}}\beta_{0,Q}
                   +\hat{\gamma}_{qg}^{(0)}\hat{\gamma}_{gq}^{(1)}
                                  \Biggr)
\N\\ &&
                +\frac{1}{\ep} \Biggl(
                       \frac{\hat{\tilde{\gamma}}_{qq}^{(2),{\sf PS}}}{3}
                      +\hat{\gamma}_{qg}^{(0)}a_{gq,Q}^{(2)}
         -\frac{\hat{\gamma}_{qg}^{(0)}\gamma_{gq}^{(0)}\beta_{0,Q}\zeta_2}{4}
                                \Biggr)
                +\frac{a_{qq,Q}^{(3), {\sf PS}}}{n_f}
                                                     \Biggr]~. 
                   \label{Ahhhqq3PSQ}
   \end{eqnarray}
   The renormalized terms are given in the ${\MS}$--scheme by
   \begin{eqnarray}
    A_{Qq}^{(3),{\sf PS}, \MS}&=&
      \frac{\hat{\gamma}_{qg}^{(0)}\gamma_{gq}^{(0)}}{48}
                  \Biggl\{
                         \gamma_{gg}^{(0)}
                        -\gamma_{qq}^{(0)}
                        +6\beta_0
                        +16\beta_{0,Q}
                  \Biggr\}
              \ln^3 \Bigl(\frac{m^2}{\mu^2}\Bigr)
  +    \frac{1}{8}\Biggl\{
                         -4\hat{\gamma}_{qq}^{(1),{\sf PS}}
                               \Bigl(
                                 \beta_0
                                +\beta_{0,Q}
                               \Bigr)
\N\\ &&
                        +\hat{\gamma}_{qg}^{(0)}
                               \Bigl(
                                 \hat{\gamma}_{gq}^{(1)}
                                -\gamma_{gq}^{(1)}
                               \Bigr)
                        -\gamma_{gq}^{(0)}\hat{\gamma}_{qg}^{(1)}
                  \Biggr\}
              \ln^2 \Bigl(\frac{m^2}{\mu^2}\Bigr)
  +   \frac{1}{16}\Biggl\{
                         8\hat{\gamma}_{qq}^{(2),{\sf PS}}
                        -8n_f\hat{\tilde{\gamma}}_{qq}^{(2),{\sf PS}}
\N\\ &&
                        -32a_{Qq}^{(2),{\sf PS}}(\beta_0+\beta_{0,Q})
                        +8\hat{\gamma}_{qg}^{(0)}a_{gq,Q}^{(2)}
                        -8\gamma_{gq}^{(0)}a_{Qg}^{(2)}
                        -\hat{\gamma}_{qg}^{(0)}\gamma_{gq}^{(0)}\zeta_2\
                          \Bigl(
                                 \gamma_{gg}^{(0)}
                                -\gamma_{qq}^{(0)}
\N\\ &&
                                +6\beta_0
                                +8\beta_{0,Q}
                          \Bigr)
                  \Biggr\}
              \ln \Bigl(\frac{m^2}{\mu^2}\Bigr)
    +4(\beta_0+\beta_{0,Q})\overline{a}_{Qq}^{(2),{\sf PS}}
    +\gamma_{gq}^{(0)}\overline{a}_{Qg}^{(2)}
    -\hat{\gamma}_{qg}^{(0)}\overline{a}_{gq,Q}^{(2)}
\N\\ &&
    +\frac{\gamma_{gq}^{(0)}\hat{\gamma}_{qg}^{(0)}\zeta_3}{48}
                  \Bigl(
                         \gamma_{gg}^{(0)}
                        -\gamma_{qq}^{(0)}
                        +6\beta_0
                  \Bigr)
    +\frac{\hat{\gamma}_{qg}^{(0)}\gamma_{gq}^{(1)}\zeta_2}{16}
    -\delta m_1^{(1)} \hat{\gamma}_{qg}^{(0)}
                        \gamma_{gq}^{(0)}
    +\delta m_1^{(0)} \hat{\gamma}_{qq}^{(1),{\sf PS}}
\N\\ &&
    +2 \delta m_1^{(-1)} a_{Qq}^{(2),{\sf PS}}
    +a_{Qq}^{(3),{\sf PS}}~.     \label{AQq3PSMSren} \\
    A_{qq,Q}^{(3),{\sf PS}, \MS}&=&n_f\Biggl\{
              \frac{\gamma_{gq}^{(0)}\hat{\gamma}_{qg}^{(0)}\beta_{0,Q}}{12}
                         \ln^3 \Bigl(\frac{m^2}{\mu^2}\Bigr)
             +\frac{1}{8}\Bigl(
                     4\hat{\gamma}_{qq}^{(1), {\sf PS}}\beta_{0,Q}
                    +\hat{\gamma}_{qg}^{(0)}\hat{\gamma}_{gq}^{(1)}
              \Bigr)\ln^2 \Bigl(\frac{m^2}{\mu^2}\Bigr)
\N\\ &&
             +\frac{1}{4}\Bigl(
                    2\hat{\tilde{\gamma}}_{qq}^{(2), {\sf PS}}
                   +\hat{\gamma}_{qg}^{(0)}\Bigl\{
                                   2a_{gq,Q}^{(2)}
                                  -\gamma_{gq}^{(0)}\beta_{0,Q}\zeta_2
                                                     \Bigr\}
              \Bigr)\ln \Bigl(\frac{m^2}{\mu^2}\Bigr)
\N\\ &&
              -\hat{\gamma}_{qg}^{(0)}\overline{a}_{gq,Q}^{(2)}
              +\frac{\gamma_{gq}^{(0)}\hat{\gamma}_{qg}^{(0)}
                         \beta_{0,Q}\zeta_3}{12}
              -\frac{\hat{\gamma}_{qq}^{(1), {\sf PS}}\beta_{0,Q}\zeta_2}{4}
                                             \Biggr\}
              +a_{qq,Q}^{(3), {\sf PS}}~. 
              \label{Aqq3PSQMSren}
   \end{eqnarray}
%
%
%
 \subsection{\boldmath $A_{Qg}$ and $A_{qg,Q}$}
  \label{SubSec-AQqg}
  The OME $A_{Qg}$ is the most complex expression. As in the 
${\sf PS}$--case, there are two different contributions, depending on whether the 
  operator couples to a light quark line, denoted by $A_{qg,Q}$, or to a
  heavy quark line, given by $A_{Qg}$,
  \begin{eqnarray}
   A_{Qg}&=&
                        a_s  A_{Qg}^{(1)}
                       +a_s^2A_{Qg}^{(2)}
                       +a_s^3A_{Qg}^{(3)}
                       +O(a_s^4)~. \label{AQgpert}\\
   A_{qg,Q}&=&
                        a_s^3A_{qg,Q}^{(3)}
                       +O(a_s^4)~. \label{AqgQpert}
  \end{eqnarray}
  In the {\sf MOM}--scheme
  the $1$-- and $2$--loop contributions obey the following relations 
  \begin{eqnarray} 
    A_{Qg}^{(1), \MOM}&=&
                    \hat{A}_{Qg}^{(1), \MOM}
                   +Z^{-1,(1)}_{qg}(n_f+1)
                   -Z^{-1,(1)}_{qg}(n_f) ~, \\
    A_{Qg}^{(2), \MOM}&=&
                    \hat{A}_{Qg}^{(2), \MOM}
                   +Z^{-1,(2)}_{qg}(n_f+1)
                   -Z^{-1,(2)}_{qg}(n_f)
                   +Z^{-1,(1)}_{qg}(n_f+1)\hat{A}_{gg,Q}^{(1), \MOM}
 \N\\ &&
                   +Z^{-1,(1)}_{qq}(n_f+1)\hat{A}_{Qg}^{(1), \MOM}
                   +\Bigl[ \hat{A}_{Qg}^{(1), \MOM}
                          +Z_{qg}^{-1,(1)}(n_f+1)
\N\\ &&
                          -Z_{qg}^{-1,(1)}(n_f)
                    \Bigr]\Gamma^{-1,(1)}_{gg}(n_f)~.
  \end{eqnarray}
  The unrenormalized terms are given by 
  \begin{eqnarray}
   \Ahathat_{Qg}^{(1)}&=&
                        \Bigl(\frac{\hat{m}^2}{\mu^2}\Bigr)^{\ep/2}
                        \frac{\hat{\gamma}_{qg}^{(0)}}{\ep}
                          \exp \Bigl(\sum_{i=2}^{\infty}\frac{\zeta_i}{i}
                          \Bigl(\frac{\ep}{2}\Bigr)^{i}\Bigr)~, 
                          \label{AhhhQg1} 
\end{eqnarray}\begin{eqnarray}
   \Ahathat_{Qg}^{(2)}&=&
                  \Bigl(\frac{\hat{m}^2}{\mu^2}\Bigr)^{\ep}
                     \Biggl[
                            -\frac{\hat{\gamma}_{qg}^{(0)}}{2\ep^2}
                                \Bigl(
                                       \gamma_{gg}^{(0)}
                                      -\gamma_{qq}^{(0)}
                                      +2\beta_{0}
                                      +4\beta_{0,Q}
                                \Bigr)
                            +\frac{ \hat{\gamma}_{qg}^{(1)}
                                   -2\delta m_1^{(-1)} \hat{\gamma}_{qg}^{(0)}}
                                  {2\ep}
                            +a_{Qg}^{(2)}
                           -\delta m_1^{(0)} \hat{\gamma}_{qg}^{(0)}
\N\\ &&
                           -\frac{\hat{\gamma}_{qg}^{(0)}\beta_{0,Q}\zeta_2}{2}
                            +\ep\Bigl( 
                               \overline{a}_{Qg}^{(2)}
                        -\delta m_1^{(1)} \hat{\gamma}_{qg}^{(0)}
                        -\frac{\hat{\gamma}_{qg}^{(0)}\beta_{0,Q}\zeta_2}{12}
                                \Bigr)
                         \Biggr]
                         ~.\label{AhhhQg2}
  \end{eqnarray}
   Note that we have already made the
   one--particle reducible contributions to Eq. (\ref{AhhhQg2}) explicit, which
   are given by the 1--loop contribution multiplied
   by the 1--loop term of the gluon--self energy, cf. Eq. (\ref{GluSelf1}).
   Furthermore, Eq. (\ref{AhhhQg2}) already contains terms which result from
   mass renormalization in the $O(\ep^0)$ and $O(\ep)$ expressions.  At this
   stage of the renormalization procedure, they should not be present,
   however, we have included them here in order to have the same notation as
   in Refs. \cite{Buza:1995ie,Buza:1996wv} at the $2$--loop level.  The
   renormalized terms then become in the $\MS$--scheme
   \begin{eqnarray}
    A_{Qg}^{(1), \MS}&=&
                  \frac{\hat{\gamma}_{qg}^{(0)}}{2}
                  \ln \Bigl(\frac{m^2}{\mu^2}\Bigr)
                  ~, \label{AQg1MSren} \\
    A_{Qg}^{(2), \MS}&=&
                  -\frac{\hat{\gamma}_{qg}^{(0)}}{8}
                     \Biggl[
                             \gamma_{gg}^{(0)}
                            -\gamma_{qq}^{(0)}
                            +2\beta_{0}
                            +4\beta_{0,Q}
                     \Biggr]
                     \ln^2 \Bigl(\frac{m^2}{\mu^2}\Bigr)
                  +\frac{\hat{\gamma}_{qg}^{(1)}}{2}
                      \ln \Bigl(\frac{m^2}{\mu^2}\Bigr)
 \N\\ &&
                  +a_{Qg}^{(2)}
                  +\frac{\hat{\gamma}_{qg}^{(0)}\zeta_2}{8}
                              \Bigl(
                                     \gamma_{gg}^{(0)}
                                    -\gamma_{qq}^{(0)}
                                    +2\beta_{0}
                              \Bigr)~. \label{AQg2MSren}
   \end{eqnarray}
   The generic renormalization relation at the $3$--loop level is given by
   \begin{eqnarray}
    && A_{Qg}^{(3), \MOM}+A_{qg,Q}^{(3), \MOM}
            =
                     \hat{A}_{Qg}^{(3), \MOM}
                    +\hat{A}_{qg,Q}^{(3), \MOM}
                    +Z^{-1,(3)}_{qg}(n_f+1)
                    -Z^{-1,(3)}_{qg}(n_f)
\N\\ && \phantom{abcdef} 
                    +Z^{-1,(2)}_{qg}(n_f+1)\hat{A}_{gg,Q}^{(1), \MOM}
                    +Z^{-1,(1)}_{qg}(n_f+1)\hat{A}_{gg,Q}^{(2), \MOM}
                    +Z^{-1,(2)}_{qq}(n_f+1)\hat{A}_{Qg}^{(1), \MOM}
\N\\ && \phantom{abcdef} 
                    +Z^{-1,(1)}_{qq}(n_f+1)\hat{A}_{Qg}^{(2), \MOM}
                    +\Bigl[
                            \hat{A}_{Qg}^{(1), \MOM}
                           +Z^{-1,(1)}_{qg}(n_f+1)
                           -Z^{-1,(1)}_{qg}(n_f)
                     \Bigr]\Gamma^{-1,(2)}_{gg}(n_f)
\N\\ && \phantom{abcdef} 
                    +\Bigl[ \hat{A}_{Qg}^{(2), \MOM}
                           +Z^{-1,(2)}_{qg}(n_f+1) 
                           -Z^{-1,(2)}_{qg}(n_f)
                           +Z^{-1,(1)}_{qq}(n_f+1)A_{Qg}^{(1), \MOM}
\N\\ && \phantom{abcdef} 
                           +Z^{-1,(1)}_{qg}(n_f+1)A_{gg,Q}^{(1), \MOM}
                     \Bigr]\Gamma^{-1,(1)}_{gg}(n_f)
                    +\Bigl[ \hat{A}_{Qq}^{(2), {\sf PS}, \MOM}
                           +Z^{-1,(2), {\sf PS}}_{qq}(n_f+1)
\N\\ && \phantom{abcdef} 
                           -Z^{-1,(2), {\sf PS}}_{qq}(n_f)
                     \Bigr]\Gamma^{-1,(1)}_{qg}(n_f)
                    +\Bigl[ \hat{A}_{qq,Q}^{(2), {\sf NS}, \MOM}
                           +Z^{-1,(2), {\sf NS}}_{qq}(n_f+1)
\N\\ && \phantom{abcdef} 
                           -Z^{-1,(2), {\sf NS}}_{qq}(n_f)
                     \Bigr]\Gamma^{-1,(1)}_{qg}(n_f)~.
  \end{eqnarray}
  Similar to the ${\sf PS}$--case, the different contributions can be separated
  and one obtains the following unrenormalized results
   \begin{eqnarray}
   \Ahathat_{Qg}^{(3)}&=&
                  \Bigl(\frac{\hat{m}^2}{\mu^2}\Bigr)^{3\ep/2}
                     \Biggl[
           \frac{\hat{\gamma}_{qg}^{(0)}}{6\ep^3}
             \Biggl(
                   (n_f+1)\gamma_{gq}^{(0)}\hat{\gamma}_{qg}^{(0)}
                 +\gamma_{qq}^{(0)} 
                                \Bigl[
                                        \gamma_{qq}^{(0)}
                                      -2\gamma_{gg}^{(0)}
                                      -6\beta_0
                                      -8\beta_{0,Q}
                                \Bigr]
                 +8\beta_0^2
\N\\ &&
                 +28\beta_{0,Q}\beta_0
                 +24\beta_{0,Q}^2 
                  +\gamma_{gg}^{(0)} 
                                \Bigl[
                                        \gamma_{gg}^{(0)}
                                       +6\beta_0
                                       +14\beta_{0,Q}
                                \Bigr]
             \Biggr)
          +\frac{1}{6\ep^2}
             \Biggl(
                   \hat{\gamma}_{qg}^{(1)}
                      \Bigl[
                              2\gamma_{qq}^{(0)}
                             -2\gamma_{gg}^{(0)}
                             -8\beta_0
\N\\ &&
                             -10\beta_{0,Q} 
                      \Bigr]
                  +\hat{\gamma}_{qg}^{(0)}
                      \Bigl[
                              \hat{\gamma}_{qq}^{(1), {\sf PS}}\{1-2n_f\}
                             +\gamma_{qq}^{(1), {\sf NS}}
                             +\hat{\gamma}_{qq}^{(1), {\sf NS}}
                             +2\hat{\gamma}_{gg}^{(1)}
                             -\gamma_{gg}^{(1)}
                             -2\beta_1
                             -2\beta_{1,Q}
                      \Bigr]
\N\\ 
&&
                  + 6 \delta m_1^{(-1)} \hat{\gamma}_{qg}^{(0)} 
                      \Bigl[
                              \gamma_{gg}^{(0)}
                             -\gamma_{qq}^{(0)}
                             +3\beta_0
                             +5\beta_{0,Q}
                      \Bigr]
             \Biggr)
          +\frac{1}{\ep}
             \Biggl(
                   \frac{\hat{\gamma}_{qg}^{(2)}}{3}
                  -n_f \frac{\hat{\tilde{\gamma}}_{qg}^{(2)}}{3}
                  +\hat{\gamma}_{qg}^{(0)}\Bigl[
                                    a_{gg,Q}^{(2)}
                                   -n_fa_{Qq}^{(2),{\sf PS}}
                                          \Bigr]
\nonumber
\end{eqnarray}
\begin{eqnarray}
&&
                  +a_{Qg}^{(2)}
                      \Bigl[
                              \gamma_{qq}^{(0)}
                             -\gamma_{gg}^{(0)}
                             -4\beta_0
                             -4\beta_{0,Q}
                      \Bigr]
                  +\frac{\hat{\gamma}_{qg}^{(0)}\zeta_2}{16}
                      \Bigl[
                              \gamma_{gg}^{(0)} \Bigl\{
                                                        2\gamma_{qq}^{(0)}
                                                       -\gamma_{gg}^{(0)}
                                                       -6\beta_0
                                                       +2\beta_{0,Q}
                                                \Bigr\}
\N\\ &&
                             -(n_f+1)\gamma_{gq}^{(0)}\hat{\gamma}_{qg}^{(0)}
                             +\gamma_{qq}^{(0)} \Bigl\{
                                                       -\gamma_{qq}^{(0)}
                                                       +6\beta_0
                                                \Bigr\}
                             -8\beta_0^2
                             +4\beta_{0,Q}\beta_0
                             +24\beta_{0,Q}^2
                      \Bigr]
                  + \frac{\delta m_1^{(-1)}}{2}
                      \Bigl[
                              -2\hat{\gamma}_{qg}^{(1)}
\N\\ &&
                              +3\delta m_1^{(-1)}\hat{\gamma}_{qg}^{(0)}
                              +2\delta m_1^{(0)}\hat{\gamma}_{qg}^{(0)}
                      \Bigr]
                  + \delta m_1^{(0)}\hat{\gamma}_{qg}^{(0)}
                       \Bigl[
                               \gamma_{gg}^{(0)}
                              -\gamma_{qq}^{(0)}
                              +2\beta_0
                              +4\beta_{0,Q}
                      \Bigr]
                  -\delta m_2^{(-1)}\hat{\gamma}_{qg}^{(0)}
             \Biggr)
\N\\ &&
                 +a_{Qg}^{(3)}
                  \Biggr]~. \label{AhhhQg3} \\
   \Ahathat_{qg,Q}^{(3)}&=&
                   n_f\Bigl(\frac{\hat{m}^2}{\mu^2}\Bigr)^{3\ep/2}
                     \Biggl[
           \frac{\hat{\gamma}_{qg}^{(0)}}{6\ep^3}
             \Biggl(
                    \gamma_{gq}^{(0)}\hat{\gamma}_{qg}^{(0)}
                   +2\beta_{0,Q}\Bigl[
                                  \gamma_{gg}^{(0)}
                                 -\gamma_{qq}^{(0)}
                                 +2\beta_0 
                                \Bigr]
             \Biggr)
          +\frac{1}{\ep^2}
             \Biggl(
                   \frac{\hat{\gamma}_{qg}^{(0)}}{6} \Bigl[
                                     2\hat{\gamma}_{gg}^{(1)}
\N\\ &&
                                    +\hat{\gamma}_{qq}^{(1), {\sf PS}}
                                    -2\hat{\gamma}_{qq}^{(1), {\sf NS}}
                                    +4\beta_{1,Q}
                             \Bigr]
                   +\frac{\hat{\gamma}_{qg}^{(1)}\beta_{0,Q}}{3}
             \Biggr)
          +\frac{1}{\ep}
             \Biggl(
                   \frac{\hat{\tilde{\gamma}}_{qg}^{(2)}}{3}
                  +\hat{\gamma}_{qg}^{(0)}\Bigl[
                                            a_{gg,Q}^{(2)}
                                           -a_{qq,Q}^{(2),{\sf NS}}
\N\\ &&
                                           +\beta_{1,Q}^{(1)}
                                          \Bigr]
                  -\frac{\hat{\gamma}_{qg}^{(0)}\zeta_2}{16}\Bigl[
                                    \gamma_{gq}^{(0)}\hat{\gamma}_{qg}^{(0)}
                                   +2\beta_{0,Q}\Bigl\{
                                         \gamma_{gg}^{(0)}
                                        -\gamma_{qq}^{(0)}
                                        +2\beta_0
                                                \Bigr\}
                                                           \Bigr]
            \Biggr)
         +\frac{a_{qg,Q}^{(3)}}{n_f}
                         \Biggr]~.\label{Ahhhqg3Q}
  \end{eqnarray}  
  The renormalized expressions are
   \begin{eqnarray}
   A_{Qg}^{(3), \MS}&=&
                  \frac{\hat{\gamma}_{qg}^{(0)}}{48}
                     \Biggl\{
                             (n_f+1)\gamma_{gq}^{(0)}\hat{\gamma}_{qg}^{(0)}
                            +\gamma_{gg}^{(0)}\Bigl(
                                      \gamma_{gg}^{(0)}
                                     -2\gamma_{qq}^{(0)}
                                     +6\beta_0
                                     +14\beta_{0,Q}
                                              \Bigr)
                            +\gamma_{qq}^{(0)}\Bigl(
                                      \gamma_{qq}^{(0)}
                                     -6\beta_0
\N\\ &&
                                     -8\beta_{0,Q}
                                              \Bigr)
                            +8\beta_0^2
                            +28\beta_{0,Q}\beta_0
                            +24\beta_{0,Q}^2
                     \Biggr\}
                     \ln^3 \Bigl(\frac{m^2}{\mu^2}\Bigr)
                    +\frac{1}{8}\Biggl\{
                            \hat{\gamma}_{qg}^{(1)}
                                \Bigl(
                                        \gamma_{qq}^{(0)}
                                       -\gamma_{gg}^{(0)}
                                       -4\beta_0
\N\\ &&
                                       -6\beta_{0,Q}
                                \Bigr)
                           +\hat{\gamma}_{qg}^{(0)}
                                \Bigl(
                                        \hat{\gamma}_{gg}^{(1)}
                                       -\gamma_{gg}^{(1)}
                                       +(1-n_f) \hat{\gamma}_{qq}^{(1), {\sf PS}}
                                       +\gamma_{qq}^{(1), {\sf NS}}
                                      +\hat{\gamma}_{qq}^{(1), {\sf NS}}
                                      -2\beta_1
\N\\ &&
                                      -2\beta_{1,Q}
                                \Bigr)
                     \Biggr\}
                     \ln^2 \Bigl(\frac{m^2}{\mu^2}\Bigr)
                    +\Biggl\{
                            \frac{\hat{\gamma}_{qg}^{(2)}}{2}
                           -n_f\frac{\hat{\tilde{\gamma}}_{qg}^{(2)}}{2}
                           +\frac{a_{Qg}^{(2)}}{2}
                                \Bigl(
                                        \gamma_{qq}^{(0)}
                                       -\gamma_{gg}^{(0)}
                                       -4\beta_0
                                       -4\beta_{0,Q}
                                \Bigr)
\N\\ &&
                           +\frac{\hat{\gamma}_{qg}^{(0)}}{2}
                                \Bigl(
                                       a_{gg,Q}^{(2)}
                                      -n_fa_{Qq}^{(2), {\sf PS}}
                                \Bigr)
                           +\frac{\hat{\gamma}_{qg}^{(0)}\zeta_2}{16}
                                \Bigl(
                                       -(n_f+1)\gamma_{gq}^{(0)}
                                             \hat{\gamma}_{qg}^{(0)}
                                       +\gamma_{gg}^{(0)}\Bigl[
                                                2\gamma_{qq}^{(0)}
                                               -\gamma_{gg}^{(0)}
                                               -6\beta_0
\N\\ &&
                                               -6\beta_{0,Q}
                                                         \Bigr]
                                       -4\beta_0[2\beta_0+3\beta_{0,Q}]
                                       +\gamma_{qq}^{(0)}\Bigl[
                                               -\gamma_{qq}^{(0)}
                                               +6\beta_0
                                               +4\beta_{0,Q}
                                                         \Bigr]
                                \Bigr)
                     \Biggr\}
                     \ln \Bigl(\frac{m^2}{\mu^2}\Bigr)
                           +\overline{a}_{Qg}^{(2)}
                                \Bigl(
                                        \gamma_{gg}^{(0)}
\N\\ 
&&
                                       -\gamma_{qq}^{(0)}
                                       +4\beta_0
                                       +4\beta_{0,Q}
                                \Bigr)
                           +\hat{\gamma}_{qg}^{(0)}\Bigl(
                                        n_f\overline{a}_{Qq}^{(2), {\sf PS}}
                                       -\overline{a}_{gg,Q}^{(2)}
                                                   \Bigr)
                           +\frac{\hat{\gamma}_{qg}^{(0)}\zeta_3}{48}
                                \Bigl(
                                        (n_f+1)\gamma_{gq}^{(0)}
                                                \hat{\gamma}_{qg}^{(0)}
\N\\ &&
                                       +\gamma_{gg}^{(0)}\Bigl[
                                                 \gamma_{gg}^{(0)}
                                               -2\gamma_{qq}^{(0)}
                                               +6\beta_0
                                               -2\beta_{0,Q}
                                                         \Bigr]
                                       +\gamma_{qq}^{(0)}\Bigl[
                                                \gamma_{qq}^{(0)}
                                               -6\beta_0
                                                         \Bigr]
                                       +8\beta_0^2
                                       -4\beta_0\beta_{0,Q}
                                       -24\beta_{0,Q}^2
                                \Bigr)
\N\\ 
&&
                           +\frac{\hat{\gamma}_{qg}^{(1)}\beta_{0,Q}\zeta_2}{8}
                           +\frac{\hat{\gamma}_{qg}^{(0)}\zeta_2}{16}
                                \Bigl(
                                        \gamma_{gg}^{(1)}
                                       -\hat{\gamma}_{qq}^{(1), {\sf NS}}
                                       -\gamma_{qq}^{(1), {\sf NS}}
                                       -\hat{\gamma}_{qq}^{(1),{\sf PS}}
                                       +2\beta_1
                                       +2\beta_{1,Q}
                                \Bigr)
\N\\ &&
                           +\frac{\delta m_1^{(-1)}}{8}
                                \Bigl(
                                       16 a_{Qg}^{(2)}
                                  +\hat{\gamma}_{qg}^{(0)}\Bigl[
                                           -24 \delta m_1^{(0)}
                                           -8 \delta m_1^{(1)}
                                           -\zeta_2\beta_0
                                           -9\zeta_2\beta_{0,Q}
                                                          \Bigr]
                                \Bigr)
                           +\frac{\delta m_1^{(0)}}{2}
                                \Bigl(
                                       2\hat{\gamma}_{qg}^{(1)}
\N\\ &&
                                      -\delta m_1^{(0)}
                                       \hat{\gamma}_{qg}^{(0)} 
                                \Bigr)
                           +\delta m_1^{(1)}\hat{\gamma}_{qg}^{(0)}
                                \Bigl(
                                        \gamma_{qq}^{(0)}
                                       -\gamma_{gg}^{(0)}
                                       -2\beta_0
                                       -4 \beta_{0,Q}
                                \Bigr)
                           +\delta m_2^{(0)}\hat{\gamma}_{qg}^{(0)}
                           +a_{Qg}^{(3)}~. \label{AQg3MSren} 
\end{eqnarray}\begin{eqnarray}
    A_{qg,Q}^{(3), \MS}&=&n_f\Biggl[
         \frac{\hat{\gamma}_{qg}^{(0)}}{48}\Biggl\{
                            \gamma_{gq}^{(0)}\hat{\gamma}_{qg}^{(0)}
                           +2\beta_{0,Q}\Bigl(
                                      \gamma_{gg}^{(0)}
                                     -\gamma_{qq}^{(0)}
                                     +2\beta_0
                                        \Bigr)
                                           \Biggr\}
                         \ln^3 \Bigl(\frac{m^2}{\mu^2}\Bigr) 
        +\frac{1}{8}\Biggl\{
                     2\hat{\gamma}_{qg}^{(1)}\beta_{0,Q}
\N\\ &&
                           +\hat{\gamma}_{qg}^{(0)} 
                                      \Bigl(
                                 \hat{\gamma}_{qq}^{(1), {\sf PS}}
                                -\hat{\gamma}_{qq}^{(1), {\sf NS}}
                                +\hat{\gamma}_{gg}^{(1)}
                                +2\beta_{1,Q}
                                        \Bigr)
                   \Biggr\}
                         \ln^2 \Bigl(\frac{m^2}{\mu^2}\Bigr)
        +\frac{1}{2}\Biggl\{
                       \hat{\tilde{\gamma}}_{qg}^{(2)}
                      +\hat{\gamma}_{qg}^{(0)}  \Bigl(
                                a_{gg,Q}^{(2)}
\N\\ &&
                               -a_{qq,Q}^{(2),{\sf NS}}
                               +\beta_{1,Q}^{(1)}
                                        \Bigr)
                      -\frac{\hat{\gamma}_{qg}^{(0)}}{8}\zeta_2 \Bigl(
                                     \gamma_{gq}^{(0)}\hat{\gamma}_{qg}^{(0)}
                                    +2\beta_{0,Q}\Bigl[
                                                 \gamma_{gg}^{(0)}
                                                -\gamma_{qq}^{(0)}
                                                +2\beta_0
                                                 \Bigr]
                                        \Bigr)
                  \Biggr\}
                         \ln \Bigl(\frac{m^2}{\mu^2}\Bigr)
\N\\ &&
           +\hat{\gamma}_{qg}^{(0)}\Bigl(
                              \overline{a}_{qq,Q}^{(2),{\sf NS}}
                             -\overline{a}_{gg,Q}^{(2)}
                             -\beta_{1,Q}^{(2)}
                                   \Bigr)
           +\frac{\hat{\gamma}_{qg}^{(0)}}{48}\zeta_3\Bigl(
                            \gamma_{gq}^{(0)}\hat{\gamma}_{qg}^{(0)}
                                    +2\beta_{0,Q}\Bigl[
                                                 \gamma_{gg}^{(0)}
                                                -\gamma_{qq}^{(0)}
                                                +2\beta_0
                                                 \Bigr]
                                   \Bigr)
\N\\ &&
           -\frac{\zeta_2}{16}\Bigl(
                     \hat{\gamma}_{qg}^{(0)}\hat{\gamma}_{qq}^{(1), {\sf PS}}
                    +2\hat{\gamma}_{qg}^{(1)}\beta_{0,Q}
                                   \Bigr)
           +\frac{a_{qg,Q}^{(3)}}{n_f}
              \Biggr]~. \label{Aqg3QMSren}
   \end{eqnarray}
%
%
%
 \subsection{\boldmath $A_{gq,Q}$}
  \label{SubSec-AgqQ}
  The $gq$--contributions start at $O(a_s^2)$,
  \begin{eqnarray}
   A_{gq,Q}&=&
                       a_s^2A_{gq,Q}^{(2)}
                       +a_s^3A_{gq,Q}^{(3)}
                       +O(a_s^4)~. \label{AgqQpert}
  \end{eqnarray}
  The renormalization formulae in the {\sf MOM}--scheme read
  \begin{eqnarray}
    A_{gq,Q}^{(2),\MOM}&=&
                     \hat{A}_{gq,Q}^{(2),\MOM}
                    +Z_{gq}^{-1,(2)}(n_f+1)
                    -Z_{gq}^{-1,(2)}(n_f)
\N\\ &&
                    +\Bigl(
                           \hat{A}_{gg,Q}^{(1),\MOM}
                          +Z_{gg}^{-1,(1)}(n_f+1)
                          -Z_{gg}^{-1,(1)}(n_f)
                      \Bigr)\Gamma_{gq}^{-1,(1)}~, \\
    A_{gq,Q}^{(3),\MOM}&=& 
                        \hat{A}_{gq,Q}^{(3),\MOM}
                       +Z^{-1,(3)}_{gq}(n_f+1)
                       -Z^{-1,(3)}_{gq}(n_f)
                       +Z^{-1,(1)}_{gg}(n_f+1)\hat{A}_{gq,Q}^{(2),\MOM}
\N\\ &&
                       +Z^{-1,(1)}_{gq}(n_f+1)\hat{A}_{qq}^{(2),\MOM}
                       +\Bigl[ \hat{A}_{gg,Q}^{(1),\MOM}
                              +Z^{-1,(1)}_{gg}(n_f+1)
\N\\ &&
                              -Z^{-1,(1)}_{gg}(n_f)
                        \Bigr]
                              \Gamma^{-1,(2)}_{gq}(n_f)
                       +\Bigl[ \hat{A}_{gq,Q}^{(2),\MOM}
                              +Z^{-1,(2)}_{gq}(n_f+1)
                              -Z^{-1,(2)}_{gq}(n_f)
                        \Bigr]
                              \Gamma^{-1,(1)}_{qq}(n_f)
\N\\ &&
                       +\Bigl[ \hat{A}_{gg,Q}^{(2),\MOM}
                              +Z^{-1,(2)}_{gg}(n_f+1)
                              -Z^{-1,(2)}_{gg}(n_f)
                              +Z^{-1,(1)}_{gg}(n_f+1)\hat{A}_{gg,Q}^{(1),\MOM}
\N\\ &&
                              +Z^{-1,(1)}_{gq}(n_f+1)\hat{A}_{Qg}^{(1),\MOM}
                        \Bigr]
                              \Gamma^{-1,(1)}_{gq}(n_f)
                      \label{AgqQRen1}~,
   \end{eqnarray}
   while the unrenormalized expressions are
   \begin{eqnarray}
    \Ahathat_{gq,Q}^{(2)}&=&\Bigl(\frac{\hat{m}^2}{\mu^2}\Bigr)^{\ep}\Biggl[
                     \frac{2\beta_{0,Q}}{\ep^2}\gamma_{gq}^{(0)}
                    +\frac{\hat{\gamma}_{gq}^{(1)}}{2\ep}
                    +a_{gq,Q}^{(2)}
                    +\overline{a}_{gq,Q}^{(2)}\ep
                        \Biggr]~, \label{Ahhhgq2Q} \\
    \Ahathat_{gq,Q}^{(3)}&=&
        \Bigl(\frac{\hat{m}^2}{\mu^2}\Bigr)^{3\ep/2}\Biggl\{
               -\frac{\gamma_{gq}^{(0)}}{3\ep^3}
                   \Biggl(
                           \gamma_{gq}^{(0)}\hat{\gamma}_{qg}^{(0)}
                          +\Bigl[
                                 \gamma_{qq}^{(0)}
                                -\gamma_{gg}^{(0)}
                                +10\beta_0
                                +24\beta_{0,Q}
                                      \Bigr]\beta_{0,Q}
                   \Biggr)
               +\frac{1}{\ep^2}
                   \Biggl(
                      \gamma_{gq}^{(1)}\beta_{0,Q}
\N\\ &&
                     +\frac{\hat{\gamma}_{gq}^{(1)}}{3}\Bigl[
                          \gamma_{gg}^{(0)}
                         -\gamma_{qq}^{(0)}
                         -4\beta_0
                         -6\beta_{0,Q}
                                                        \Bigr]
                     +\frac{\gamma_{gq}^{(0)}}{3}\Bigl[
                                \hat{\gamma}_{qq}^{(1), {\sf NS}}
                               +\hat{\gamma}_{qq}^{(1), {\sf PS}}
                               -\hat{\gamma}_{gg}^{(1)}
                               +2\beta_{1,Q}
                                                        \Bigr]
\N\\ &&
                     -4\delta m_1^{(-1)}\beta_{0,Q}\gamma_{gq}^{(0)} 
                   \Biggr)
               +\frac{1}{\ep}
                   \Biggl(
                                \frac{\hat{\gamma}_{gq}^{(2)}}{3}
                               +a_{gq,Q}^{(2)}\Bigl[
                                       \gamma_{gg}^{(0)}
                                      -\gamma_{qq}^{(0)}
                                      -6\beta_{0,Q}
                                      -4\beta_0
                                              \Bigr]
                               +\gamma_{gq}^{(0)}\Bigl[
                                     a_{qq,Q}^{(2),{\sf NS}}
                                    +a_{Qq}^{(2),{\sf PS}}
\N\\ &&
                                    -a_{gg,Q}^{(2)}
                                              \Bigr]
                              +\gamma_{gq}^{(0)}\beta_{1,Q}^{(1)}
                               +\frac{\gamma_{gq}^{(0)}\zeta_2}{8}\Bigl[
                                     \gamma_{gq}^{(0)}\hat{\gamma}_{qg}^{(0)}
                                    +\beta_{0,Q} ( 
                                        \gamma_{qq}^{(0)}
                                        -\gamma_{gg}^{(0)}
                                        +10\beta_0
                                                 )
                                              \Bigr]
                               -\delta m_1^{(-1)}\hat{\gamma}_{gq}^{(1)}
\nonumber
\end{eqnarray}\begin{eqnarray}
&&
                               -4\delta m_1^{(0)}\beta_{0,Q}\gamma_{gq}^{(0)}
                   \Biggr)
                +a_{gq,Q}^{(3)}
                        \Biggr\}~. \label{AhhhgqQ3}
   \end{eqnarray}
   The contributions to the renormalized operator matrix element are given by
   \begin{eqnarray}
    A_{gq,Q}^{(2), \MS}&=&\frac{\beta_{0,Q}\gamma_{gq}^{(0)}}{2}
    \ln^2 \Bigl(\frac{m^2}{\mu^2}\Bigr)
    +\frac{\hat{\gamma}_{gq}^{(1)}}{2} \ln \Bigl(\frac{m^2}{\mu^2}\Bigr)
    +a_{gq,Q}^{(2)}-\frac{\beta_{0,Q}\gamma_{gq}^{(0)}}{2}\zeta_2~,
    \label{Agq2QMSren} \\
    A_{gq,Q}^{(3), \MS}&=&
                     -\frac{\gamma_{gq}^{(0)}}{24}
                      \Biggl\{
                          \gamma_{gq}^{(0)}\hat{\gamma}_{qg}^{(0)}
                         +\Bigl(
                              \gamma_{qq}^{(0)}
                             -\gamma_{gg}^{(0)}
                             +10\beta_0
                             +24\beta_{0,Q}
                                     \Bigr)\beta_{0,Q}
                      \Biggr\}
                           \ln^3 \Bigl(\frac{m^2}{\mu^2}\Bigr)
                     +\frac{1}{8}\Biggl\{
                         6\gamma_{gq}^{(1)}\beta_{0,Q}
\N\\ &&
                        +\hat{\gamma}_{gq}^{(1)}\Bigl(
                                      \gamma_{gg}^{(0)}
                                     -\gamma_{qq}^{(0)}
                                     -4\beta_0
                                     -6\beta_{0,Q}
                                                 \Bigr)
                        +\gamma_{gq}^{(0)}\Bigl(
                                       \hat{\gamma}_{qq}^{(1), {\sf NS}}
                                      +\hat{\gamma}_{qq}^{(1), {\sf PS}}
                                      -\hat{\gamma}_{gg}^{(1)}
                                      +2\beta_{1,Q}
                                                 \Bigr)
                      \Biggr\}
                           \ln^2 \Bigl(\frac{m^2}{\mu^2}\Bigr)
\N\\ &&
                     +\frac{1}{8}\Biggl\{
                              4\hat{\gamma}_{gq}^{(2)}
                            + 4a_{gq,Q}^{(2)}         \Bigl(
                                    \gamma_{gg}^{(0)}
                                   -\gamma_{qq}^{(0)}
                                   -4\beta_0
                                   -6\beta_{0,Q}
                                                       \Bigr)
                            + 4\gamma_{gq}^{(0)}       \Bigl(
                                      a_{qq,Q}^{(2),{\sf NS}}
                                     +a_{Qq}^{(2),{\sf PS}}
                                     -a_{gg,Q}^{(2)}
\N\\ &&
                                     +\beta_{1,Q}^{(1)}
                                                       \Bigr)
                            + \gamma_{gq}^{(0)}\zeta_2 \Bigl(
                               \gamma_{gq}^{(0)}\hat{\gamma}_{qg}^{(0)}
                               +\Bigl[
                                        \gamma_{qq}^{(0)}
                                       -\gamma_{gg}^{(0)}
                                       +12\beta_{0,Q}
                                       +10\beta_0
                                            \Bigr]\beta_{0,Q}
                                                       \Bigr)
                      \Biggr\}
                           \ln \Bigl(\frac{m^2}{\mu^2}\Bigr)
\N\\ &&
                  + \overline{a}_{gq,Q}^{(2)} \Bigl(
                                       \gamma_{qq}^{(0)}
                                      -\gamma_{gg}^{(0)}
                                      +4\beta_0
                                      +6\beta_{0,Q}
                                             \Bigr)
                  + \gamma_{gq}^{(0)} \Bigl(
                                       \overline{a}_{gg,Q}^{(2)}
                                      -\overline{a}_{Qq}^{(2),{\sf PS}}
                                      -\overline{a}_{qq,Q}^{(2),{\sf NS}}
                                             \Bigr)
                -\gamma_{gq}^{(0)}\beta_{1,Q}^{(2)}
\N\\ && 
                -\frac{\gamma_{gq}^{(0)}\zeta_3}{24} \Bigl(
                           \gamma_{gq}^{(0)}\hat{\gamma}_{qg}^{(0)}
                          +\Bigl[
                                   \gamma_{qq}^{(0)}
                                  -\gamma_{gg}^{(0)}
                                  +10\beta_0
                                      \Bigr]\beta_{0,Q}
                                             \Bigr)
                -\frac{3\gamma_{gq}^{(1)}\beta_{0,Q}\zeta_2}{8}
                +2 \delta m_1^{(-1)} a_{gq,Q}^{(2)}
\N\\ &&
                +\delta m_1^{(0)} \hat{\gamma}_{gq}^{(1)}
                +4 \delta m_1^{(1)} \beta_{0,Q} \gamma_{gq}^{(0)}
                +a_{gq,Q}^{(3)}~. \label{Agq3QMSren}
    \label{AgqQ3Ren1}
   \end{eqnarray}
%
%
%
 \subsection{\boldmath $A_{gg,Q}$}
  \label{SubSec-AggQ}
  The $gg$--contributions start at $O(a_s)$,
  \begin{eqnarray}
   A_{gg,Q}&=&
                        a_sA_{gg,Q}^{(1)}
                       +a_s^2A_{gg,Q}^{(2)}
                       +a_s^3A_{gg,Q}^{(3)}
                       +O(a_s^4)~. \label{AggQpert}
  \end{eqnarray}
  The corresponding renormalization formulae read in the {\sf MOM}--scheme
  \begin{eqnarray}
    A_{gg,Q}^{(1), \MOM}&=&
                    \hat{A}_{gg,Q}^{(1), \MOM}
                   +Z^{-1,(1)}_{gg}(n_f+1)
                   -Z^{-1,(1)}_{gg}(n_f)
     ~, \label{AggQ1ren1} \\ 
    A_{gg,Q}^{(2), \MOM}&=&
                    \hat{A}_{gg,Q}^{(2), \MOM}
                   +Z^{-1,(2)}_{gg}(n_f+1)
                   -Z^{-1,(2)}_{gg}(n_f)
                   +Z^{-1,(1)}_{gg}(n_f+1)\hat{A}_{gg,Q}^{(1), \MOM}
 \N\\ &&
                   +Z^{-1,(1)}_{gq}(n_f+1)\hat{A}_{Qg}^{(1), \MOM}
 \N\\ &&
                   +\Bigl[ \hat{A}_{gg,Q}^{(1), \MOM}
                          +Z_{gg}^{-1,(1)}(n_f+1)
                          -Z_{gg}^{-1,(1)}(n_f)
                    \Bigr]\Gamma^{-1,(1)}_{gg}(n_f)
    ~, \label{AggQ1ren2} \\
    A_{gg,Q}^{(3), \MOM}&=&
                     \hat{A}_{gg,Q}^{(3), \MOM}
                    +Z^{-1,(3)}_{gg}(n_f+1)
                    -Z^{-1,(3)}_{gg}(n_f)
                    +Z^{-1,(2)}_{gg}(n_f+1)\hat{A}_{gg,Q}^{(1), \MOM}
\N\\ &&
                    +Z^{-1,(1)}_{gg}(n_f+1)\hat{A}_{gg,Q}^{(2), \MOM}
                    +Z^{-1,(2)}_{gq}(n_f+1)\hat{A}_{Qg}^{(1), \MOM}
                    +Z^{-1,(1)}_{gq}(n_f+1)\hat{A}_{Qg}^{(2), \MOM}
\N\\ &&
                    +\Bigl[
                            \hat{A}_{gg,Q}^{(1), \MOM}
                           +Z^{-1,(1)}_{gg}(n_f+1)
                           -Z^{-1,(1)}_{gg}(n_f)
                     \Bigr]\Gamma^{-1,(2)}_{gg}(n_f)
                    +\Bigl[ \hat{A}_{gg,Q}^{(2), \MOM}
  \N\\ &&
                           +Z^{-1,(2)}_{gg}(n_f+1)
                           -Z^{-1,(2)}_{gg}(n_f)
                           +Z^{-1,(1)}_{gq}(n_f+1)A_{Qg}^{(1), \MOM}
\N\\ &&
                           +Z^{-1,(1)}_{gg}(n_f+1)A_{gg,Q}^{(1), \MOM}
                     \Bigr]\Gamma^{-1,(1)}_{gg}(n_f)
\N\\ &&
                    +\Bigl[ \hat{A}_{gq,Q}^{(2), \MOM}
                           +Z^{-1,(2)}_{gq}(n_f+1)
                           -Z^{-1,(2)}_{gq}(n_f)
                     \Bigr]\Gamma^{-1,(1)}_{qg}(n_f)
         ~.\label{AggQ1ren3}
  \end{eqnarray}
  The general structure of the unrenormalized $1$--loop result
  is then given by
  \begin{eqnarray}
    \Ahathat_{gg,Q}^{(1)}&=&
             \Bigl(\frac{\hat{m}^2}{\mu^2}\Bigr)^{\ep/2}\Biggl(
                          \frac{\hat{\gamma}_{gg}^{(0)}}{\ep}
                         +a_{gg,Q}^{(1)}
                         +\ep\overline{a}_{gg,Q}^{(1)}
                         +\ep^2\overline{\overline{a}}_{gg,Q}^{(1)}
                        \Biggr)
                        ~. \label{AggQ1unren1}
   \end{eqnarray}
   An explicit calculation reveals
   \begin{eqnarray}
    \Ahathat_{gg,Q}^{(1)}&=&
                \Bigl(\frac{\hat{m}^2}{\mu^2}\Bigr)^{\ep/2}
                \Bigl(-\frac{2\beta_{0,Q}}{\ep}\Bigr)
                \exp \Bigl(\sum_{i=2}^{\infty}\frac{\zeta_i}{i}
                       \Bigl(\frac{\ep}{2}\Bigr)^{i}\Bigr)
                        ~. \label{AggQ1unren2}
   \end{eqnarray}
   Using Eq. (\ref{AggQ1unren2}), the $2$--loop term
   is given by
   \begin{eqnarray}
    \Ahathat_{gg,Q}^{(2)}&=&
             \Bigl(\frac{\hat{m}^2}{\mu^2}\Bigr)^{\ep}
                  \Biggl[
                          \frac{1}{2\ep^2}
                             \Bigl\{
                                \gamma_{gq}^{(0)}\hat{\gamma}_{qg}^{(0)}
                               +2\beta_{0,Q}
                                    \Bigl(
                                           \gamma_{gg}^{(0)}
                                          +2\beta_0
                                          +4\beta_{0,Q}
                                    \Bigr)
                             \Bigr\}
                         +\frac{\hat{\gamma}_{gg}^{(1)}
                                  +4\delta m_1^{(-1)}\beta_{0,Q}}{2\ep}
 \N\\ &&
                         +a_{gg,Q}^{(2)}
                         +2\delta m_1^{(0)}\beta_{0,Q}
                         +\beta_{0,Q}^2\zeta_2
                         +\ep\Bigl[\overline{a}_{gg,Q}^{(2)}
                                   +2\delta m_1^{(1)}\beta_{0,Q}
                                   +\frac{\beta_{0,Q}^2\zeta_3}{6}
                             \Bigr]
                 \Biggr]~. \label{AhhhggQ2}
  \end{eqnarray}
   For Eq. (\ref{AhhhggQ2}) the same as for Eq. (\ref{AhhhQg2}) holds.
   We have already included one--particle reducible contributions 
   and terms stemming from mass renormalization in order to refer to  
   the notation of Refs. \cite{Buza:1995ie,Buza:1996wv}. The $3$--loop contribution 
   becomes
   \begin{eqnarray}
    \Ahathat_{gg,Q}^{(3)}&=&
                  \Bigl(\frac{\hat{m}^2}{\mu^2}\Bigr)^{3\ep/2}
                     \Biggl[
           \frac{1}{\ep^3}
             \Biggl(
                  -\frac{\gamma_{gq}^{(0)}\hat{\gamma}_{qg}^{(0)}}{6}
                                \Bigl[
                                        \gamma_{gg}^{(0)}
                                       -\gamma_{qq}^{(0)}
                                       +6\beta_0
                                       +4n_f\beta_{0,Q}
                                       +10\beta_{0,Q}
                                \Bigr]
\N\\ &&
                  -\frac{2\gamma_{gg}^{(0)}\beta_{0,Q}}{3}
                                \Bigl[
                                        2\beta_0
                                       +7\beta_{0,Q}
                                \Bigr]
                  -\frac{4\beta_{0,Q}}{3}
                                \Bigl[
                                        2\beta_0^2
                                       +7\beta_{0,Q}\beta_0
                                       +6\beta_{0,Q}^2
                                \Bigr]
             \Biggr)
\N\\ &&
          +\frac{1}{\ep^2}
             \Biggl(
                   \frac{\hat{\gamma}_{qg}^{(0)}}{6}
                                \Bigl[
                                        \gamma_{gq}^{(1)}
                                       -(2n_f-1)\hat{\gamma}_{gq}^{(1)}
                                \Bigr]
                  +\frac{\gamma_{gq}^{(0)}\hat{\gamma}_{qg}^{(1)}}{3}
                  -\frac{\hat{\gamma}_{gg}^{(1)}}{3}                 
                                \Bigl[
                                        4\beta_0
                                       +7\beta_{0,Q}
                                \Bigr]
\N\\ &&
                  +\frac{2\beta_{0,Q}}{3}                 
                                \Bigl[
                                        \gamma_{gg}^{(1)}
                                       +\beta_1
                                       +\beta_{1,Q}
                                \Bigr]
                  +\frac{2\gamma_{gg}^{(0)}\beta_{1,Q}}{3}
                           +\delta m_1^{(-1)}
                                \Bigl[
                                    -\hat{\gamma}_{qg}^{(0)}\gamma_{gq}^{(0)}
                                    -2\beta_{0,Q}\gamma_{gg}^{(0)}
                                    -10\beta_{0,Q}^2
\N\\ &&
                                    -6\beta_{0,Q}\beta_0
                               \Bigr]
             \Biggr)
          +\frac{1}{\ep}
             \Biggl(
                   \frac{\hat{\gamma}_{gg}^{(2)}}{3}
                  -2(2\beta_0+3\beta_{0,Q})a_{gg,Q}^{(2)}
                  -n_f\hat{\gamma}_{qg}^{(0)}a_{gq,Q}^{(2)}
                  +\gamma_{gq}^{(0)}a_{Qg}^{(2)}
                  +\beta_{1,Q}^{(1)} \gamma_{gg}^{(0)}
\N\\ &&
                  +\frac{\gamma_{gq}^{(0)}\hat{\gamma}_{qg}^{(0)}\zeta_2}{16}
                                \Bigl[
                                         \gamma_{gg}^{(0)}
                                       - \gamma_{qq}^{(0)}
                                       +2(2n_f+1)\beta_{0,Q}
                                       +6\beta_0
                                \Bigr]
                  +\frac{\beta_{0,Q}\zeta_2}{4}
                                \Bigl[
                                       \gamma_{gg}^{(0)}
                                      \{2\beta_0-\beta_{0,Q}\}
                                       +4\beta_0^2
\N\\ &&
                                       -2\beta_{0,Q}\beta_0
                                       -12\beta_{0,Q}^2
                                \Bigr]
                           +\delta m_1^{(-1)}
                                \Bigl[
                                   -3\delta m_1^{(-1)}\beta_{0,Q}
                                   -2\delta m_1^{(0)}\beta_{0,Q}
                                   -\hat{\gamma}_{gg}^{(1)} 
                                \Bigr]
\N\\ &&
                           +\delta m_1^{(0)}
                                \Bigl[
                                   -\hat{\gamma}_{qg}^{(0)}\gamma_{gq}^{(0)} 
                                   -2\gamma_{gg}^{(0)}\beta_{0,Q}
                                   -4\beta_{0,Q}\beta_0
                                   -8\beta_{0,Q}^2
                                \Bigr]
                           +2 \delta m_2^{(-1)} \beta_{0,Q}
             \Biggr)
           +a_{gg,Q}^{(3)}
                  \Biggr]~. \label{Ahhhgg3Q}
   \end{eqnarray}
   The renormalized results are
   \begin{eqnarray}
    A_{gg,Q}^{(1), \MS}&=& - \beta_{0,Q} 
\ln\left(\frac{m^2}{\mu^2}\right)~,\\
    A_{gg,Q}^{(2), \MS}&=&
      \frac{1}{8}\Biggl\{
                          2\beta_{0,Q}
                             \Bigl(
                                    \gamma_{gg}^{(0)}
                                   +2\beta_0 
                             \Bigr)
                         +\gamma_{gq}^{(0)}\hat{\gamma}_{qg}^{(0)}
                         +8\beta_{0,Q}^2
                 \Biggr\}
                    \ln^2 \Bigl(\frac{m^2}{\mu^2}\Bigr)
                +\frac{\hat{\gamma}_{gg}^{(1)}}{2}
                     \ln \Bigl(\frac{m^2}{\mu^2}\Bigr)
\N\\ &&
                -\frac{\zeta_2}{8}\Bigl[
                                         2\beta_{0,Q}
                                            \Bigl(
                                                   \gamma_{gg}^{(0)}
                                                  +2\beta_0
                                            \Bigr)
                                        +\gamma_{gq}^{(0)}
                                            \hat{\gamma}_{qg}^{(0)}
                                  \Bigr]
                +a_{gg,Q}^{(2)}~,   \label{AggQ2MSren} 
\end{eqnarray}\begin{eqnarray} 
  A_{gg,Q}^{(3), \MS}&=&
                    \frac{1}{48}\Biggl\{
                            \gamma_{gq}^{(0)}\hat{\gamma}_{qg}^{(0)}
                                \Bigl(
                                        \gamma_{qq}^{(0)}
                                       -\gamma_{gg}^{(0)}
                                       -6\beta_0
                                       -4n_f\beta_{0,Q}
                                       -10\beta_{0,Q}
                                \Bigr)
                           -4
                                \Bigl(
                                        \gamma_{gg}^{(0)}\Bigl[
                                            2\beta_0
                                           +7\beta_{0,Q}
                                                         \Bigr]
\N\\ &&
                                       +4\beta_0^2
                                       +14\beta_{0,Q}\beta_0
                                       +12\beta_{0,Q}^2
                                \Bigr)\beta_{0,Q}
                     \Biggr\}
                     \ln^3 \Bigl(\frac{m^2}{\mu^2}\Bigr)
                    +\frac{1}{8}\Biggl\{
                            \hat{\gamma}_{qg}^{(0)}
                                \Bigl(
                                        \gamma_{gq}^{(1)}
                                       +(1-n_f)\hat{\gamma}_{gq}^{(1)}
                                \Bigr)
\N\\ &&
                           +\gamma_{gq}^{(0)}\hat{\gamma}_{qg}^{(1)}
                           +4\gamma_{gg}^{(1)}\beta_{0,Q}
                           -4\hat{\gamma}_{gg}^{(1)}[\beta_0+2\beta_{0,Q}]
                           +2\gamma_{gg}^{(0)}\beta_{1,Q}
                           +4[\beta_1+\beta_{1,Q}]\beta_{0,Q}
                     \Biggr\}
                     \ln^2 \Bigl(\frac{m^2}{\mu^2}\Bigr)
\N\\ &&
                    +\frac{1}{16}\Biggl\{
                            8\hat{\gamma}_{gg}^{(2)}
                           -8n_fa_{gq,Q}^{(2)}\hat{\gamma}_{qg}^{(0)}
                           -16a_{gg,Q}^{(2)}(2\beta_0+3\beta_{0,Q})
                           +8\gamma_{gq}^{(0)}a_{Qg}^{(2)}
                           +8\gamma_{gg}^{(0)}\beta_{1,Q}^{(1)}
\N\\ &&
                   +\gamma_{gq}^{(0)}\hat{\gamma}_{qg}^{(0)}\zeta_2
                                \Bigl(
                                        \gamma_{gg}^{(0)}
                                       -\gamma_{qq}^{(0)}
                                       +6\beta_0
                                       +4n_f\beta_{0,Q}
                                       +6\beta_{0,Q}
                                \Bigr)
\N\\ &&
                   +4\beta_{0,Q}\zeta_2
                                \Bigl( 
                                       \gamma_{gg}^{(0)}
                                      +2\beta_0
                                \Bigr)
                                \Bigl(
                                       2\beta_0
                                      +3\beta_{0,Q}
                                \Bigr)
                     \Biggr\}
                     \ln \Bigl(\frac{m^2}{\mu^2}\Bigr)
                   +2(2\beta_0+3\beta_{0,Q})\overline{a}_{gg,Q}^{(2)}
\N\\ &&
                   +n_f\hat{\gamma}_{qg}^{(0)}\overline{a}_{gq,Q}^{(2)}
                   -\gamma_{gq}^{(0)}\overline{a}_{Qg}^{(2)}
                   -\beta_{1,Q}^{(2)} \gamma_{gg}^{(0)}
                   +\frac{\gamma_{gq}^{(0)}\hat{\gamma}_{qg}^{(0)}\zeta_3}{48}
                                \Bigl(
                                        \gamma_{qq}^{(0)}
                                       -\gamma_{gg}^{(0)}
                                       -2[2n_f+1]\beta_{0,Q}
\N\\ &&
                                       -6\beta_0
                                \Bigr)
                   +\frac{\beta_{0,Q}\zeta_3}{12}
                                \Bigl(
                                        [\beta_{0,Q}-2\beta_0]\gamma_{gg}^{(0)}
                                       +2[\beta_0+6\beta_{0,Q}]\beta_{0,Q}
                                       -4\beta_0^2
                                \Bigr)
\N\\ &&
                   -\frac{\hat{\gamma}_{qg}^{(0)}\zeta_2}{16}
                                \Bigl(
                                        \gamma_{gq}^{(1)}
                                       +\hat{\gamma}_{gq}^{(1)}
                                \Bigr)
                   +\frac{\beta_{0,Q}\zeta_2}{8}
                                \Bigl(
                                        \hat{\gamma}_{gg}^{(1)}
                                      -2\gamma_{gg}^{(1)}
                                      -2\beta_1
                                      -2\beta_{1,Q}
                                \Bigr)
                           +\frac{\delta m_1^{(-1)}}{4}
                                \Bigl(
                                     8 a_{gg,Q}^{(2)}
\N\\ &&
                                    +24 \delta m_1^{(0)} \beta_{0,Q}
                                    +8 \delta m_1^{(1)} \beta_{0,Q} 
                                    +\zeta_2 \beta_{0,Q} \beta_0
                                    +9 \zeta_2 \beta_{0,Q}^2
                                \Bigr)
                           +\delta m_1^{(0)}
                                \Bigl(
                                     \beta_{0,Q} \delta m_1^{(0)}
                                    +\hat{\gamma}_{gg}^{(1)}
                                \Bigr)
\N\\ &&
                           +\delta m_1^{(1)}
                                \Bigl(
                                     \hat{\gamma}_{qg}^{(0)} \gamma_{gq}^{(0)}
                                    +2 \beta_{0,Q} \gamma_{gg}^{(0)}
                                    +4 \beta_{0,Q} \beta_0
                                    +8 \beta_{0,Q}^2
                                \Bigr)
                           -2 \delta m_2^{(0)} \beta_{0,Q}
                 +a_{gg,Q}^{(3)}~. \label{Agg3QMSren}
   \end{eqnarray}

\section{The Calculation of the Operator Matrix Elements}
\renewcommand{\theequation}{\thesection.\arabic{equation}}
\setcounter{equation}{0}
\label{sec-calc}

\begin{figure}[htb]
\begin{center}
\includegraphics[angle=0, width=2cm]{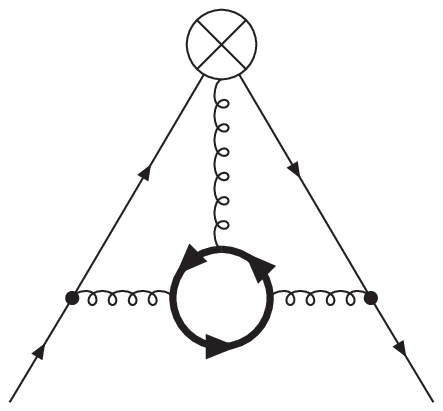}
\includegraphics[angle=0, width=2cm]{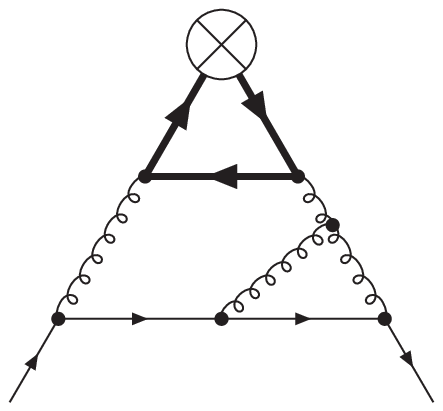}
\includegraphics[angle=0, width=2cm]{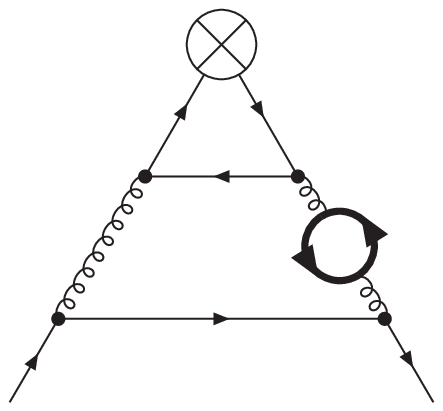}
\includegraphics[angle=0, width=2cm]{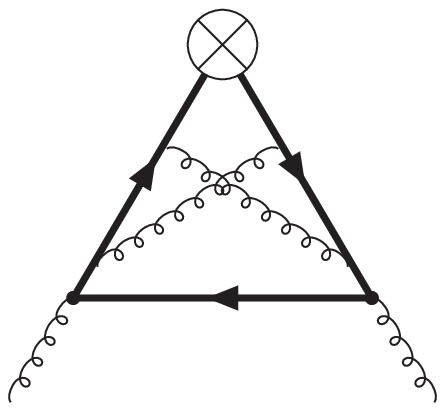}
\includegraphics[angle=0, width=2cm]{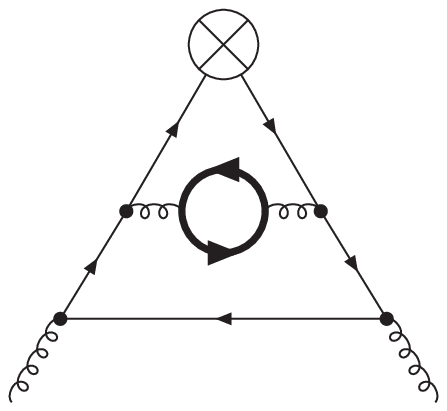}
\includegraphics[angle=0, width=2cm]{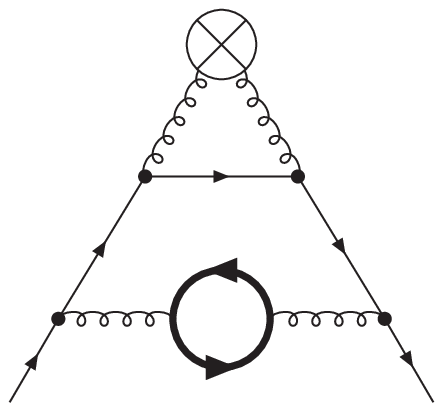}
\includegraphics[angle=0, width=2cm]{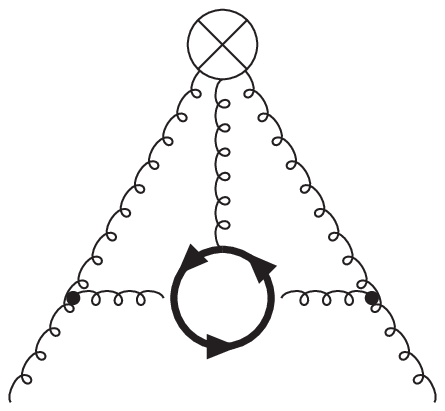}
\includegraphics[angle=0, width=2cm]{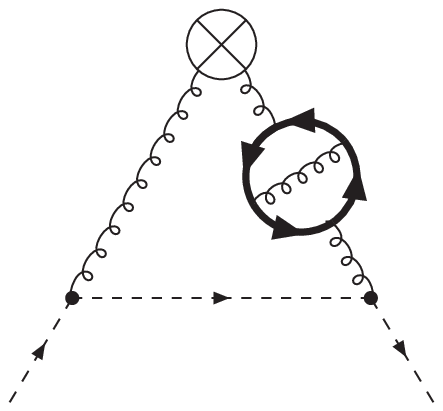}
\end{center}
{\small
\hspace*{5mm}
($\sf NS$) \hspace{1cm}
($\sf PS_H$) \hspace{1cm}
($\sf PS_l$) \hspace{1.1cm}
($\sf qg_H$) \hspace{1.1cm}
($\sf qg_l$) \hspace{1cm}
($\sf gq$) \hspace{1.1cm}
($\sf gg$) \hspace{1.1cm}
{\sf ghost}} 
\begin{center} 
\caption{\sf 
Examples for 3--loop diagrams contributing to the massive operator matrix 
elements: NS - non--singlet, ${\sf PS_{H,l}}$ - pure--singlet, singlet ${\sf 
qg_{H,l}}$, {\sf gq}, gg and ghost contributions. Here the coupling 
of the gauge boson to a heavy or light fermion line is labeled by {\sf H} 
and {\sf l},respectively. Thick lines: heavy quarks, curly lines:
gluons, full lines:  quarks, dashed lines: ghosts.}
\label{diaex}
\end{center}
\end{figure}

\vspace{1mm}
\noindent
In this chapter, we describe the computation of the 3--loop corrections to the
massive operator matrix elements in detail. Typical Feynman diagrams
contributing to the different channels are shown in Figure~\ref{diaex}, where
$\otimes$ denotes the corresponding composite operator insertions,
(\ref{COMP1}--\ref{COMP3}). The generation of these diagrams with the {\sf
  FORTRAN}--based program {\sf QGRAF}, cf. \cite{Nogueira:1991ex}, is
described in Section~\ref{SubSec-3LGen} along with the subsequent steps to
prepare the input for the {\sf FORM}--based program {\sf MATAD},
\cite{Steinhauser:2000ry}.  The latter allows the calculation of massive
tadpole integrals in $D$ dimensions up to three loops and relies on the {\sf
  MINCER} algorithm, \cite{Gorishnii:1989gt,Larin:1991fz}. The use of {\sf
  MATAD} and the projection onto fixed moments are explained in 
Section~\ref{SubSec-3LMatad}. Finally, we present our results for the fixed moments of
the $3$--loop OMEs and the fermionic contributions to the anomalous dimensions
in Section~\ref{SubSec-3LResUn}. The calculation is mainly performed by
using {\sf FORM} programs, \cite{Vermaseren:2000nd}, while in a few cases
codes have also been written in {\sf MAPLE}.
\noindent

  \subsection{\bf\boldmath Generation of Diagrams}
   \label{SubSec-3LGen}
  
  {\sf QGRAF} is a quite general program to generate Feynman diagrams and  
  allows to specify various kinds of particles and interactions. 
  Our main issue is to generate diagrams which contain composite
  operator insertions, cf. (\ref{COMP1}--\ref{COMP3}) and 
  appendix~\ref{App-FeynRules}, as special vertices.  

\begin{figure}[b]
  \begin{center}
  \includegraphics[angle=0, width=14.0cm]{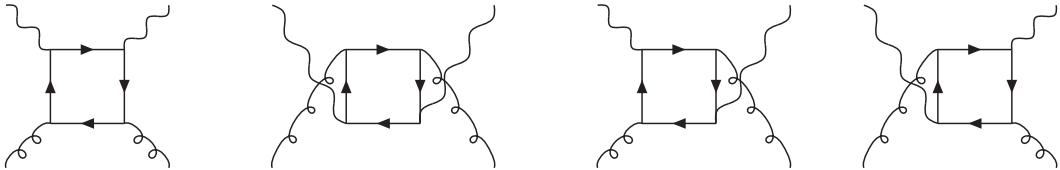}
   \end{center}
   \begin{center} 
    \caption{\sf Diagrams contributing to $A_{Qg}^{(1)}$ via 
                 the optical theorem. Wavy lines denote photons; for the 
    other lines, see Figure~\ref{diaex}. } 
     \label{GENOPINS6}
   \end{center}
 \end{figure}

\begin{figure}[t]
  \begin{center}
    \includegraphics[angle=0, width=14.0cm]{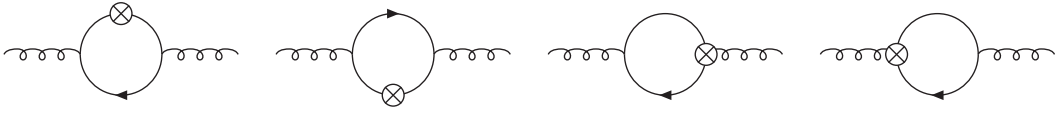}
   \end{center}
   \begin{center} 
    \caption{\sf Diagrams contributing to $A_{Qg}^{(1)}$.}
     \label{GENOPINS7}
   \end{center}
 \end{figure}

\begin{figure}[t]
  \begin{center}
   \includegraphics[angle=0, width=8.0cm]{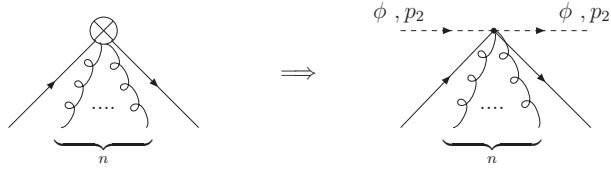}
   \end{center}
   \begin{center} 
    \caption{\sf Generation of the operator insertion.}
     \label{GENOPINS1}
   \end{center}
 \end{figure}

\begin{figure}[t]
   \begin{center}
   \includegraphics[angle=0, width=10.0cm]{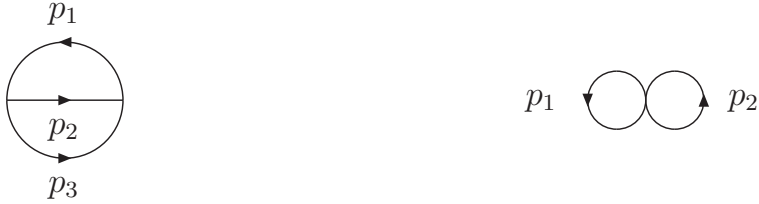}
   \end{center}
   \begin{center} 
    \caption{\sf $2$--Loop topologies, indicating labeling 
                 of momenta.}
     \label{MATADTOP1}
   \end{center}
   \end{figure}

\begin{figure}[t]
     \begin{center}
   \includegraphics[angle=0, width=4.0cm]{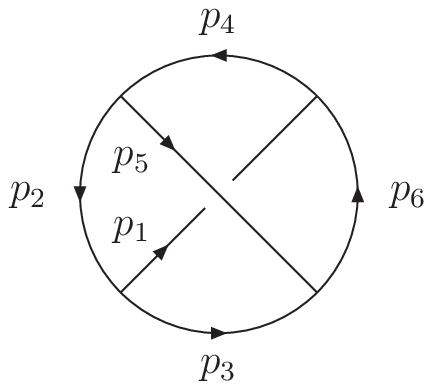}
     \end{center}
     \begin{center} 
      \caption{\sf Master $3$--loop topology for MATAD, indicating labeling 
                   of momenta.}
        \label{MATADTOP2}
    \end{center}
  \end{figure}

\begin{figure}[t]
     \begin{center}
    \includegraphics[angle=0, width=12.0cm]{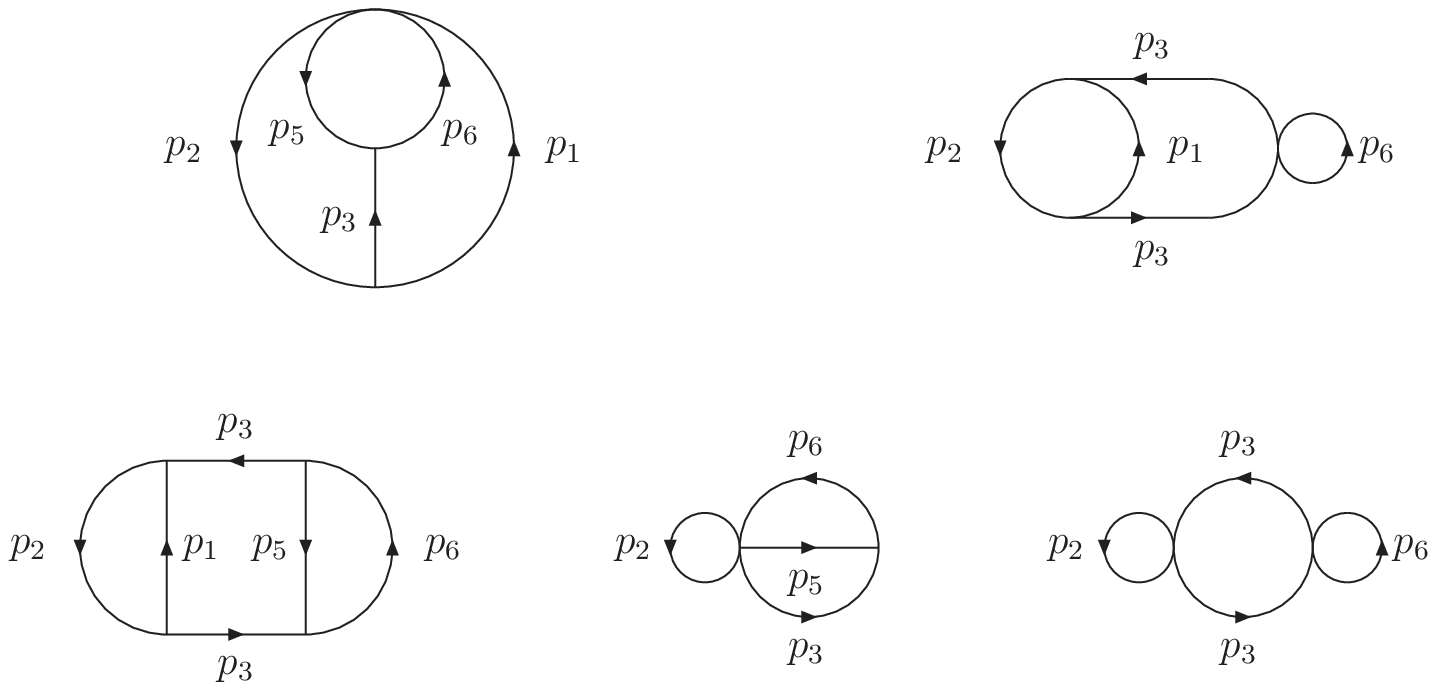}
     \end{center}
     \begin{center} 
      \caption{\sf Additional topologies contributing at the $3$--loop level.}
        \label{MATADTOP3}
    \end{center}
  \end{figure}

  To give an example, let us consider the contributions to $A_{Qg}^{(1)}$. 
  Within the light--cone expansion,
  \cite{LCE}, this term derives from the Born diagrams squared of the
  photon--gluon fusion process shown in Figure~\ref{GENOPINS6}.
  \noindent
  After expanding these diagrams with respect to the virtuality of the photon,
  the mass effects are then given by the diagrams 
  in Figure~\ref{GENOPINS7}. These are obtained 
  by contracting the lines between the external photons. 
  \noindent
  Thus, one may think of the operator insertion as being coupled to two
  external particles, an incoming and an outgoing one, which carry the same
  momentum. Therefore, one defines in the model file of {\sf QGRAF} vertices
  which resemble the operator insertions in this manner, using a scalar field
  $\phi$, which shall not propagate in order to ensure that there is only one
  of these vertices for each diagram. For the quarkonic operators, one defines
  the vertices
  \begin{eqnarray}
   \phi+\phi+q+\overline{q}+n~g~~, \hspace*{3mm} 0 \le n \le 3~, \label{phiquark}
  \end{eqnarray}
  which is illustrated in Figure~\ref{GENOPINS1}. 
  \noindent
  The same procedure can be used for the purely gluonic interactions 
  and one defines in this case
  \begin{eqnarray}
   \phi+\phi+n~g~~, \hspace*{3mm} 0 \le n \le 4~. \label{phigluon}
  \end{eqnarray}
  The number of diagrams we obtain contributing to each OME 
  is shown in Table~1. 
  \begin{table}[h]
  \label{table:numdiags}
  \begin{center}
  \newcommand{\m}{\hphantom{ }}
  \newcommand{\cc}[1]{\multicolumn{1}{c}{#1}}
  \renewcommand{\arraystretch}{2.0} 
  \begin{tabular}{|llllllll|}
   \hline\hline
    Term & \phantom{00}\#  & Term & \phantom{00}\# & Term & \phantom{00}\#  & Term & \phantom{00}\#  \\
   \hline\hline
     $A_{Qg}^{(3)}$             & 1358 
   & $A_{qg,Q}^{(3)}$           & \phantom{0}140
   & $A_{Qq}^{(3),{\sf PS}}$    & \phantom{0}125 
   & $A_{qq,Q}^{(3),{\sf PS}}$  & \phantom{000}8
 \\[0.75em]
     $A_{qq,Q}^{(3),{\sf NS}}$ & \phantom{0}129 
   & $A_{gq,Q}^{(3)}$          & \phantom{00}89
   & $A_{gg,Q}^{(3)}$          & \phantom{0}886
   &                           &               
  \\
   [3mm]
   \hline\hline
  \end{tabular}\\[2pt]
  \caption{\sf Number of diagrams contributing to the $3$--loop  
               heavy OMEs. }
  \end{center}
  \end{table}
  \renewcommand{\arraystretch}{1.0} 
  \noindent
  The next step consists in rewriting the output provided by {\sf QGRAF} in
  such a way, that the Feynman rules given in Appendix~\ref{App-FeynRules} can
  be inserted. Thus, one has to introduce Lorentz and color indices and align
  the fermion lines. Additionally, the integration momenta have to be written
  in such a way that {\sf MATAD} can handle them. For the latter step, all
  information on the types of particles, the operator insertion and the
  external momentum are irrelevant, leading to only two basic topologies to be
  considered at the $2$--loop level, which are shown in Figure~\ref{MATADTOP1}.
  \noindent
  Note, that in the case at hand the topology on the right--hand side of 
  Figure~\ref{MATADTOP1} always yields zero after integration.  At the $3$--loop
  level, the master topology is given in Figure~\ref{MATADTOP2}.
  \noindent
  From this topology, five types of diagrams are derived by shrinking
  various lines. These diagrams are shown in Figure~\ref{MATADTOP3}.
  \noindent
  After assigning the loop momenta, the Feynman rules are inserted.
  The computation of the Green's functions, which are associated to the 
  respective operator matrix elements, still contain trace terms and require
  the symmetrization of the Lorentz indices. It is convenient to project 
  these terms out by multiplying with an external source 
   \begin{eqnarray}
    J_N \equiv \Delta_{\mu_1}...\Delta_{\mu_N}~, \label{Jsource}
   \end{eqnarray}
   with $\Delta_{\mu}$ being a light-like vector, $\Delta^2 = 0$. 
   Additionally, one has to amputate the external field.\footnote{
   Note that we choose to renormalize the mass and the coupling
   multiplicatively and thus have to include self--energy insertions 
   containing a massive line on external legs.}
   The Green's functions in momentum space corresponding 
   to the local operators defined in Eqs. (\ref{COMP2}, \ref{COMP3}) 
   between gluonic states are then given by
   \begin{eqnarray} 
    \epsilon^\mu(p) G^{ab}_{Q,\mu\nu} \epsilon^\nu(p)&=&
    J_N \bra{A^a_{\mu}(p)} O^{\mu_1 ... \mu_N}_{Q} (0) \ket{A^b_{\nu}(p)}~, 
        \label{GabmnQgdef} \\
    \epsilon^\mu(p) G^{ab}_{q,Q,\mu\nu} \epsilon^\nu(p)&=&
    J_N \bra{A^a_{\mu}(p)} O^{\mu_1 ... \mu_N}_{q} (0) \ket{A^b_{\nu}(p)}_Q~,
     \label{GabmnqgQdef} \\
    \epsilon^\mu(p) G^{ab}_{g,Q,\mu\nu} \epsilon^\nu(p)&=&
    J_N \bra{A^a_{\mu}(p)} O^{\mu_1 ... \mu_N}_{g} (0) \ket{A^b_{\nu}(p)}~,
        \label{GabmnggQdef}
   \end{eqnarray}
   cf.~\cite{Buza:1995ie}, with $A^{a}_{\mu}$ an external gluon field with 
   color index $a$, Lorentz 
   index $\mu$, momentum $p$, and  $\epsilon^{\mu}(p)$ the 
   gluon polarization vector. 
   In the flavor non--singlet case, 
   Eq. (\ref{COMP1}), only one term contributes 
   \begin{eqnarray} 
    \overline{u}(p,s) G^{ij, {\sf NS}}_{q} \lambda_r u(p,s)&=&
    J_N\bra{\overline{\Psi}_i(p)}O^{\mu_1...\mu_N}_{q,r} (0)\ket{\Psi^j(p)}_Q~
     \label{GijNS}~,
   \end{eqnarray}
   with $u(p,s),~\overline{u}(p,s)$ being the bi--spinors of the 
   external quark and anti--quark, respectively. 
   The remaining singlet and pure--singlet 
   Green's functions with an external quark are given by, 
   \cite{Buza:1995ie}, 
   \begin{eqnarray} 
    \overline{u}(p,s) G^{ij, {\sf S}}_{Q}  u(p,s)&=&
    J_N\bra{\overline{\Psi}_i(p)} O^{\mu_1 ... \mu_N}_{Q} (0) \ket{\Psi^j(p)}, 
      \label{GijQqPS} \\ 
    \overline{u}(p,s) G^{ij, {\sf S}}_{q,Q}  u(p,s)&=&
    J_N\bra{\overline{\Psi}_i(p)}O^{\mu_1...\mu_N}_{q} (0) \ket{\Psi^j(p)}_Q
       \label{GijqqQPS} ~, \\
    \overline{u}(p,s) G^{ij, {\sf S}}_{g,Q}  u(p,s)&=&
    J_N\bra{\overline{\Psi}_i(p)}O^{\mu_1...\mu_N}_{g} (0)\ket{\Psi^j(p)}_Q
      \label{GijgqQ}~. 
   \end{eqnarray}
   Note, that in the quarkonic case the fields $\overline{\Psi},~\Psi$ 
   with color indices $i,j$ stand for the external light quarks only. 
   The above tensors have the general form, cf. 
   \cite{Buza:1995ie,Matiounine:1998ky}, 
   \begin{eqnarray} 
    G^{ab}_{l,\mu\nu}&=&
                       \Ahathat_{lg}\Bigl(\frac{\hat{m}^2}{\mu^2},\ep,N\Bigr)
                           \delta^{ab}
                          (\Delta \cdot p)^N 
                      \Big [- g_{\mu\nu}
                            +\frac{p_{\mu}\Delta_{\nu}+\Delta_{\mu}p_{\nu}}
                                  {\Delta \cdot p}
                      \Big ] ~, \quad l=Q,g,q~, \label{omeGluOp} \\
   \hat{G}_{l}^{r,ij}     &=&
                    \Ahathat_{lq}^{r}\Bigl(\frac{\hat{m}^2}{\mu^2},\ep,N\Bigr)
                    \delta^{ij}  
                          (\Delta \cdot p)^{N-1}
                           \adag \Delta ~, \quad l=Q,g,q~,~
                           \quad r={\sf S,~NS,~PS}~. \label{omelqproj}
   \end{eqnarray}

   Here, $\Ahathat_{ij}$ are the massive OMEs which we will
   calculate. In order to simplify this calculation, it is useful to define
   projection operators, which, applied to the Green's function, yield the
   corresponding OME. In the gluonic case, one defines
   \begin{eqnarray}
    P^{(1), \mu\nu}_{ab; g} G^{ab}_{l,\mu\nu}
               &\equiv& 
            - \frac{\delta_{ab}}{N_c^2-1} \frac{g^{\mu\nu}}{D-2}
               (\Delta\cdot p)^{-N} G^{ab}_{l,\mu\nu} ~, \label{projG1} \\
    P^{(2), \mu\nu}_{ab; g} G^{ab}_{l,\mu\nu} &\equiv& 
              \frac{\delta_{ab} }{N_c^2-1} \frac{1}{D-2}
              (\Delta\cdot p)^{-N}   
              \Bigl(-g^{\mu\nu}
                 +\frac{p^{\mu}\Delta^{\nu}+p^{\nu}\Delta^{\mu}}{\Delta\cdot p}
              \Bigr)G^{ab}_{l,\mu\nu}
              ~. \label{projG2}
   \end{eqnarray}
   In the quarkonic case, there is only one projector
   \begin{eqnarray} 
    P_{ij;q} G^{ij}_{l} &\equiv& 
              \frac{\delta_{ij}}{N_c} ( \Delta\cdot p)^{-N} 
               \frac{1}{4} {\sf Tr}[ \adag p G^{ij}_{l}] ~. \label{projQ}
   \end{eqnarray}
   The unrenormalized OMEs are given by
   \begin{eqnarray}
    \Ahathat_{lg}\Bigl(\frac{\hat{m}^2}{\mu^2},\ep,N\Bigr)
                      &=&P_{ab; g}^{(1,2), \mu\nu}G^{ab}_{l,\mu\nu}~, 
\label{proGa}\\
    \Ahathat_{lq}\Bigl(\frac{\hat{m}^2}{\mu^2},\ep,N\Bigr)
                      &=&P_{ij; q} G^{ij}_{l}~. \label{proQa}
   \end{eqnarray}
   These projections yield the advantage that one does not have to resort to 
   complicated tensorial reductions. In perturbation theory, the expressions 
   (\ref{proGa}, \ref{proQa}) can then be evaluated order by 
   order in the coupling constant by applying the
   Feynman rules given in Appendix~\ref{App-FeynRules}.  
   While the projector (\ref{projG1}) includes unphysical transverse 
   gluon states, which have to be compensated adding the corresponding 
   ghost-diagrams, (\ref{projG2}) projects onto the physical states.

  To calculate the color factor of each diagram, we use the program provided
  in Ref. \cite{vanRitbergen:1998pn}. Up to this point, all operations have
  been performed for general values of Mellin $N$ and the dimensional
  parameter $\ep$. The integrals do not contain any Lorentz or color indices
  anymore.  In order to use {\sf MATAD}, one now has to assign to $N$ a
  specific value. Additionally, the unphysical momentum $\Delta$ has to be
  eliminated by applying a suitable projector, which we define in the
  following section.
  \subsection{\bf\boldmath Calculation of Fixed $3$--Loop Moments 
                           Using {\sf MATAD}}
   \label{SubSec-3LMatad}
   We consider integrals of the type
   \begin{eqnarray}
    I_l(p,m,n_1,...,n_j) &\equiv&
                      \int \frac{d^Dk_1}{(2\pi)^D}...
                      \int \frac{d^Dk_l}{(2\pi)^D}
                      (\Delta.q_1)^{n_1}...
                      (\Delta.q_j)^{n_j}
                      f(k_1,...,k_l,p,m)~. \label{ExInt1}
   \end{eqnarray}
   Here $p$ denotes the external momentum, $p^2=0$, $m$ is the heavy quark
   mass, and $\Delta$ is a light--like vector, $\Delta^2=0$.  The momenta
   $q_{i}$ are given by any linear combination of the loop momenta $k_i$ and
   external momentum $p$. The exponents $n_i$ are integers or possibly sums
   of integers, see the Feynman rules in Appendix \ref{App-FeynRules}. Their
   sum is given by
   \begin{eqnarray}
    \sum_{i=1}^j n_i = N~.
   \end{eqnarray}
   The function $f$ in Eq. (\ref{ExInt1}) contains propagators, of which at
   least one is massive, dot-products of its arguments and powers of $m$. If
   one sets $N=0$, (\ref{ExInt1}) becomes
   \begin{eqnarray}
    I_l(p,m,0,...,0)=I_l(m)=
                      \int \frac{d^Dk_1}{(2\pi)^D}...
                      \int \frac{d^Dk_l}{(2\pi)^D}
                      f(k_1,...,k_l,m)~. \label{ExInt2}
   \end{eqnarray}
   From $p^2=0$ it follows, that the result can not depend on $p$ anymore.
   The above integral is a massive tadpole integral and thus of the
   type {\sf MATAD} can process. Additionally, {\sf MATAD} can calculate the
   integral up to a given order as a power series in $p^2/m^2$. Let us return
   to the general integral given in Eq. (\ref{ExInt1}).  One notes, that for
   fixed moments of $N$, each integral of this type splits up into one or more
   integrals of the same type with $n_i$ being just integers. At this
   point, it is useful to recall that the auxiliary vector $\Delta$ has only
   been introduced to get rid of the trace terms of the expectation values of
   the composite operators and has no physical significance.  By undoing the
   contraction with $\Delta$, these terms appear again. Consider as an
   example
   \begin{eqnarray}
    I_l(p,m,2,1) 
                    &=&
                      \int \frac{d^Dk_1}{(2\pi)^D}...
                      \int \frac{d^Dk_l}{(2\pi)^D}
                      (\Delta.q_1)^2 
                      (\Delta.q_2) 
                      f(k_1,...,k_l,p,m)  \label{ExInt5} \\
                    &=&\Delta^{\mu_1}\Delta^{\mu_2}\Delta^{\mu_3}
                      \int \frac{d^Dk_1}{(2\pi)^D}...
                      \int \frac{d^Dk_l}{(2\pi)^D}
                      q_{1,\mu_1}q_{1,\mu_2}q_{2,\mu_3}
                      f(k_1,...,k_l,p,m)~. \label{ExInt3}
   \end{eqnarray}
   One notices that the way of distributing the indices in 
   Eq. (\ref{ExInt3}) is somewhat arbitrary, since due to the 
   contraction with the totally symmetric tensor
   $\Delta^{\mu_1}\Delta^{\mu_2}\Delta^{\mu_3}$, the result of
   the corresponding tensor integral can be taken to be fully 
   symmetric as well. This is achieved by distributing the indices 
   among the $q_i$ in all possible ways and dividing by the number 
   of permutations one has used. Thus Eq. (\ref{ExInt3}) becomes
   \begin{eqnarray}
    I_l(p,m,2,1) 
                    &=&\Delta^{\mu_1}\Delta^{\mu_2}\Delta^{\mu_3}
                       \frac{1}{3}
                      \int \frac{d^Dk_1}{(2\pi)^D}...
                      \int \frac{d^Dk_l}{(2\pi)^D} (
                      q_{1,\mu_2}q_{1,\mu_3}q_{2,\mu_1}
                     +q_{1,\mu_1}q_{1,\mu_3}q_{2,\mu_2} 
\N \\ &&
                     +q_{1,\mu_1}q_{1,\mu_2}q_{2,\mu_3}
                                                   )
                      f(k_1,...,k_l,p,m)~. \label{ExInt4}
   \end{eqnarray}
   Generally speaking, the symmetrization of the tensor 
   resulting from 
   \begin{eqnarray}
                     \prod_{i=1}^j (\Delta.q_1)^{n_i}
   \end{eqnarray}
   can be achieved by shuffling indices, 
\cite{Borwein:1999js,Blumlein:1998if,Vermaseren:1998uu,Remiddi:1999ew,
      Moch:2001zr,Blumlein:2003gb}, and dividing 
   by the number of terms.
   The shuffle product is given by
   \begin{eqnarray}
    C \left[\underbrace{(k_1, \ldots, k_1)}_{ \small n_1} \SHU 
        \underbrace{(k_2, \ldots, k_2)}_{ \small n_2}  \SHU \ldots 
         \SHU \underbrace{(k_I, \ldots, k_I)}_{ \small n_I} \right]~,
   \end{eqnarray}
   where $C$ is the normalization constant
   \begin{eqnarray}
    C = \binom{N}{n_1, \ldots, n_I}^{-1}~. 
   \end{eqnarray}
   As an example, the symmetrization of
   \begin{eqnarray}
    q_{1,\mu_1} q_{1,\mu_2} q_{2,\mu_3}
   \end{eqnarray}
   can be inferred from Eq. (\ref{ExInt4}).  After undoing the contraction
   with $\Delta$ in integral (\ref{ExInt1}) and shuffling the indices, one may
   make the following Ansatz for the result, which follows from the necessity
   of complete symmetry in the Lorentz indices
   \begin{eqnarray}
    R_{\{\mu_1...\mu_N\}} &\equiv&\sum_{j=1}^{[N/2]+1} A_j
                  \Bigl(\prod_{k=1}^{j-1} g_{\{\mu_{2k}\mu_{2k-1}} \Bigr)
                  \Bigl(\prod_{l=2j-1}^N p_{\mu_l\}} \Bigr)
                   ~. \label{GenResInt}
   \end{eqnarray}
   In the above equation, $[~~]$ denotes the Gauss--bracket and $\{\}$
   symmetrization with respect to the indices enclosed and dividing by the
   number of terms, as outlined above. The first few terms are then given by
   \begin{eqnarray}
    R_0 &\equiv& 1~, \\
    R_{\{\mu_1\}} 
                  &=&      A_1 p_{\mu_1}~, \\
    R_{\{\mu_1\mu_2\}} 
                  &=&      A_1 p_{\mu_1}p_{\mu_2}+A_2 g_{\mu_1\mu_2} ~, \\
    R_{\{\mu_1\mu_2\mu_3\}} 
                  &=&       
                  A_1 p_{\mu_1}p_{\mu_2}p_{\mu_3}
                 +A_2 g_{\{\mu_1\mu_2}p_{\mu_3\}} ~. 
   \end{eqnarray}
   The scalars $A_j$ have in general different mass dimensions. 
   By contracting again with $\Delta$, all trace terms vanish 
   and one obtains 
   \begin{eqnarray}
    I_l(p,m,n_1,...,n_j) &=&\Delta^{\mu_1}...\Delta^{\mu_N}
                         R_{\{\mu_1...\mu_N\}}  \\
                     &=& A_1 (\Delta.p)^N
   \end{eqnarray}
   and thus the coefficient $A_1$ in Eq. (\ref{GenResInt}) gives the desired 
   result.
   To obtain it, one constructs a different projector, 
   which is made up only of the external momentum $p$ and the metric tensor. 
   By making a general Ansatz for this projector, applying 
   it to Eq. (\ref{GenResInt}) and demanding that the result 
   shall be equal to $A_1$, the coefficients 
   of the different Lorentz structures can be 
   determined. The projector reads
   \begin{eqnarray}
    \Pi_{\mu_1...\mu_N}&=&F(N)
                              \sum_{i=1}^{[N/2]+1}C(i,N)
                              \Bigl(\prod_{l=1}^{[N/2]-i+1}
                                   \frac{g_{\mu_{2l-1}\mu_{2l}}}{p^2} 
                              \Bigr)
                              \Bigl(\prod_{k=2[N/2]-2i+3}^N
                                   \frac{p_{\mu_k}}{p^2}
                              \Bigr)~. \label{Proj1}
   \end{eqnarray}
   For the overall pre-factors $F(N)$ and the coefficients 
   $C(i,N)$, one has to distinguish between even and odd values of $N$, 
   \begin{eqnarray}
    C^{odd}(k,N)&=&(-1)^{N/2+k+1/2}
                 \frac{2^{2k-N/2-3/2}\Gamma(N+1)\Gamma(D/2+N/2+k-3/2)}
                      {\Gamma(N/2-k+3/2)\Gamma(2k)\Gamma(D/2+N/2-1/2)}~, \\
    F^{odd}(N)  &=&\frac{2^{3/2-N/2}\Gamma(D/2+1/2)}
                        {(D-1)\Gamma(N/2+D/2-1)}~, \\
    C^{even}(k,N)&=&(-1)^{N/2+k+1}
                    \frac{2^{2k-N/2-2}\Gamma(N+1)\Gamma(D/2+N/2-2+k)}
                         {\Gamma(N/2-k+2)\Gamma(2k-1)\Gamma(D/2+N/2-1)}~, \\
    F^{even}(N)  &=&\frac{2^{1-N/2}\Gamma(D/2+1/2)}
                         {(D-1)\Gamma(N/2+D/2-1/2)}~.
   \end{eqnarray}
   The projector obeys the normalization 
   condition
   \begin{eqnarray}
    \Pi_{\mu_1,...,\mu_N}R^{\mu_1,...,\mu_N} &=&A_1 ~, 
   \end{eqnarray}
   which implies 
   \begin{eqnarray}
    \Pi_{\mu_1...\mu_N}p^{\mu_1}...p^{\mu_N}=1~. \\
   \end{eqnarray}
   As an example for the above procedure, we consider the case $N=3$,
   \begin{eqnarray}
    \Pi_{\mu_1\mu_2\mu_3}
                       &=&\frac{1}{D-1}
                              \Bigl(
                             -3\frac{g_{\mu_{1}\mu_{2}}p_{\mu_3}}{p^4}
                             +(D+2)
                                   \frac{p_{\mu_1}p_{\mu_2}p_{\mu3}}{p^6}
                              \Bigr)~.
   \end{eqnarray}  
   Applying this term to (\ref{ExInt4}) yields
   \begin{eqnarray}
    I_l(p,m,2,1) 
                    &=&
                       \frac{1}{(D-1)p^6}
                      \int \frac{d^Dk_1}{(2\pi)^D}...
                      \int \frac{d^Dk_l}{(2\pi)^D} 
                        (
                         -2 p^2 q_1.q_2 p.q_1  
\N\\[1em] &&
                         -p^2 q_1^2 p.q_2
                         +(D+2) (q_1.p)^2 q_2.p
                                                   )
                      f(k_1,...,k_l,p,m)~. \label{ExInt6}
   \end{eqnarray}
   It is important to keep $p$ artificially off--shell until the end of the
   calculation.  By construction, the overall result will not contain any term
   $\propto~1/p^2$, since the integral one starts with cannot contain such a
   term. Thus, at the end, these terms have to cancel, one can set
   $p^2=0$ and the remaining constant term in $p^2$ is the desired result.

   The above projectors are similar to the harmonic projectors used in the 
   ${\sf MINCER}$--program, cf. \cite{Larin:1991fz,Vermaseren:mincer}. These 
   are, however,
   applied to the virtual forward Compton--amplitude to determine the
   anomalous dimensions and the moments of the massless Wilson coefficients
   up to 3--loop order. 

   The calculation was in general performed in Feynman gauge. For the external
   quark and gluon lines, the projectors (\ref{projQ}, \ref{projG1}) are
   applied, which requires to include the ghost terms into the calculation. We
   also performed part of the calculation keeping the gauge parameter in 
   $R_\xi$--gauges, in particular for the moments $N=2,4$ in the singlet case and
   $N=1,2,3,4$ in the non--singlet case, yielding agreement with the results  
   being obtained using Feynman--gauge.
   In  addition, for the moments $N=2,4$ in the terms with external gluons, we
   applied the physical projector in Eq. (\ref{projG2}), which serves as
   another verification of our results. The computation of the more complicated 
   diagrams
   was performed on various 32/64 Gb machines using {\sf FORM} and for part of
   the calculation {\sf TFORM},~\cite{Tentyukov:2007mu}, spending about 250
   days of computational time.
  \subsection{\bf\boldmath Results}
   \label{SubSec-3LResUn}

\vspace{1mm}\noindent
We calculated the unrenormalized operator matrix elements treating the 
1PI-contributions explicitly. They contribute to $A_{Qg}^{(3)}, 
A_{gg,Q}^{(3)}$ 
and $A_{qq,Q}^{(3), {\sf NS}}$. One obtains the following representations
   \begin{eqnarray}
     \Ahathat_{Qg}^{(3)}&=&
            \Ahathat_{Qg}^{(3), \mbox{\small \sf irr}}
           -~\Ahathat_{Qg}^{(2), \mbox{\small \sf irr}}
            \hat{\Pi}^{(1)}\Bigl(0,\frac{\hat{m}^2}{\mu^2}\Bigr)
           -~\Ahathat_{Qg}^{(1)}
            \hat{\Pi}^{(2)}\Bigl(0,\frac{\hat{m}^2}{\mu^2}\Bigr)
\N\\ &&
           +~\Ahathat_{Qg}^{(1)}
            \hat{\Pi}^{(1)}\Bigl(0,\frac{\hat{m}^2}{\mu^2}\Bigr)
            \hat{\Pi}^{(1)}\Bigl(0,\frac{\hat{m}^2}{\mu^2}\Bigr)~,\\
     \Ahathat_{gg,Q}^{(3)}&=&
            \Ahathat_{gg,Q}^{(3), \mbox{\small \sf irr}}
           -\hat{\Pi}^{(3)}\Bigl(0,\frac{\hat{m}^2}{\mu^2}\Bigr)
           -~\Ahathat_{gg,Q}^{(2), \mbox{\small \sf irr}}
            \hat{\Pi}^{(1)}\Bigl(0,\frac{\hat{m}^2}{\mu^2}\Bigr)
\N\\ &&
           -2~\Ahathat_{gg,Q}^{(1)}
            \hat{\Pi}^{(2)}\Bigl(0,\frac{\hat{m}^2}{\mu^2}\Bigr)
           +~\Ahathat_{gg,Q}^{(1)}
            \hat{\Pi}^{(1)}\Bigl(0,\frac{\hat{m}^2}{\mu^2}\Bigr)
            \hat{\Pi}^{(1)}\Bigl(0,\frac{\hat{m}^2}{\mu^2}\Bigr)~,
\\
     \Ahathat_{qq,Q}^{(3), {\sf NS}}&=&
            \Ahathat_{qq,Q}^{(3), {\sf NS}, \mbox{\small \sf irr}}
           -~\hat{\Sigma}^{(3)}\Bigl(0,\frac{\hat{m}^2}{\mu^2}\Bigr)~.
   \end{eqnarray}
The self-energies are given in
Eqs.~(\ref{eqPI1}, \ref{eqPI2}, \ref{Pia}, \ref{Sig3}). The calculation of 
the
one-particle irreducible 3--loop contributions is performed using {\sf
  MATAD}.~\footnote{Partial results of the calculation were presented in
  \cite{Bierenbaum:2008dk,Bierenbaum:2008tt}.} The amount of moments, which
could be calculated, depended on the available computer resources
w.r.t. memory and computational time, as well as possible parallelization
using {\sf TFORM}.  Increasing the Mellin moment by two demands both a 
factor of 6--8 larger memory and CPU time. We have 
calculated the even
moments $N = 2, \ldots, 10$ for $A_{Qg}^{(3)}$, $A_{gg,Q}^{(3)}$, and
$A_{qg,Q}^{(3)}$, for $A_{Qq}^{(3), \rm PS}$ up to $ N = 12$, and for 
$A_{qq,Q}^{(3), \rm NS}, A_{qq,Q}^{(3),\rm PS}, A_{gq,Q}^{(3)}$ up to $N=14$.

\vspace*{7mm}\noindent
\underline {\large \sf $(i)$ Anomalous Dimensions :}

\vspace*{2mm}\noindent 
The pole terms of the unrenormalized
OMEs emerging in the calculation agree with the general structure we 
presented in 
Eqs.~(\ref{Ahhhqq3NSQ},
\ref{AhhhQq3PS}, \ref{Ahhhqq3PSQ}, \ref{AhhhQg3}, \ref{Ahhhqg3Q},
\ref{AhhhgqQ3}, \ref{Ahhhgg3Q}). Using lower order renormalization
coefficients and the constant terms of the $2$--loop results,
\cite{Buza:1995ie,Buza:1996xr,Bierenbaum:2007qe,Bierenbaum:2009zt}, allows to
determine the fixed moments of the 2--loop anomalous dimensions and the
contributions $\propto T_F$ of the $3$--loop anomalous dimensions,
cf. Appendix~\ref{App-ANDIM}.  All our results agree with the results of
Refs.~\cite{Gracey:1993nn,
  Larin:1996wd,Retey:2000nq,Moch:2002sn,Moch:2004pa,Vogt:2004mw}. The
anomalous dimensions $\gamma_{qg}^{(2)}$ and $\gamma_{qq}^{(2), {\sf PS}}$ are
obtained completely. The present calculation is fully independent both in the
algorithms and codes compared to
Refs.~\cite{Larin:1996wd,Retey:2000nq,Moch:2002sn, Moch:2004pa,Vogt:2004mw}
and thus provides a stringent check on these results.

\vspace*{7mm}\noindent
\underline {\large \sf $(ii)$ The constant terms $a_{ij}^{(3)}(N)$:}

\vspace*{2mm}\noindent 
The constant terms in Eq.~(\ref{epAhat}) at $O(a_s^3)$,
(\ref{Ahhhqq3NSQ},
\ref{AhhhQq3PS}, \ref{Ahhhqq3PSQ}, \ref{AhhhQg3}, \ref{Ahhhqg3Q},
\ref{AhhhgqQ3}, \ref{Ahhhgg3Q}),
are the new contributions to the non--logarithmic part of the 3--loop massive
operator matrix elements, which can not be constructed by other
renormalization constants calculated previously. They are given in
Appendix~\ref{SubSec-3LResUnHigh}. All other contributions to the heavy flavor
Wilson coefficients in the region $Q^2 \gg m^2$ are known for general values
of $N$, cf. Sections~\ref{sec-forma},~\ref{SubSec-RENPred}. The functions
$a_{ij}^{(3)}(N)$ still contain coefficients $\propto \zeta_2$ and we will see
below, under which circumstances these terms will contribute to the heavy
flavor contributions to the deep--inelastic structure functions. The constant
${\sf B_4}$, (\ref{eqB4}), emerges as in other massive single--scale
calculations~\cite{B4cite}.

\vspace*{7mm}\noindent
\underline {\large \sf $(iii)$ Moments of the Constant Terms of the $3$--loop 
                                                                Massive OMEs}

\vspace*{2mm}\noindent
  The logarithmic terms of the renormalized $3$--loop massive OMEs 
  are determined by known renormalization constants and can be inferred from 
  Eqs. (\ref{Aqq3NSQMSren}, \ref{AQq3PSMSren}, \ref{Aqq3PSQMSren}, 
  \ref{AQg3MSren}, \ref{Aqg3QMSren}, \ref{Agq3QMSren}, \ref{Agg3QMSren}).
  In the following, we consider as examples the non--logarithmic 
  contributions to the second moments of the renormalized massive OMEs. 
  We refer to coupling constant renormalization in the $\MS$--scheme and
  compare the results performing the mass renormalization in the on--shell
  scheme $(m)$ and the $\MS$--scheme $(\overline{m})$.

  For the matrix elements with external gluons, we obtain~:
\begin{eqnarray}
A_{Qg}^{(3), \MS}(\mu^2=m^2,2) &=&
T_FC_A^2
      \Biggl( 
                 \frac{174055}{4374}
                -\frac{88}{9}{\sf B_4}+72\zeta_4
                -\frac{29431}{324}\zeta_3
      \Biggr)
\N \\ \N \\ &&
+T_FC_FC_A
      \Biggl( 
                -\frac{18002}{729}
                +\frac{208}{9}{\sf B_4}-104\zeta_4
                +\frac{2186}{9}\zeta_3
                -\frac{64}{3}\zeta_2+64\zeta_2\ln(2)
      \Biggr)
\N \\ \N \\ &&
+T_FC_F^2
      \Biggl( 
                -\frac{8879}{729}
                -\frac{64}{9}{\sf B_4}+32\zeta_4
                -\frac{701}{81}\zeta_3+80\zeta_2-128\zeta_2\ln(2)
      \Biggr)
\N \\ \N \\ &&
+T_F^2C_A
      \Biggl( 
                -\frac{21586}{2187}
                +\frac{3605}{162}\zeta_3
      \Biggr)
+T_F^2C_F
      \Biggl( 
                -\frac{55672}{729}
                +\frac{889}{81}\zeta_3
                +\frac{128}{3}\zeta_2
      \Biggr)
\N \\ \N \\ &&
+n_fT_F^2C_A
      \Biggl( 
                -\frac{7054}{2187}
                -\frac{704}{81}\zeta_3
      \Biggr)
+n_fT_F^2C_F
      \Biggl( 
                -\frac{22526}{729}
                +\frac{1024}{81}\zeta_3
                -\frac{64}{3}\zeta_2
      \Biggr)
\nonumber\\
~. \label{AQg3N2ONMS} \\
A_{Qg}^{(3), \MS}(\mu^2=\overline{m}^2,2) &=&
T_FC_A^2
      \Biggl( 
                 \frac{174055}{4374}
                -\frac{88}{9}{\sf B_4}+72\zeta_4
                -\frac{29431}{324}\zeta_3
      \Biggr)
+T_FC_FC_A
      \Biggl( 
                -\frac{123113}{729}
\N \\ \N \\ &&
                +\frac{208}{9}{\sf B_4}-104\zeta_4
                +\frac{2330}{9}\zeta_3
      \Biggr)
+T_FC_F^2
      \Biggl( 
                -\frac{8042}{729}
                -\frac{64}{9}{\sf B_4}+32\zeta_4
\N \\ \N \\ 
&&
                -\frac{3293}{81}\zeta_3
      \Biggr)
+T_F^2C_A
      \Biggl( 
                -\frac{21586}{2187}
                +\frac{3605}{162}\zeta_3
      \Biggr)
+T_F^2C_F
      \Biggl( 
                -\frac{9340}{729}
                +\frac{889}{81}\zeta_3
      \Biggr)
\N \\ \N \\ 
&&
+n_fT_F^2C_A
      \Biggl( 
                -\frac{7054}{2187}
                -\frac{704}{81}\zeta_3
      \Biggr)
+n_fT_F^2C_F
      \Biggl( 
                 \frac{478}{729}
                +\frac{1024}{81}\zeta_3
      \Biggr)
~. \label{AQg3N2MSMS}
\\
A_{qg,Q}^{(3), \MS}(\mu^2=m^2,2) &=&
n_fT_F^2C_A
      \Biggl( 
                 \frac{64280}{2187}
                -\frac{704}{81}\zeta_3
      \Biggr)
+n_fT_F^2C_F
      \Biggl( 
                -\frac{7382}{729}
                +\frac{1024}{81}\zeta_3
      \Biggr)
~. \label{Aqg3QN2ONMS}
\\
A_{gg,Q}^{(3), \MS}(\mu^2=m^2,2) &=&
T_FC_A^2
      \Biggl( 
                -\frac{174055}{4374}
                +\frac{88}{9}{\sf B_4}-72\zeta_4
                +\frac{29431}{324}\zeta_3
      \Biggr)
\\
&&
+T_FC_FC_A
      \Biggl( 
                 \frac{18002}{729}
                -\frac{208}{9}{\sf B_4}+104\zeta_4
                -\frac{2186}{9}\zeta_3
                +\frac{64}{3}\zeta_2-64\zeta_2\ln(2)
      \Biggr)
\nonumber
\end{eqnarray}\begin{eqnarray}
&&
+T_FC_F^2
      \Biggl( 
                 \frac{8879}{729}
                +\frac{64}{9}{\sf B_4}-32\zeta_4
                +\frac{701}{81}\zeta_3-80\zeta_2+128\zeta_2\ln(2)
      \Biggr)
\N \\ \N \\ 
&&
+T_F^2C_A
      \Biggl( 
                 \frac{21586}{2187}
                -\frac{3605}{162}\zeta_3
      \Biggr)
+T_F^2C_F
      \Biggl( 
                 \frac{55672}{729}
                -\frac{889}{81}\zeta_3
                -\frac{128}{3}\zeta_2
      \Biggr)
\N \\ \N \\ &&
+n_fT_F^2C_A
      \Biggl( 
                -\frac{57226}{2187}
                +\frac{1408}{81}\zeta_3
      \Biggr)
+n_fT_F^2C_F
      \Biggl( 
                 \frac{29908}{729}
                -\frac{2048}{81}\zeta_3
                +\frac{64}{3}\zeta_2
      \Biggr)
~. \label{Agg3QN2ONMS}
\nonumber\\
\\
A_{gg,Q}^{(3), \MS}(\mu^2=\overline{m}^2,2) &=&
T_FC_A^2
      \Biggl( 
                -\frac{174055}{4374}
                +\frac{88}{9}{\sf B_4}-72\zeta_4
                +\frac{29431}{324}\zeta_3
      \Biggr)
\N \\ \N \\ &&
+T_FC_FC_A
      \Biggl( 
                 \frac{123113}{729}
                -\frac{208}{9}{\sf B_4}+104\zeta_4
                -\frac{2330}{9}\zeta_3
      \Biggr)
+T_FC_F^2
      \Biggl( 
                 \frac{8042}{729}
                +\frac{64}{9}{\sf B_4}
\N \\ \N \\ &&
                -32\zeta_4
                +\frac{3293}{81}\zeta_3
      \Biggr)
+T_F^2C_A
      \Biggl( 
                 \frac{21586}{2187}
                -\frac{3605}{162}\zeta_3
      \Biggr)
+T_F^2C_F
      \Biggl( 
                 \frac{9340}{729}
                -\frac{889}{81}\zeta_3
      \Biggr)
\N \\ \N \\ &&
+n_fT_F^2C_A
      \Biggl( 
                -\frac{57226}{2187}
                +\frac{1408}{81}\zeta_3
      \Biggr)
+n_fT_F^2C_F
      \Biggl( 
                 \frac{6904}{729}
                -\frac{2048}{81}\zeta_3
      \Biggr)
~. \label{Agg3QN2MSMS}
\end{eqnarray} 

Comparing the operator matrix elements in case of the on--shell scheme and
$\MS$--scheme, one notices that the terms $\ln(2) \zeta_2$, $\zeta_2$ are
absent in the latter. The $\zeta_2$ terms, which contribute to
$a_{ij}^{(3)}(N)$, are canceled by other contributions through
renormalization. Although the present process is massive, this observation
resembles the known result that $\zeta_2$--terms do not contribute in
space--like massless higher order calculations in even dimensions,~\cite{BROAD}. This
behaviour is found for all calculated moments. 
In addition, $\zeta_4$-terms occur, which may partly cancel 
with those in the $3$--loop light Wilson coefficients, 
\cite{Vermaseren:2005qc}. 
Note, that
Eq.~(\ref{Aqg3QN2ONMS}) is not sensitive to mass renormalization due to the
structure of the contributing diagrams.

An additional check is provided by the sum rule, \cite{Buza:1996wv},
   \begin{eqnarray}
\label{CONS1}
      A_{Qg}^{(3)}(N=2)
     +A_{qg,Q}^{(3)}(N=2)
     +A_{gg,Q}^{(3)}(N=2) &=&0~,
   \end{eqnarray}
which is fulfilled in all renormalization schemes and as well as on the 
unrenormalized level.

Unlike the operator matrix element with external gluons, 
the second moments of the quarkonic OMEs emerge for the first time at
$O(a_s^2)$. 
To 3--loop order, the quarkonic
OMEs do not contain terms $\propto \zeta_2$. Due to their simpler structure,
mass renormalization in the on--shell--scheme does not give rise to terms
$\propto \zeta_2, \ln(2) \zeta(2)$. Only the rational contribution in the
color factor $\propto T_F C_F^2$ turns out to be different and $A_{qq,Q}^{\sf
  PS, (3)}$, (\ref{eqqqQ3}), is not affected at all. This holds 
again for all moments
we calculated. The non--logarithmic contributions are given by
\begin{eqnarray}
A_{Qq}^{(3), \MS, {\sf PS}}(\mu^2=m^2,2) &=&
T_FC_FC_A
      \Biggl( 
                 \frac{830}{2187}
                +\frac{64}{9}{\sf B_4}-64\zeta_4
                +\frac{1280}{27}\zeta_3
      \Biggr)
+T_FC_F^2
      \Biggl( 
                 \frac{95638}{729}
\N \\ \N \\ &&
                -\frac{128}{9}{\sf B_4}+64\zeta_4
                -\frac{9536}{81}\zeta_3
      \Biggr)
+T_F^2C_F
      \Biggl( 
                 \frac{53144}{2187}
                -\frac{3584}{81}\zeta_3
      \Biggr)
\nonumber
\end{eqnarray}
\begin{eqnarray}
&&
+n_fT_F^2C_F
      \Biggl( 
                -\frac{34312}{2187}
                +\frac{1024}{81}\zeta_3
      \Biggr)
~.
\nonumber
\end{eqnarray}\begin{eqnarray}
A_{Qq}^{(3), \MS, {\sf PS}}(\mu^2=\overline{m}^2,2) &=&
T_FC_FC_A
      \Biggl( 
                 \frac{830}{2187}
                +\frac{64}{9}{\sf B_4}-64\zeta_4
                +\frac{1280}{27}\zeta_3
      \Biggr)
+T_FC_F^2
      \Biggl( 
                 \frac{78358}{729}
\N \\ \N \\ &&
                -\frac{128}{9}{\sf B_4}+64\zeta_4
                -\frac{9536}{81}\zeta_3
      \Biggr)
+T_F^2C_F
      \Biggl( 
                 \frac{53144}{2187}
                -\frac{3584}{81}\zeta_3
      \Biggr)
\N \\ \N \\ &&
+n_fT_F^2C_F
      \Biggl( 
                -\frac{34312}{2187}
                +\frac{1024}{81}\zeta_3
      \Biggr)
~.
\\
A_{qq,Q}^{(3), \MS, {\sf PS}}(\mu^2=m^2,2) &=&
n_fT_F^2C_F
      \Biggl( 
                -\frac{52168}{2187}
                +\frac{1024}{81}\zeta_3
      \Biggr)
~.
\label{eqqqQ3}
\\
A_{qq,Q}^{(3), \MS, {\sf NS}}(\mu^2=m^2,2) &=&
T_FC_FC_A
      \Biggl( 
                -\frac{101944}{2187}
                +\frac{64}{9}{\sf B_4}-64\zeta_4
                +\frac{4456}{81}\zeta_3
      \Biggr)
+T_FC_F^2
      \Biggl( 
                 \frac{283964}{2187}
\N \\ \N \\ &&
                -\frac{128}{9}{\sf B_4}+64\zeta_4
                -\frac{848}{9}\zeta_3
      \Biggr)
+T_F^2C_F
      \Biggl( 
                 \frac{25024}{2187}
                -\frac{1792}{81}\zeta_3
      \Biggr)
\N \\ \N \\ &&
+n_fT_F^2C_F
      \Biggl( 
                -\frac{46336}{2187}
                +\frac{1024}{81}\zeta_3
      \Biggr)
~.
\\
A_{qq,Q}^{(3), \MS, {\sf NS}}(\mu^2=\overline{m}^2,2) &=&
T_FC_FC_A
      \Biggl( 
                -\frac{101944}{2187}
                +\frac{64}{9}{\sf B_4}-64\zeta_4
                +\frac{4456}{81}\zeta_3
      \Biggr)
+T_FC_F^2
      \Biggl( 
                 \frac{201020}{2187}
\N \\ \N \\ &&
                -\frac{128}{9}{\sf B_4}+64\zeta_4
                -\frac{848}{9}\zeta_3
      \Biggr)
+T_F^2C_F
      \Biggl( 
                 \frac{25024}{2187}
                -\frac{1792}{81}\zeta_3
      \Biggr)
\N \\ \N \\ &&
+n_fT_F^2C_F
      \Biggl( 
                -\frac{46336}{2187}
                +\frac{1024}{81}\zeta_3
      \Biggr)
~.
\\
A_{gq,Q}^{(3), \MS}(\mu^2=m^2,2) &=&
T_FC_FC_A
      \Biggl( 
                 \frac{101114}{2187}
                -\frac{128}{9}{\sf B_4}+128\zeta_4
                -\frac{8296}{81}\zeta_3
      \Biggr)
+T_FC_F^2
      \Biggl( 
                -\frac{570878}{2187}
\N \\ \N \\ 
&&
                +\frac{256}{9}{\sf B_4}-128\zeta_4
                +\frac{17168}{81}\zeta_3
      \Biggr)
+T_F^2C_F
      \Biggl( 
                -\frac{26056}{729}
                +\frac{1792}{27}\zeta_3
      \Biggr)
\N \\ \N \\ &&
+n_fT_F^2C_F
      \Biggl( 
                 \frac{44272}{729}
                -\frac{1024}{27}\zeta_3
      \Biggr)
~.
\end{eqnarray}\begin{eqnarray}
A_{gq,Q}^{(3), \MS}(\overline{m}^2,2) &=&
T_FC_FC_A
      \Biggl( 
                 \frac{101114}{2187}
                -\frac{128}{9}{\sf B_4}+128\zeta_4
                -\frac{8296}{81}\zeta_3
      \Biggr)
+T_FC_F^2
      \Biggl( 
                -\frac{436094}{2187}
\nonumber
\end{eqnarray}\begin{eqnarray}
&&
                +\frac{256}{9}{\sf B_4}-128\zeta_4
                +\frac{17168}{81}\zeta_3
      \Biggr)
+T_F^2C_F
      \Biggl( 
                -\frac{26056}{729}
                +\frac{1792}{27}\zeta_3
      \Biggr)
\N \\ \N \\ &&
+n_fT_F^2C_F
      \Biggl( 
                 \frac{44272}{729}
                -\frac{1024}{27}\zeta_3
      \Biggr)
~.
\end{eqnarray}
   Finally, the sum rule, \cite{Buza:1996wv}, 
   \begin{eqnarray}
\label{CONS2}
      A_{Qq}^{(3), {\sf PS}}(N=2)
     +A_{qq,Q}^{(3), {\sf PS}}(N=2)
     +A_{qq,Q}^{(3), {\sf NS}}(N=2)
     +A_{gq,Q}^{(3)}(N=2) &=&0
   \end{eqnarray}
   holds on the unrenormalized level, as well as for the renormalized
   expressions in all schemes considered.

{\sf FORM}--codes for the constant terms $a_{ij}^{(3)}(N)$, 
Appendix~\ref{SubSec-3LResUnHigh}, and the corresponding  moments
of the renormalized massive operator matrix elements, both for the mass renormalization 
carried out in the on--shell-- and $\MS$--scheme, are attached to this paper and can be 
obtained upon request. Phenomenological studies of the 3--loop heavy flavor Wilson coefficients 
in the region $Q^2 \gg m^2$ will be given elsewhere \cite{BK9}.
\section{Heavy Quark Parton Densities}
\renewcommand{\theequation}{\thesection.\arabic{equation}}
\setcounter{equation}{0}
\label{sec-HQPD}

\vspace{1mm}
\noindent
In the kinematic region in which the factorization relation (\ref{FAC1})
holds, one may redefine the results obtained in the fixed flavor number
scheme, which allows for a partonic description at the level of $(n_f+1)$ flavors.
As before, we consider $n_f$ massless and one heavy quark flavor.  Since parton
distributions are process independent quantities, we define the parton
distributions for $(n_f+1)$ flavors from the light--flavor parton
distribution functions and the massive operator matrix elements for $n_f$
light flavors. Also in case of the structure functions associated to 
transverse virtual gauge boson polarizations, like $F_2(x,Q^2)$, the 
factorization (\ref{FAC1}) only occurs far above threshold, $Q^2
\sim 4 m^2 x/(1-x)$, and at even larger scales for $F_L(x,Q^2)$.  The 
following set of parton densities is obtained, cf.
\cite{Buza:1996wv}~:
\begin{eqnarray}
\label{HPDF1}
f_k(n_f+1,\mu^2,m^2,N) + f_{\overline{k}}(n_f+1,\mu^2,m^2,N)
&=& A_{qq,Q}^{\rm NS}\left(n_f,\frac{\mu^2}{m^2},N\right)
\cdot \bigl[f_k(n_f,\mu^2,N) 
\nonumber\\ && \hspace*{3.8cm}
+ f_{\overline{k}}(n_f,\mu^2,N)\bigr]
\nonumber\\ 
& & \hspace*{-4mm} + \tilde{A}_{qq,Q}^{\rm 
PS}\left(n_f,\frac{\mu^2}{m^2},N\right)
\cdot \Sigma(n_f,\mu^2,N)
\nonumber\\ 
& & \hspace*{-4mm} + \tilde{A}_{qg,Q}\left(n_f,\frac{\mu^2}{m^2},N\right)
\cdot G(n_f,\mu^2,N),
\\
\label{fQQB}
f_Q(n_f+1,\mu^2,m^2,N) + f_{\overline{Q}}(n_f+1,\mu^2,m^2,N)
&=&
{A}_{Qq}^{\rm PS}\left(n_f,\frac{\mu^2}{m^2},N\right)
\cdot \Sigma(n_f,\mu^2,N)
\nonumber\\ && \hspace*{-4mm}
+ {A}_{Qg}\left(n_f,\frac{\mu^2}{m^2},N\right)
\cdot G(n_f,\mu^2,N)~.
\end{eqnarray}
Here, $f_k (f_{\bar{k}})$ denote the light quark and anti--quark densities, 
$f_Q (f_{\bar{Q}})$ the heavy quark densities, and  $G$ is the gluon 
density.
The flavor singlet, non--singlet and gluon densities for $(n_f+1)$ flavors are 
given by
\begin{eqnarray}
\Sigma(n_f+1,\mu^2,m^2,N) 
&=& \Biggl[A_{qq,Q}^{\rm NS}\left(n_f, \frac{\mu^2}{m^2},N\right) +
          n_f \tilde{A}_{qq,Q}^{\rm PS}\left(n_f, \frac{\mu^2}{m^2},N\right)
  \nonumber \\ &&
         + {A}_{Qq}^{\rm PS}\left(n_f, \frac{\mu^2}{m^2},N\right)
        \Biggr]
\cdot \Sigma(n_f,\mu^2,N) \nonumber
\\
& & \hspace*{-3mm} + \left[n_f \tilde{A}_{qg,Q}\left(n_f, 
\frac{\mu^2}{m^2},N\right) +
          {A}_{Qg}\left(n_f, \frac{\mu^2}{m^2},N\right) 
\right]
\cdot G(n_f,\mu^2,N) 
\nonumber\\
\\
\Delta(n_f+1,\mu^2,m^2,N) &=& 
  f_k(n_f+1,\mu^2,N)
+ f_{\overline{k}}(n_f+1,\mu^2,m^2,N) 
\nonumber\\ &&
- \frac{1}{n_f+1} 
\Sigma(n_f+1,\mu^2,m^2,N)\\
\label{HPDF2}
G(n_f+1,\mu^2,m^2,N) &=& A_{gq,Q}\left(n_f, 
\frac{\mu^2}{m^2},N\right) 
                    \cdot \Sigma(n_f,\mu^2,N)
\nonumber\\
&& 
+ A_{gg,Q}\left(n_f, \frac{\mu^2}{m^2},N\right) 
                    \cdot G(n_f,\mu^2,N)~.
\end{eqnarray}
Note, that the {\sf new} parton densities depend on the renormalized heavy 
quark mass $m^2$. As outlined above, the corresponding relations for the 
operator matrix elements depend on the mass--renormalization scheme. 
Furthermore, $m = m(a_s(\mu^2))$. This has to be taken into account in 
QCD-analyzes, in particular $m^2$ cannot be chosen constant. 

The normalization of the quarkonic and gluonic operators obtained in the
light--cone expansion can be chosen arbitrarily. It is, however, convenient to
select the relative factor such, that the non-perturbative
nucleon-state expectation values, $\Sigma(n_f,\mu^2,N)$ and $G(n_f,\mu^2,N)$,
obey
\begin{eqnarray}
\Sigma(n_f,\mu^2,N=2)+G(n_f,\mu^2,N=2) = 1
\end{eqnarray}
due to 4-momentum conservation. As a consequence, the OMEs fulfill the 
relations (\ref{CONS1}, \ref{CONS2}).
The parton densities (\ref{HPDF1}--\ref{HPDF2}) can be applied in other 
hard--scattering reactions at high energy colliders in kinematic regions where
the corresponding power corrections $\propto (m^2/Q^2)^k,~~k \geq 1$ 
can also be safely disregarded.

Conversely, one may extend the kinematic regime for deep-inelastic scattering
to define the distribution functions (\ref{HPDF1}--\ref{HPDF2}) upon knowing 
the power corrections which occur in the heavy flavor Wilson coefficients 
${\sf H}_i^j(x,Q^2/\mu^2,m^2/\mu^2)$. This is the case for 2-loop order. 
We separate   
\begin{eqnarray}
{\sf H}_i^j\left(x,\frac{Q^2}{\mu^2},\frac{m^2}{\mu^2}\right) = 
{\sf H}_i^{j, \sf asymp}\left(x,\frac{Q^2}{\mu^2},\frac{m^2}{\mu^2}\right)
+ {\sf H}_i^{j, \sf power}\left(x,\frac{Q^2}{\mu^2},\frac{m^2}{\mu^2}\right)~,
\end{eqnarray}
where ${\sf H}_i^{j, \sf asymp}(x,Q^2/\mu^2,m^2/\mu^2)$ denotes the part of 
the 
Wilson coefficient given in Eq.~(\ref{FAC1}). If one accounts for
${\sf H}_i^{j, \sf power}(x,Q^2/\mu^2,m^2/\mu^2)$ in the fixed flavor number 
scheme,
Eqs.~(\ref{HPDF1}--\ref{HPDF2}) are still valid, but they do not necessarily 
yield the dominant contributions. In the region closer to threshold,
the kinematics of heavy quarks is by far not collinear, which is the main 
reason that a partonic description has to fail. Moreover, relation 
Eq.~(\ref{eqtim}) may be violated. In any case, it is not possible to use the 
partonic description (\ref{HPDF1}--\ref{HPDF2}) alone for other hard processes 
in a kinematic domain with significant power corrections.
\section{Conclusions}
\renewcommand{\theequation}{\thesection.\arabic{equation}}
\setcounter{equation}{0}
\label{sec-Concl}

\vspace{1mm}
\noindent
We calculated the 3--loop massive operator matrix elements, which form
the heavy flavor Wilson coefficients, (\ref{eqWIL1}--\ref{eqWIL5}), 
together with the known massless Wilson coefficients in the region 
$Q^2 \gg m^2$ due to the factorization theorem (\ref{FAC1}). All but the
power--suppressed contributions are obtained in this way. Furthermore, all
operator matrix elements needed to derive massive quark--distributions at the
3--loop level were calculated. We presented in detail the renormalization of
the massive operator matrix elements, leading to an intermediary
representation in a defined {\sf MOM}--scheme. This is necessary to maintain
the partonic description required for the factorization of the heavy flavor
Wilson coefficients into OMEs and the light flavor Wilson coefficients. The
representation of the heavy flavor Wilson coefficients in the asymptotic
region, effectively reached for the structure function $F_2(x,Q^2)$ for
$Q^2/m^2 \simeq 10$, is available for general values of $N$ in analytic
form, up to the constant parts $a_{ij}^{(3)}(N)$ of the unrenormalized 3--loop
OMEs. A number of fixed values of Mellin moments $N$ for these constant parts
were calculated, reaching up to $N = 10, 12, 14$, depending on the complexity
of the corresponding operator matrix element.  Although general methods are
available to reconstruct the recurrence formulae for anomalous
dimensions and Wilson-coefficients as a function of $N$ by a finite number of
moments, \cite{BKKS}, the number of moments calculated for $a_{ij}^{(3)}(N)$ 
is still far
too low. Through the renormalization of the massive OMEs, the corresponding
moments of the complete 2-loop anomalous dimensions and the $T_F$--terms of
the 3--loop anomalous dimensions are obtained, as are the moments of the
complete anomalous dimensions $\gamma_{qq}^{(2), \sf PS}(N)$ and
$\gamma_{qg}^{(2)}(N)$, which agree with the literature.

The results were presented performing the coupling constant renormalization of the OMEs
in the $\MS$--scheme and the mass renormalization in the on--shell scheme. 
After a transformation to the $\MS$--mass, the $\zeta_2$--terms
are canceled completely. Although being a
massive calculation, which is indicated by the emergence of the number ${\sf
  B_4}$, the use of the $\MS$--scheme moves the structure of the result towards
those observed in massless 3--loop calculations.

\vspace{5mm}\noindent
{\bf Acknowledgments.}~~We would like to thank K. Chetyrkin, J. Smith, M. Steinhauser and 
J. Vermaseren for  useful discussions. This work was supported in part by DFG 
Sonderforschungsbereich Transregio 9, Computergest\"utzte Theoretische 
Teilchenphysik, Studienstiftung des Deutschen Volkes, the European Commission MRTN
HEPTOOLS under Contract No. MRTN-CT-2006-035505, the Ministerio de Ciencia e 
Innovacion under Grant No. FPA2007-60323, CPAN (Grant No. CSD2007-00042), the 
Generalitat Valenciana under Grant No. PROMETEO/2008/069, and by the European 
Commission MRTN FLAVIAnet under Contract No. MRTN-CT-2006-035482. We thank 
both IT groups of DESY providing us access to special facilities to perform 
the present calculation.

\newpage
\section{Appendix}
\renewcommand{\theequation}{\thesection.\arabic{equation}}
\setcounter{equation}{0}
\label{sec-App}

\vspace{1mm}
\noindent
  \subsection{Feynman Rules}
   \label{App-FeynRules}
\begin{figure}[htb]
     \begin{center}
      \includegraphics[angle=0, height=14cm]{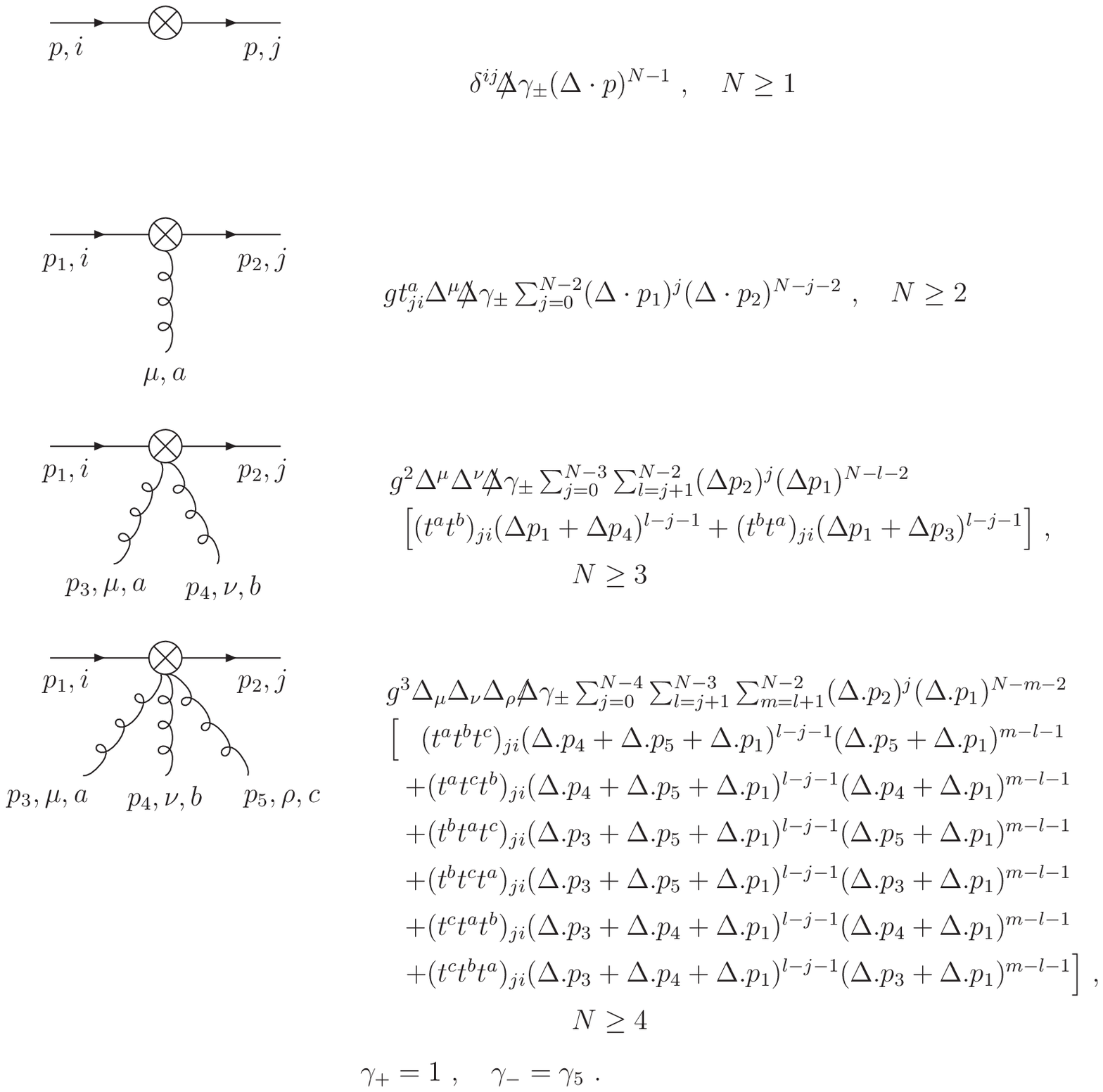}
     \end{center}
     \begin{center} 
      \caption[{\sf Feynman rules for quarkonic composite operators.}]
     {\sf Feynman rules for quarkonic composite operators. $\Delta$
     denotes  a light-like $4$-vector, $\Delta^2=0$;\\
              \phantom{Figure 2:} $N$ is an integer.}
        \label{feynrulescompqua}
      \noindent
      \small
     \end{center}
     \normalsize
    \end{figure}

\begin{figure}[htb]
     \begin{center}
      \includegraphics[angle=0, height=15cm]{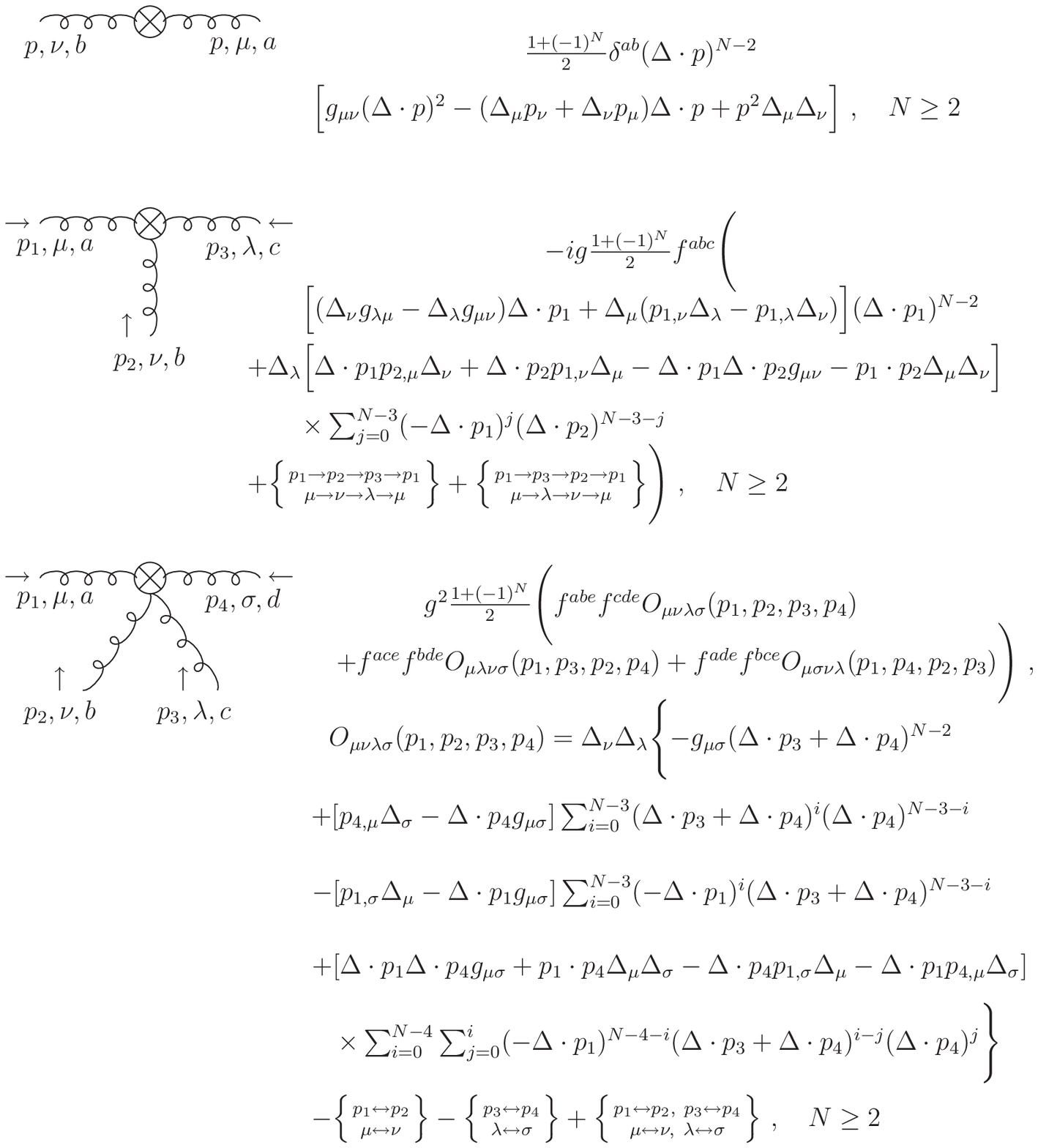}
     \end{center}
     \begin{center} 
      \caption[{\sf Feynman rules for gluonic composite operators.}]
     {\sf Feynman rules for gluonic composite operators. $\Delta$
     denotes  a light-like $4$-vector, $\Delta^2=0$; $N$ is an integer.}
        \label{feynrulescompglu}
      \noindent
      \small
     \end{center}
     \normalsize
    \end{figure}

    For the Feynman rules of QCD, we follow the convention of Ref.~\cite{YND}.
    $D$--dimensional momenta are denoted by $p_i$ and Lorentz indices by Greek
    letters. Color indices are denoted by $a,b,...$, and $i,j$ are indices of
    the color matrices.  Solid lines represent fermions and curly lines gluons.
    The Feynman rules for the quarkonic composite operators are given 
    in Figure~\ref{feynrulescompqua}. 
Up to $O(g^2)$ they can be found 
   in Refs. \cite{Floratos:1977au} and \cite{Mertig:1995ny}. Note that the 
   $O(g)$ term in the former reference contains a typographical error. We 
have checked these 
   terms and agree up to normalization factors, which may be partly due to 
   a different convention in the standard Feynman rules. We newly derived the 
   rule with three external gluons. The terms $\gamma_{\pm}$ refer to the 
   unpolarized ($+$) and polarized ($-$)  calculation, respectively.
   Gluon momenta are taken to be incoming. 
   The Feynman rules for the unpolarized gluonic composite operators are given 
   in Figure~\ref{feynrulescompglu}. Up to $O(g^2)$, they can be found 
   in Refs.~\cite{Floratos:1978ny} and \cite{Hamberg:1991qt}. 
   We have checked these terms and agree up to $O(g^0)$. At $O(g)$, 
   we agree with \cite{Floratos:1978ny}, but not with \cite{Hamberg:1991qt} 
   and \cite{YND}. 
   At $O(g^2)$, we do not agree with either of these results, which even  
   differ from each other~\footnote{We would like to thank J. Smith for the 
   possibility to compare with their {\sf FORM}--code used in Refs. 
   \cite{Buza:1995ie,Buza:1996xr,Matiounine:1998re,Matiounine:1998ky}, to 
   which we agree.}.  

\newpage
\subsection{The 3--loop Anomalous Dimensions}
   \label{App-ANDIM}
The 3--loop anomalous dimensions $\gamma_{qq}^{\sf PS}(N)$ and $\gamma_{qg}(N)$
and the contributions $\propto T_F$ to $\gamma_{qq}^{+, \sf NS}(N), 
\gamma_{gq}(N)$
and $\gamma_{gg}(N)$ are obtained from the single pole terms in 
the present calculation for even values of $N$ and for $\gamma_{qq}^{-, \sf 
NS}(N)$ for odd values of $N$. In the latter case, also $\gamma_{qq}^{s, \sf
NS}(N)$ with $\gamma_{qq}^{v, \sf NS}(N) = \gamma_{qq}^{-, \sf NS}(N) 
+ \gamma_{qq}^{s, \sf NS}(N)$ can be obtained, which will be considered 
elsewhere \cite{BK9}. 
The anomalous dimensions are~:

\vspace*{2mm}\noindent
\underline{$(i)$~~~\large $\hat{\gamma}_{qq}^{(2), \sf PS}(N)$}
%
%
\begin{eqnarray}
 {\hat{\gamma}_{qq}^{(2),{\sf PS}}(2)} &=&
{T_FC_F} \Biggl[
                         -(1+2n_f)T_F\frac{5024}{243}
                         +\frac{256}{3}\Bigl(C_F-C_A\Bigl)\zeta_3
                         +\frac{10136}{243}C_A
                         -\frac{14728}{243}C_F\Biggr]
\nonumber\\
\\
 {\hat{\gamma}_{qq}^{(2),{\sf PS}}(4)}&=&
T_F C_F \Biggl[
                         -(1+2n_f)T_F\frac{618673}{151875}
                         +\frac{968}{75}\Bigl(C_F-C_A\Bigl)\zeta_3
                         +\frac{2485097}{506250}C_A
\N\\ \N\\ &&
                         -\frac{2217031}{675000}C_F \Biggr]
\\
 {\hat{\gamma}_{qq}^{(2),{\sf PS}}(6)}&=& T_F C_F \Biggl[
                         -(1+2n_f)T_F\frac{126223052}{72930375}
                         +\frac{3872}{735}\Bigl(C_F-C_A\Bigl)\zeta_3
                         +\frac{1988624681}{4084101000}C_A
\N\\ \N\\ &&
                         +\frac{11602048711}{10210252500}C_F\Biggr]
\\
 {\hat{\gamma}_{qq}^{(2),{\sf PS}}(8)} &=& T_F C_F \Biggl[ 
                         -(1+2n_f)T_F\frac{13131081443}{13502538000}
                         +\frac{2738}{945}\Bigl(C_F-C_A\Bigl)\zeta_3
                         -\frac{343248329803}{648121824000}C_A
\N\\ \N\\ &&
                         +\frac{39929737384469}{22684263840000}C_F\Biggr]
\\
 {\hat{\gamma}_{qq}^{(2),{\sf PS}}(10)}&=& T_F C_F \Biggl[
                          -(1+2n_f)T_F\frac{265847305072}{420260754375}
                         +\frac{50176}{27225}\Bigl(C_F-C_A\Bigl)\zeta_3
\N\\ \N\\ &&
                         -\frac{1028766412107043}{1294403123475000}C_A
                         +\frac{839864254987192}{485401171303125}C_F\Biggr]
\\
 {\hat{\gamma}_{qq}^{(2),{\sf PS}}(12)}&=& T_F C_F \Biggl[
                          -(1+2n_f)T_F\frac{2566080055386457}{5703275664286200}
                         +\frac{49928}{39039}\Bigl(C_F-C_A\Bigl)\zeta_3
\N\\ \N\\ &&
                         -\frac{69697489543846494691}{83039693672007072000}C_A
                         +\frac{86033255402443256197}{54806197823524667520}C_F
\Biggr]~.
\end{eqnarray}

\vspace*{2mm}\noindent
\underline{$(ii)$~~~\large $\hat{\gamma}_{qg}^{(2)}(N)$}
%
%
\begin{eqnarray}
  {\hat{\gamma}_{qg}^{(2)}(2)}&=& T_F \Biggl[
                          (1+2n_f)T_F \Bigl(
                             \frac{8464}{243}C_A
                            -\frac{1384}{243}C_F
                                        \Bigr)
                          +\frac{\zeta_3}{3} \Bigl(
                           -416{C_AC_F}
                           +288{C_A^2}
                           +128{C_F^2}
                          \Bigr)
\N\\ \N\\ &&
                         -\frac{7178}{81}{C_A^2}
                         +\frac{556}{9}{C_AC_F}
                         -\frac{8620}{243}{C_F^2} \Biggr]
\\
  {\hat{\gamma}_{qg}^{(2)}(4)}&=& T_F \Biggl[
                          (1+2n_f)T_F \Bigl(
                             \frac{4481539}{303750}C_A
                            +\frac{9613841}{3037500}C_F
                                        \Bigr)
                          +\frac{\zeta_3}{25} \Bigl(
                             2832{C_A^2}
                            -3876{C_AC_F}
\N\\ \N\\ &&
                            +1044{C_F^2}
                         \Bigr)
                         -\frac{295110931}{3037500}{C_A^2}
                         +\frac{278546497}{2025000}{C_AC_F}
                         -\frac{757117001}{12150000}{C_F^2}\Biggr]
\\
  {\hat{\gamma}_{qg}^{(2)}(6)}&=& T_F \Biggl[
                          (1+2n_f)T_F \Bigl(
                             \frac{86617163}{11668860}C_A
                            +\frac{1539874183}{340341750}C_F
                                        \Bigr)
                          +\frac{\zeta_3}{735} \Bigl(
                              69864{C_A^2}
                             -94664{C_AC_F}
\N\\ \N\\ &&
                             +24800{C_F^2}
                          \Bigr)
                         -\frac{58595443051}{653456160}{C_A^2}
                         +\frac{1199181909343}{8168202000}{C_AC_F}
\N\\ \N\\ &&
                         -\frac{2933980223981}{40841010000}{C_F^2}\Biggr]
\\
  {\hat{\gamma}_{qg}^{(2)}(8)}&=& T_F \Biggl[
                          (1+2n_f)T_F \Bigl(
                          \frac{10379424541}{2755620000}C_A
                         +\frac{7903297846481}{1620304560000}C_F
                                        \Bigr)
                          +\zeta_3 \Bigl(
                          \frac{128042}{1575}{C_A^2}
\N\\ \N\\ &&
                         -\frac{515201}{4725}{C_AC_F}
                         +\frac{749}{27}{C_F^2}
                          \Bigr)
                         -\frac{24648658224523}{289340100000}{C_A^2}
\N\\ \N\\ &&
                         +\frac{4896295442015177}{32406091200000}{C_AC_F}
                         -\frac{4374484944665803}{56710659600000}{C_F^2}\Biggr]
\\
  {\hat{\gamma}_{qg}^{(2)}(10)}&=& T_F \Biggl[
                          (1+2n_f)T_F \Bigl(
                              \frac{1669885489}{988267500}C_A
                             +\frac{1584713325754369}{323600780868750}C_F
               \Bigr)
                          +\zeta_3 \Bigl(
                          \frac{1935952}{27225}{C_A^2}
\N\\ \N\\ &&
                         -\frac{2573584}{27225}{C_AC_F}
                         +\frac{70848}{3025}{C_F^2}
                          \Bigr)
                         -\frac{21025430857658971}{255684567600000}{C_A^2}
\N\\ \N\\ &&
                         +\frac{926990216580622991}{6040547909550000}{C_AC_F}
                       -\frac{1091980048536213833}{13591232796487500}{C_F^2}
\Biggr]~.
\end{eqnarray}

\vspace*{2mm}\noindent
\underline{$(iii)$~~~\large $\hat{\gamma}_{gq}^{(2)}(N)$}
%
%
\begin{eqnarray}
 {\hat{\gamma}_{gq}^{(2)}(2)}&=& T_F C_F \Biggl[
                          (1+2n_f)T_F\frac{2272}{81}
                         +\frac{512}{3}\Bigl(C_A-C_F\Bigl)\zeta_3
                         +\frac{88}{9}C_A
                         +\frac{28376}{243}C_F\Biggr]
\end{eqnarray}
\begin{eqnarray}
  {\hat{\gamma}_{gq}^{(2)}(4)}&=& T_F C_F \Biggl[
                          (1+2n_f)T_F\frac{109462}{10125}
                         +\frac{704}{15}\Bigl(C_A-C_F\Bigl)\zeta_3
                         -\frac{799}{12150}C_A
\N\\ \N\\ &&
                         +\frac{14606684}{759375}C_F \Biggr]
\\
  {\hat{\gamma}_{gq}^{(2)}(6)}&=& T_F C_F \Biggl[
                          (1+2n_f)T_F\frac{22667672}{3472875}
                         +\frac{2816}{105}\Bigl(C_A-C_F\Bigl)\zeta_3
                         -\frac{253841107}{145860750}C_A
\N\\ \N\\ &&
                         +\frac{20157323311}{2552563125}C_F\Biggr]
\\
  {\hat{\gamma}_{gq}^{(2)}(8)}&=& T_F C_F \Biggl[
                          (1+2n_f)T_F\frac{339184373}{75014100}
                         +\frac{1184}{63}\Bigl(C_A-C_F\Bigl)\zeta_3
                         -\frac{3105820553}{1687817250}C_A
\N\\ \N\\ &&
                         +\frac{8498139408671}{2268426384000}C_F\Biggr]
\\ 
  {\hat{\gamma}_{gq}^{(2)}(10)}&=& T_F C_F \Biggl[
                          (1+2n_f)T_F\frac{1218139408}{363862125}
                         +\frac{7168}{495}\Bigl(C_A-C_F\Bigl)\zeta_3
                         -\frac{18846629176433}{11767301122500} C_A
\N\\ \N\\ &&
                         +\frac{529979902254031}{323600780868750}C_F \Biggr]
\\
 {\hat{\gamma}_{gq}^{(2)}(12)} &=& T_F C_F \Biggl [
                          (1+2n_f)T_F\frac{13454024393417}{5222779912350}
                         +\frac{5056}{429}\Bigl(C_A-C_F\Bigl)\zeta_3
\N\\ \N\\ &&
                         -\frac{64190493078139789}{48885219979596000}C_A
                         +\frac{1401404001326440151}{3495293228541114000}C_F
\Biggr]
\\
  \hat{\gamma}_{gq}^{(2)}(14)&=&T_FC_F\Biggl[
                        (1+2n_f)T_F\frac{19285002274}{9495963477}
                       +\frac{13568}{1365}\Bigl(C_A-C_F\Bigr) \zeta_3 
\N\\ &&
                       -\frac{37115284124613269}{35434552943790000}C_A
                       -\frac{40163401444446690479}{104797690331258925000}C_F
                                       \Biggr].
\end{eqnarray}

\vspace*{2mm}\noindent
\underline{$(iv)$~~~\large $\hat{\gamma}_{gg}^{(2)}(N)$}
%
%
\begin{eqnarray}
  {\hat{\gamma}_{gg}^{(2)}(2)}&=& T_F \Biggl[
                         (1+2n_f)T_F \Bigl(
                             -\frac{8464}{243}C_A
                             +\frac{1384}{243}C_F
                                       \Bigr)
                         +\frac{\zeta_3}{3} \Bigl(
                               -288{C_A^2}
                               +416C_AC_F
\N \\ \N \\ &&
                               -128{C_F^2}
                                                  \Bigr)
                        +\frac{7178}{81}{C_A^2}
                        -\frac{556}{9}C_AC_F
                        +\frac{8620}{243}{C_F^2}\Biggr]
\\
  {\hat{\gamma}_{gg}^{(2)}(4)}&=& T_F \Biggl[
                         (1+2n_f)T_F \Bigl(
                             -\frac{757861}{30375}C_A
                             -\frac{979774}{151875}C_F
                                       \Bigr)
                         +\frac{\zeta_3}{25} \Bigl(
                             -6264{C_A^2}
                             +6528C_AC_F
\nonumber
\end{eqnarray}
\begin{eqnarray}
&&
                             -264{C_F^2}
                          \Bigr)
                         +\frac{53797499}{607500}{C_A^2}
                         -\frac{235535117}{1012500}C_AC_F
                         +\frac{2557151}{759375}{C_F^2}\Biggr]
\\
  {\hat{\gamma}_{gg}^{(2)}(6)}&=& T_F \Biggl[
                         (1+2n_f)T_F \Bigl(
                             -\frac{52781896}{2083725}C_A
                             -\frac{560828662}{72930375}C_F
                                       \Bigr)
                         +\zeta_3 \Bigl(
                        -\frac{75168}{245}{C_A^2}
\N \\ \N \\ &&
                        +\frac{229024}{735}C_AC_F
                        -\frac{704}{147}{C_F^2}
                                                  \Bigr)
                         +\frac{9763460989}{116688600}{C_A^2}
                         -\frac{9691228129}{32672808}C_AC_F
\N \\ \N \\ &&
                         -\frac{11024749151}{10210252500}{C_F^2}\Biggr]
\\ 
  {\hat{\gamma}_{gg}^{(2)}(8)}&=& T_F \Biggl[
                         (1+2n_f)T_F \Bigl(
                           -\frac{420970849}{16074450}C_A
                           -\frac{6990254812}{843908625}C_F
                                       \Bigr)
                         +\zeta_3 \Bigl(
                             -\frac{325174}{945}{C_A^2}
\N \\ \N \\ &&
                             +\frac{327764}{945}C_AC_F
                             -\frac{74}{27}{C_F^2}
                          \Bigr)
                         +\frac{2080130771161}{25719120000}{C_A^2}
\N \\ \N \\ &&
                         -\frac{220111823810087}{648121824000}C_AC_F
                         -\frac{14058417959723}{5671065960000}{C_F^2} \Biggr]
\\ 
  {\hat{\gamma}_{gg}^{(2)}(10)} &=& T_F \Biggl[
                         (1+2n_f)T_F \Bigl(
                            -\frac{2752314359}{101881395}C_A
                            -\frac{3631303571944}{420260754375}C_F
                                       \Bigr)
\N \\ \N \\ &&
                         +\zeta_3 \Bigl(
                            -\frac{70985968}{190575}{C_A^2}
                            +\frac{71324656}{190575}C_AC_F
                            -\frac{5376}{3025}{C_F^2}
                                                  \Bigr)
                         +\frac{43228502203851731}{549140719050000}{C_A^2}
\N \\ \N \\ &&
                         -\frac{3374081335517123191}{9060821864325000}C_FC_A
                         -\frac{3009386129483453}{970802342606250}{C_F^2} 
\Biggr]~.
\end{eqnarray}

\vspace*{2mm}\noindent
\underline{$(v)$~~~\large $\hat{\gamma}_{qq}^{(2), \sf NS,\rm +}(N)$}
%
%
\begin{eqnarray}
 {\hat{\gamma}_{qq}^{(2),{\sf NS},+}(2)}&=& T_F C_F \Biggl[
                         -(1+2n_f)T_F\frac{1792}{243}
                         +\frac{256}{3}\Bigl(C_F-C_A\Bigl)\zeta_3
                         -\frac{12512}{243}C_A
                         -\frac{13648}{243}C_F\Biggr]
\nonumber\\
 {\hat{\gamma}_{qq}^{(2),{\sf NS},+}(4)}&=& T_F C_F \Biggl[
                         -(1+2n_f)T_F\frac{384277}{30375}
                         +\frac{2512}{15}\Bigl(C_F-C_A\Bigl)\zeta_3
                         -\frac{8802581}{121500}C_A
\N\\ \N\\ &&
                         -\frac{165237563}{1215000}C_F\Biggr]
\\
 {\hat{\gamma}_{qq}^{(2),{\sf NS},+}(6)}&=& T_F C_F \Biggl[
                         -(1+2n_f)T_F\frac{160695142}{10418625}
                         +\frac{22688}{105}\Bigl(C_F-C_A\Bigl)\zeta_3
                         -\frac{13978373}{171500}C_A
\N\\ \N\\ &&
                         -\frac{44644018231}{243101250}C_F\Biggr]
\end{eqnarray}
\begin{eqnarray}
 {\hat{\gamma}_{qq}^{(2),{\sf NS},+}(8)}&=& T_F C_F \Biggl[
                         -(1+2n_f)T_F\frac{38920977797}{2250423000}
                         +\frac{79064}{315}\Bigl(C_F-C_A\Bigl)\zeta_3
                         -\frac{1578915745223}{18003384000}C_A
\N\\ \N\\ &&
                         -\frac{91675209372043}{420078960000}C_F\Biggr]
\\
 {\hat{\gamma}_{qq}^{(2),{\sf NS},+}(10)} &=& T_F C_F \Biggl[
                         -(1+2n_f)T_F\frac{27995901056887}{1497656506500}
                         +\frac{192880}{693}\Bigl(C_F-C_A\Bigl)\zeta_3
\N\\ \N\\ &&
                         -\frac{9007773127403}{97250422500}C_A
                         -\frac{75522073210471127}{307518802668000}C_F\Biggr]
\\
 {\hat{\gamma}_{qq}^{(2),{\sf NS},+}(12)}&=& T_F C_F\Biggl[
                         -(1+2n_f)T_F\frac{65155853387858071}{3290351344780500}
                         +\frac{13549568}{45045}\Bigl(C_F-C_A\Bigl)\zeta_3
\N\\ \N\\ &&
                         -\frac{25478252190337435009}{263228107582440000}C_A
                         -\frac{35346062280941906036867}{131745667845011220000}
C_F\Biggr]
\\
 \hat{\gamma}_{qq}^{(2),{\sf NS},+}(14)&=&T_FC_F\Biggl[
                       -(1+2n_f)T_F\frac{68167166257767019}{3290351344780500}
                       +\frac{2881936}{9009}\Bigl(C_F-C_A\Bigr)\zeta_3
\N\\ &&
                       -\frac{92531316363319241549}{921298376538540000}C_A
                     -\frac{37908544797975614512733}{131745667845011220000}C_F
                                       \Biggr]~.
\end{eqnarray}

\vspace*{2mm}\noindent
\underline{$(vi)$~~~\large $\hat{\gamma}_{qq}^{(2), \sf NS,\rm -}(N)$}
%
%
\begin{eqnarray}
 {\hat{\gamma}_{qq}^{(2),{\sf NS},-}(1)}&=&  0 \\
 {\hat{\gamma}_{qq}^{(2),{\sf NS},-}(3)}&=& T_F C_F \Biggl[
                         -(1+2n_f)T_F\frac{2569}{243}
                         +\frac{400}{3}\Bigl(C_F-C_A\Bigl)\zeta_3
                         -\frac{62249}{972}C_A
                         -\frac{203627}{1944}C_F\Biggr]
\nonumber\\
\\
 {\hat{\gamma}_{qq}^{(2),{\sf NS},-}(5)}&=& T_F C_F \Biggl[
                         -(1+2n_f)T_F\frac{431242}{30375}
                         +\frac{2912}{15}\Bigl(C_F-C_A\Bigl)\zeta_3
                         -\frac{38587}{500}C_A
\N\\ \N\\ &&
                         -\frac{5494973}{33750}C_F\Biggr]
\\
 {\hat{\gamma}_{qq}^{(2),{\sf NS},-}(7)}&=& T_F C_F \Biggl[
                         -(1+2n_f)T_F\frac{1369936511}{83349000}
                         +\frac{8216}{35}\Bigl(C_F-C_A\Bigl)\zeta_3
                         -\frac{2257057261}{26671680}C_A
\N\\ \N\\ &&
                         -\frac{3150205788689}{15558480000}C_F\Biggr]
\\
 {\hat{\gamma}_{qq}^{(2),{\sf NS},-}(9)}&=& T_F C_F \Biggl[
                         -(1+2n_f)T_F\frac{20297329837}{1125211500}
                         +\frac{16720}{63}\Bigl(C_F-C_A\Bigl)\zeta_3
                         -\frac{126810403414}{1406514375}C_A
\N\\ \N\\ &&
                         -\frac{1630263834317}{7001316000}C_F\Biggr]
\end{eqnarray}
\begin{eqnarray}
 {\hat{\gamma}_{qq}^{(2),{\sf NS},-}(11)}&=& T_F C_F \Biggl[
                         -(1+2n_f)T_F\frac{28869611542843}{1497656506500}
                         +\frac{1005056}{3465}\Bigl(C_F-C_A\Bigl)\zeta_3
\N\\ \N\\ &&
                         -\frac{1031510572686647}{10892047320000}C_A
                         -\frac{1188145134622636787}{4612782040020000}C_F
\Biggr]
\end{eqnarray}
\begin{eqnarray}
 {\hat{\gamma}_{qq}^{(2),{\sf NS},-}(13)}&=& T_F C_F \Biggl [
                         -(1+2n_f)T_F\frac{66727681292862571}{3290351344780500}
                         +\frac{13995728}{45045}\Bigl(C_F-C_A\Bigl)\zeta_3
\N\\ \N\\ &&
                         -\frac{90849626920977361109}{921298376538540000}C_A
                         -\frac{36688336888519925613757}
{131745667845011220000}C_F \Biggr]~.
\end{eqnarray}
We agree with the anomalous dimensions given in
\cite{Larin:1993vu,Larin:1996wd,Retey:2000nq,Vogt:2004mw,Moch:2004pa}.

\newpage
  \subsection{\bf\boldmath The $O(\varepsilon^0)$ contributions to 
              \hspace*{0.5mm} $\hat{\hspace*{-1.7mm} \hat{A}}_{ij}^{(3)}(N)$}
   \label{SubSec-3LResUnHigh}
The constant contributions to the unrenormalized massive operator matrix 
elements at $O(a_s^3)$ read~:

\vspace*{2mm}\noindent
\underline{$(i)$~\large $a_{Qq}^{(3), \sf PS}(N)$} 
\begin{eqnarray}
a_{Qq}^{(3), {\sf PS}}(2)&=&
T_FC_FC_A
      \Biggl( 
                 \frac{117290}{2187}
                +\frac{64}{9}{\sf B_4}-64\zeta_4
                +\frac{1456}{27}\zeta_3
                +\frac{224}{81}\zeta_2
      \Biggr)
\N \\ \N \\ &&
+T_FC_F^2
      \Biggl( 
                 \frac{42458}{243}
                -\frac{128}{9}{\sf B_4}+64\zeta_4
                -\frac{9664}{81}\zeta_3
                +\frac{704}{27}\zeta_2
      \Biggr)
+T_F^2C_F
      \Biggl( 
                -\frac{36880}{2187}
\N \\ \N \\ &&
                -\frac{4096}{81}\zeta_3
                -\frac{736}{81}\zeta_2
      \Biggr)
+n_fT_F^2C_F
      \Biggl( 
                -\frac{76408}{2187}
                +\frac{896}{81}\zeta_3
                -\frac{112}{81}\zeta_2
      \Biggr)
\\
a_{Qq}^{(3), {\sf PS}}(4)&=&
T_FC_FC_A
      \Biggl( 
                 \frac{23115644813}{1458000000}
                +\frac{242}{225}{\sf B_4}
                -\frac{242}{25}\zeta_4
                +\frac{1403}{180}\zeta_3
                +\frac{283481}{270000}\zeta_2
      \Biggr)
\N \\ \N \\ &&
+T_FC_F^2
      \Biggl( 
                -\frac{181635821459}{8748000000}
                -\frac{484}{225}{\sf B_4}
                +\frac{242}{25}\zeta_4
                +\frac{577729}{40500}\zeta_3
                +\frac{4587077}{1620000}\zeta_2
      \Biggr)
\N \\ \N \\ &&
+T_F^2C_F
      \Biggl( 
                -\frac{2879939}{5467500}
                -\frac{15488}{2025}\zeta_3
                -\frac{1118}{2025}\zeta_2
      \Biggr)
\N \\ \N \\ &&
+n_fT_F^2C_F
      \Biggl( 
                -\frac{474827503}{109350000}
                +\frac{3388}{2025}\zeta_3
                -\frac{851}{20250}\zeta_2
      \Biggr)
\\
a_{Qq}^{(3), {\sf PS}}(6)&=&
T_FC_FC_A
      \Biggl( 
                 \frac{111932846538053}{10291934520000}
                +\frac{968}{2205}{\sf B_4}
                -\frac{968}{245}\zeta_4
                +\frac{2451517}{1852200}\zeta_3
                +\frac{5638039}{7779240}\zeta_2
      \Biggr)
\N \\ \N \\ &&
+T_FC_F^2
      \Biggl( 
                -\frac{238736626635539}{5145967260000}
                -\frac{1936}{2205}{\sf B_4}
                +\frac{968}{245}\zeta_4
                +\frac{19628197}{555660}\zeta_3
\N \\ \N \\ &&
                +\frac{8325229}{10804500}\zeta_2
      \Biggr)
+T_F^2C_F
      \Biggl( 
                 \frac{146092097}{1093955625}
                -\frac{61952}{19845}\zeta_3
                -\frac{7592}{99225}\zeta_2
      \Biggr)
\N \\ \N \\ &&
+n_fT_F^2C_F
      \Biggl( 
                -\frac{82616977}{45378900}
                +\frac{1936}{2835}\zeta_3
                -\frac{16778}{694575}\zeta_2
      \Biggr)
\\
a_{Qq}^{(3), {\sf PS}}(8)&=&
T_FC_FC_A
      \Biggl( 
                 \frac{314805694173451777}{32665339929600000}
                +\frac{1369}{5670}{\sf B_4}
                -\frac{1369}{630}\zeta_4
                -\frac{202221853}{137168640}\zeta_3
\N \\ \N \\ &&
                +\frac{1888099001}{3429216000}\zeta_2
      \Biggr)
+T_FC_F^2
      \Biggl( 
                -\frac{25652839216168097959}{457314759014400000}
                -\frac{1369}{2835}{\sf B_4}
                +\frac{1369}{630}\zeta_4
\N \\ \N \\ 
&&
                +\frac{2154827491}{48988800}\zeta_3
                +\frac{12144008761}{48009024000}\zeta_2
      \Biggr)
+T_F^2C_F
      \Biggl( 
                 \frac{48402207241}{272211166080}
                -\frac{43808}{25515}\zeta_3
\nonumber
\end{eqnarray}
\begin{eqnarray}
&&
                +\frac{1229}{142884}\zeta_2
      \Biggr)
+n_fT_F^2C_F
      \Biggl( 
                -\frac{16194572439593}{15122842560000}
                +\frac{1369}{3645}\zeta_3
                -\frac{343781}{14288400}\zeta_2
      \Biggr)
\\
a_{Qq}^{(3), {\sf PS}}(10)&=&
T_FC_FC_A
      \Biggl( 
                 \frac{989015303211567766373}{107642563748181000000}
                +\frac{12544}{81675}{\sf B_4}
                -\frac{12544}{9075}\zeta_4
                -\frac{1305489421}{431244000}\zeta_3
\N \\ \N \\ &&
                +\frac{2903694979}{6670805625}\zeta_2
      \Biggr)
+T_FC_F^2
      \Biggl( 
                -\frac{4936013830140976263563}{80731922811135750000}
                -\frac{25088}{81675}{\sf B_4}
                +\frac{12544}{9075}\zeta_4
\N \\ \N \\ &&
                +\frac{94499430133}{1940598000}\zeta_3
                +\frac{282148432}{4002483375}\zeta_2
      \Biggr)
+T_F^2C_F
      \Biggl( 
                 \frac{430570223624411}{2780024890190625}
                -\frac{802816}{735075}\zeta_3
\N \\ \N \\ &&
                +\frac{319072}{11026125}\zeta_2
      \Biggr)
+n_fT_F^2C_F
      \Biggl( 
                -\frac{454721266324013}{624087220246875}
                +\frac{175616}{735075}\zeta_3
\N \\ \N \\ &&
                -\frac{547424}{24257475}\zeta_2
      \Biggr)
\\
a_{Qq}^{(3), {\sf PS}}(12)&=&
T_FC_FC_A
      \Biggl( 
                 \frac{968307050156826905398206547}{107727062441920086477312000}
                +\frac{12482}{117117}{\sf B_4}
                -\frac{12482}{13013}\zeta_4
\N \\ \N \\ &&
                -\frac{64839185833913}{16206444334080}\zeta_3
                +\frac{489403711559293}{1382612282251200}\zeta_2
      \Biggr)
\N \\ \N \\ &&
+T_FC_F^2
      \Biggl( 
                -\frac{190211298439834685159055148289}{2962494217152802378126080000}
                -\frac{24964}{117117}{\sf B_4}
                +\frac{12482}{13013}\zeta_4
\N \\ \N \\ &&
                +\frac{418408135384633}{8103222167040}\zeta_3
                -\frac{72904483229177}{15208735104763200}\zeta_2
      \Biggr)
\N \\ \N \\ &&
+T_F^2C_F
      \Biggl( 
                 \frac{1727596215111011341}{13550982978344011200}
                -\frac{798848}{1054053}\zeta_3
                +\frac{11471393}{347837490}\zeta_2
      \Biggr)
\N \\ \N \\ &&
+n_fT_F^2C_F
      \Biggl( 
                -\frac{6621557709293056160177}{12331394510293050192000}
                +\frac{24964}{150579}\zeta_3
\N \\ \N \\ &&
                -\frac{1291174013}{63306423180}\zeta_2
      \Biggr)
~.
\end{eqnarray} 

\vspace*{2mm}\noindent
\underline{$(ii)$~\large $a_{qq,Q}^{(3), \sf PS}(N)$} 
\begin{eqnarray}
a_{qq,Q}^{(3), {\sf PS}}(2)&=&
n_fT_F^2C_F
      \Biggl( 
                -\frac{100096}{2187}
                +\frac{896}{81}\zeta_3
                -\frac{256}{81}\zeta_2
      \Biggr)
\\
a_{qq,Q}^{(3), {\sf PS}}(4)&=&
n_fT_F^2C_F
      \Biggl( 
                -\frac{118992563}{21870000}
                +\frac{3388}{2025}\zeta_3
                -\frac{4739}{20250}\zeta_2
      \Biggr)
\end{eqnarray}
\begin{eqnarray}
a_{qq,Q}^{(3), {\sf PS}}(6)&=&
n_fT_F^2C_F
      \Biggl( 
                -\frac{17732294117}{10210252500}
                +\frac{1936}{2835}\zeta_3
                -\frac{9794}{694575}\zeta_2
      \Biggr)
\\
a_{qq,Q}^{(3), {\sf PS}}(8)&=&
n_fT_F^2C_F
      \Biggl( 
                -\frac{20110404913057}{27221116608000}
                +\frac{1369}{3645}\zeta_3
                +\frac{135077}{4762800}\zeta_2
      \Biggr)
\\
a_{qq,Q}^{(3), {\sf PS}}(10)&=&
n_fT_F^2C_F
      \Biggl( 
                -\frac{308802524517334}{873722108345625}
                +\frac{175616}{735075}\zeta_3
                +\frac{4492016}{121287375}\zeta_2
      \Biggr)
\\
a_{qq,Q}^{(3), {\sf PS}}(12)&=&
n_fT_F^2C_F
      \Biggl( 
                -\frac{6724380801633998071}{38535607844665781850}
                +\frac{24964}{150579}\zeta_3
                +\frac{583767694}{15826605795}\zeta_2
      \Biggr)
\\
a_{qq,Q}^{(3), {\sf PS}}(14)&=&
n_fT_F^2C_F
      \Biggl( 
                -\frac{616164615443256347333}{7545433703850642600000}
                +\frac{22472}{184275}\zeta_3
                +\frac{189601441}{5533778250}\zeta_2
      \Biggr)~.
\nonumber\\
\end{eqnarray}

\vspace*{2mm}\noindent
\underline{$(iii)$~\large $a_{Qg}^{\rm (3)}(N)$} 
\begin{eqnarray}
a_{Qg}^{(3)}(2)&=&
T_FC_A^2
      \Biggl( 
                 \frac{170227}{4374}
                -\frac{88}{9}{\sf B_4}
                +72\zeta_4
                -\frac{31367}{324}\zeta_3
                +\frac{1076}{81}\zeta_2
      \Biggr)
+T_FC_FC_A
      \Biggl( 
                -\frac{154643}{729}
\N \\ \N \\ &&
                +\frac{208}{9}{\sf B_4}
                -104\zeta_4
                +\frac{7166}{27}\zeta_3
                -54\zeta_2
      \Biggr)
+T_FC_F^2
      \Biggl( 
                -\frac{15574}{243}
                -\frac{64}{9}{\sf B_4}+32\zeta_4
\N \\ \N \\ &&
                -\frac{3421}{81}\zeta_3
                +\frac{704}{27}\zeta_2
      \Biggr)
+T_F^2C_A
      \Biggl( 
                -\frac{20542}{2187}
                +\frac{4837}{162}\zeta_3
                -\frac{670}{81}\zeta_2
      \Biggr)
+T_F^2C_F
      \Biggl( 
                 \frac{11696}{729}
\N \\ \N \\ &&
                +\frac{569}{81}\zeta_3
                +\frac{256}{9}\zeta_2
      \Biggr)
                -\frac{64}{27}T_F^3\zeta_3
+n_fT_F^2C_A
      \Biggl( 
                -\frac{6706}{2187}
                -\frac{616}{81}\zeta_3
                -\frac{250}{81}\zeta_2
      \Biggr)
\N \\ \N \\ &&
+n_fT_F^2C_F
      \Biggl( 
                 \frac{158}{243}
                +\frac{896}{81}\zeta_3
                +\frac{40}{9}\zeta_2
      \Biggr)
\\
a_{Qg}^{(3)}(4)&=&
T_FC_A^2
      \Biggl( 
                -\frac{425013969083}{2916000000}
                -\frac{559}{50}{\sf B_4}
                +\frac{2124}{25}\zeta_4
                -\frac{352717109}{5184000}\zeta_3
                -\frac{4403923}{270000}\zeta_2
      \Biggr)
\N \\ \N \\ &&
+T_FC_FC_A
      \Biggl( 
                -\frac{95898493099}{874800000}
                +\frac{646}{25}{\sf B_4}
                -\frac{2907}{25}\zeta_4
                +\frac{172472027}{864000}\zeta_3
                -\frac{923197}{40500}\zeta_2
      \Biggr)
\N \\ \N \\ &&
+T_FC_F^2
      \Biggl( 
                -\frac{87901205453}{699840000}
                -\frac{174}{25}{\sf B_4}
                +\frac{783}{25}\zeta_4
                +\frac{937829}{12960}\zeta_3
                +\frac{62019319}{3240000}\zeta_2
      \Biggr)
\N \\ \N \\ &&
+T_F^2C_A
      \Biggl( 
                 \frac{960227179}{29160000}
                +\frac{1873781}{51840}\zeta_3
                +\frac{120721}{13500}\zeta_2
      \Biggr)
+T_F^2C_F
      \Biggl( 
                -\frac{1337115617}{874800000}
\N \\ \N \\ &&
                +\frac{73861}{324000}\zeta_3
                +\frac{8879111}{810000}\zeta_2
      \Biggr)
                -\frac{176}{135}T_F^3\zeta_3
+n_fT_F^2C_A
      \Biggl( 
                 \frac{947836283}{72900000}
                -\frac{18172}{2025}\zeta_3
\nonumber
\end{eqnarray}\begin{eqnarray}
&&
                -\frac{11369}{13500}\zeta_2
      \Biggr)
+n_fT_F^2C_F
      \Biggl( 
                 \frac{8164734347}{4374000000}
                +\frac{130207}{20250}\zeta_3
                +\frac{1694939}{810000}\zeta_2
      \Biggr)
\\
a_{Qg}^{(3)}(6)&=&
T_FC_A^2
      \Biggl( 
                -\frac{48989733311629681}{263473523712000}
                -\frac{2938}{315}{\sf B_4}
                +\frac{17466}{245}\zeta_4
                -\frac{748603616077}{11379916800}\zeta_3
\N \\ \N \\ &&
                -\frac{93013721}{3457440}\zeta_2
      \Biggr)
+T_FC_FC_A
      \Biggl( 
                 \frac{712876107019}{55319040000}
                +\frac{47332}{2205}{\sf B_4}
                -\frac{23666}{245}\zeta_4
\N \\ \N \\ &&
                +\frac{276158927731}{1896652800}\zeta_3
                +\frac{4846249}{11113200}\zeta_2
      \Biggr)
+T_FC_F^2
      \Biggl( 
                -\frac{38739867811364113}{137225793600000}
\N \\ \N \\ &&
                -\frac{2480}{441}{\sf B_4}
                +\frac{1240}{49}\zeta_4
                +\frac{148514798653}{711244800}\zeta_3
                +\frac{4298936309}{388962000}\zeta_2
      \Biggr)
\N \\ \N \\ &&
+T_F^2C_A
      \Biggl( 
                 \frac{706058069789557}{18819537408000}
                +\frac{3393002903}{116121600}\zeta_3
                +\frac{6117389}{555660}\zeta_2
      \Biggr)
\N \\ \N \\ &&
+T_F^2C_F
      \Biggl( 
                -\frac{447496496568703}{54890317440000}
                -\frac{666922481}{284497920}\zeta_3
                +\frac{49571129}{9724050}\zeta_2
      \Biggr)
\N \\ \N \\ &&
                -\frac{176}{189}T_F^3\zeta_3
+n_fT_F^2C_A
      \Biggl( 
                 \frac{12648331693}{735138180}
                -\frac{4433}{567}\zeta_3
                +\frac{23311}{111132}\zeta_2
      \Biggr)
\N \\ \N \\ &&
+n_fT_F^2C_F
      \Biggl( 
                -\frac{8963002169173}{1715322420000}
                +\frac{111848}{19845}\zeta_3
                +\frac{11873563}{19448100}\zeta_2
      \Biggr)
\\
a_{Qg}^{(3)}(8)&=&
T_FC_A^2
      \Biggl( 
                -\frac{358497428780844484961}{2389236291993600000}
                -\frac{899327}{113400}{\sf B_4}
                +\frac{64021}{1050}\zeta_4
\N \\ \N \\ &&
                -\frac{12321174818444641}{112368549888000}\zeta_3
                -\frac{19581298057}{612360000}\zeta_2
      \Biggr)
+T_FC_FC_A
      \Biggl( 
                 \frac{941315502886297276939}{8362327021977600000}
\N \\ \N \\ &&
                +\frac{515201}{28350}{\sf B_4}
                -\frac{515201}{6300}\zeta_4
                +\frac{5580970944338269}{56184274944000}\zeta_3
                +\frac{495290785657}{34292160000}\zeta_2
      \Biggr)
\N \\ \N \\ &&
+T_FC_F^2
      \Biggl( 
                -\frac{23928053971795796451443}{36585180721152000000}
                -\frac{749}{162}{\sf B_4}
                +\frac{749}{36}\zeta_4
                +\frac{719875828314061}{1404606873600}\zeta_3
\N \\ \N \\ &&
                +\frac{2484799653079}{480090240000}\zeta_2
      \Biggr)
+T_F^2C_A
      \Biggl( 
                 \frac{156313300657148129}{4147979673600000}
                +\frac{58802880439}{2388787200}\zeta_3
\N \\ \N \\ &&
                +\frac{46224083}{4082400}\zeta_2
      \Biggr)
+T_F^2C_F
      \Biggl( 
                -\frac{986505627362913047}{87107573145600000}
                -\frac{185046016777}{50164531200}\zeta_3
\N \\ \N \\ &&
                +\frac{7527074663}{3429216000}\zeta_2
      \Biggr)
                -\frac{296}{405}T_F^3\zeta_3
+n_fT_F^2C_A
      \Biggl( 
                 \frac{24718362393463}{1322697600000}
                -\frac{125356}{18225}\zeta_3
\nonumber
\end{eqnarray}\begin{eqnarray}
&&
                +\frac{2118187}{2916000}\zeta_2
      \Biggr)
+n_fT_F^2C_F
      \Biggl( 
                -\frac{291376419801571603}{32665339929600000}
                +\frac{887741}{174960}\zeta_3
\N \\ \N \\ 
&&
                -\frac{139731073}{1143072000}\zeta_2
      \Biggr)
\\
a_{Qg}^{(3)}(10)&=&
T_FC_A^2
      \Biggl( 
                 \frac{6830363463566924692253659}{685850575063965696000000}
                -\frac{563692}{81675}{\sf B_4}
                +\frac{483988}{9075}\zeta_4
\N \\ \N \\ &&
                -\frac{103652031822049723}{415451499724800}\zeta_3
                -\frac{20114890664357}{581101290000}\zeta_2
      \Biggr)
\N \\ \N \\ &&
+T_FC_FC_A
      \Biggl( 
                 \frac{872201479486471797889957487}{2992802509370032128000000}
                +\frac{1286792}{81675}{\sf B_4}
                -\frac{643396}{9075}\zeta_4
\N \\ \N \\ &&
                -\frac{761897167477437907}{33236119977984000}\zeta_3
                +\frac{15455008277}{660342375}\zeta_2
      \Biggr)
\N \\ \N \\ &&
+T_FC_F^2
      \Biggl( 
                -\frac{247930147349635960148869654541}{148143724213816590336000000}
                -\frac{11808}{3025}{\sf B_4}
                +\frac{53136}{3025}\zeta_4
\N \\ \N \\ &&
                +\frac{9636017147214304991}{7122025709568000}\zeta_3
                +\frac{14699237127551}{15689734830000}\zeta_2
      \Biggr)
\N \\ \N \\ &&
+T_F^2C_A
      \Biggl( 
                 \frac{23231189758106199645229}{633397356480430080000}
                +\frac{123553074914173}{5755172290560}\zeta_3
                +\frac{4206955789}{377338500}\zeta_2
      \Biggr)
\N \\ \N \\ &&
+T_F^2C_F
      \Biggl( 
                -\frac{18319931182630444611912149}{1410892611560158003200000}
                -\frac{502987059528463}{113048027136000}\zeta_3
\N \\ \N \\ &&
                +\frac{24683221051}{46695639375}\zeta_2
      \Biggr)
                -\frac{896}{1485}T_F^3\zeta_3
+n_fT_F^2C_A
      \Biggl( 
                 \frac{297277185134077151}{15532837481700000}
\N \\ \N \\ &&
                -\frac{1505896}{245025}\zeta_3
                +\frac{189965849}{188669250}\zeta_2
      \Biggr)
+n_fT_F^2C_F
      \Biggl( 
                -\frac{1178560772273339822317}{107642563748181000000}
\N \\ \N \\ &&
                +\frac{62292104}{13476375}\zeta_3
                -\frac{49652772817}{93391278750}\zeta_2
      \Biggr)
~.
\end{eqnarray}

\vspace*{2mm}\noindent
\underline{$(iv)$~\large $a_{qg,Q}^{\rm (3)}(N)$} 
\begin{eqnarray}
a_{qg,Q}^{(3)}(2)&=&
n_fT_F^2C_A
      \Biggl( 
                 \frac{83204}{2187}
                -\frac{616}{81}\zeta_3
                +\frac{290}{81}\zeta_2
      \Biggr)
+n_fT_F^2C_F
      \Biggl( 
                -\frac{5000}{243}
                +\frac{896}{81}\zeta_3
                -\frac{4}{3}\zeta_2
      \Biggr)
\nonumber\\
\end{eqnarray}\begin{eqnarray}
a_{qg,Q}^{(3)}(4)&=&
n_fT_F^2C_A
      \Biggl( 
                 \frac{835586311}{14580000}
                -\frac{18172}{2025}\zeta_3
                +\frac{71899}{13500}\zeta_2
      \Biggr)
\N \\ \N \\ &&
+n_fT_F^2C_F
      \Biggl( 
                -\frac{21270478523}{874800000}
                +\frac{130207}{20250}\zeta_3
                -\frac{1401259}{810000}\zeta_2
      \Biggr)
\\
a_{qg,Q}^{(3)}(6)&=&
n_fT_F^2C_A
      \Biggl( 
                 \frac{277835781053}{5881105440}
                -\frac{4433}{567}\zeta_3
                +\frac{2368823}{555660}\zeta_2
      \Biggr)
\N \\ \N \\ &&
+n_fT_F^2C_F
      \Biggl( 
                -\frac{36123762156197}{1715322420000}
                +\frac{111848}{19845}\zeta_3
                -\frac{26095211}{19448100}\zeta_2
      \Biggr)
\\
a_{qg,Q}^{(3)}(8)&=&
n_fT_F^2C_A
      \Biggl( 
                 \frac{157327027056457}{3968092800000}
                -\frac{125356}{18225}\zeta_3
                +\frac{7917377}{2268000}\zeta_2
      \Biggr)
\N \\ \N \\ &&
+n_fT_F^2C_F
      \Biggl( 
                -\frac{201046808090490443}{10888446643200000}
                +\frac{887741}{174960}\zeta_3
                -\frac{3712611349}{3429216000}\zeta_2
      \Biggr)
\\
a_{qg,Q}^{(3)}(10)&=&
n_fT_F^2C_A
      \Biggl( 
                 \frac{6542127929072987}{191763425700000}
                -\frac{1505896}{245025}\zeta_3
                +\frac{1109186999}{377338500}\zeta_2
      \Biggr)
\N \\ \N \\ &&
+n_fT_F^2C_F
      \Biggl( 
                -\frac{353813854966442889041}{21528512749636200000}
                +\frac{62292104}{13476375}\zeta_3
\N \\ \N \\ &&
                -\frac{83961181063}{93391278750}\zeta_2
      \Biggr)
~.
\end{eqnarray}

\vspace*{2mm}\noindent
\underline{$(v)$~\large $a_{gq,Q}^{\rm (3)}(N)$} 
\begin{eqnarray}
a_{gq,Q}^{(3)}(2)&=&
T_FC_FC_A
      \Biggl( 
                -\frac{126034}{2187}
                -\frac{128}{9}{\sf B_4}+128\zeta_4
                -\frac{9176}{81}\zeta_3
                -\frac{160}{81}\zeta_2
      \Biggr)
\N \\ \N \\ &&
+T_FC_F^2
      \Biggl( 
                -\frac{741578}{2187}
                +\frac{256}{9}{\sf B_4}-128\zeta_4
                +\frac{17296}{81}\zeta_3
                -\frac{4496}{81}\zeta_2
      \Biggr)
+T_F^2C_F
      \Biggl( 
                 \frac{21872}{729}
\N \\ \N \\ &&
                +\frac{2048}{27}\zeta_3
                +\frac{416}{27}\zeta_2
      \Biggr)
+n_fT_F^2C_F
      \Biggl( 
                 \frac{92200}{729}
                -\frac{896}{27}\zeta_3
                +\frac{208}{27}\zeta_2
      \Biggr)
\\
a_{gq,Q}^{(3)}(4)&=&
T_FC_FC_A
      \Biggl( 
                -\frac{5501493631}{218700000}
                -\frac{176}{45}{\sf B_4}
                +\frac{176}{5}\zeta_4
                -\frac{8258}{405}\zeta_3
                +\frac{13229}{8100}\zeta_2
      \Biggr)
\N \\ \N \\ &&
+T_FC_F^2
      \Biggl( 
                -\frac{12907539571}{145800000}
                +\frac{352}{45}{\sf B_4}
                -\frac{176}{5}\zeta_4
                +\frac{132232}{2025}\zeta_3
                -\frac{398243}{27000}\zeta_2
      \Biggr)
\N \\ \N \\ &&
+T_F^2C_F
      \Biggl( 
                 \frac{1914197}{911250}
                +\frac{2816}{135}\zeta_3
                +\frac{1252}{675}\zeta_2
      \Biggr)
\nonumber
\end{eqnarray}\begin{eqnarray}
&&
+n_fT_F^2C_F
      \Biggl( 
                 \frac{50305997}{1822500}
                -\frac{1232}{135}\zeta_3
                +\frac{626}{675}\zeta_2
      \Biggr)
\\
a_{gq,Q}^{(3)}(6)&=&
T_FC_FC_A
      \Biggl( 
                -\frac{384762916141}{24504606000}
                -\frac{704}{315}{\sf B_4}
                +\frac{704}{35}\zeta_4
                -\frac{240092}{19845}\zeta_3
                +\frac{403931}{463050}\zeta_2
      \Biggr)
\N \\ \N \\ &&
+T_FC_F^2
      \Biggl( 
                -\frac{40601579774533}{918922725000}
                +\frac{1408}{315}{\sf B_4}
                -\frac{704}{35}\zeta_4
                +\frac{27512264}{694575}\zeta_3
\N \\ \N \\ &&
                -\frac{24558841}{3472875}\zeta_2
      \Biggr)
+T_F^2C_F
      \Biggl( 
                -\frac{279734446}{364651875}
                +\frac{11264}{945}\zeta_3
                +\frac{8816}{33075}\zeta_2
      \Biggr)
\N \\ \N \\ &&
+n_fT_F^2C_F
      \Biggl( 
                 \frac{4894696577}{364651875}
                -\frac{704}{135}\zeta_3
                +\frac{4408}{33075}\zeta_2
      \Biggr)
\\
a_{gq,Q}^{(3)}(8)&=&
T_FC_FC_A
      \Biggl( 
                -\frac{10318865954633473}{816633498240000}
                -\frac{296}{189}{\sf B_4}
                +\frac{296}{21}\zeta_4
                -\frac{1561762}{178605}\zeta_3
                +\frac{30677543}{85730400}\zeta_2
      \Biggr)
\N \\ \N \\ &&
+T_FC_F^2
      \Biggl( 
                -\frac{305405135103422947}{11432868975360000}
                +\frac{592}{189}{\sf B_4}
                -\frac{296}{21}\zeta_4
                +\frac{124296743}{4286520}\zeta_3
\N \\ \N \\ &&
                -\frac{4826251837}{1200225600}\zeta_2
      \Biggr)
+T_F^2C_F
      \Biggl( 
                -\frac{864658160833}{567106596000}
                +\frac{4736}{567}\zeta_3
                -\frac{12613}{59535}\zeta_2
      \Biggr)
\N \\ \N \\ &&
+n_fT_F^2C_F
      \Biggl( 
                 \frac{9330164983967}{1134213192000}
                -\frac{296}{81}\zeta_3
                -\frac{12613}{119070}\zeta_2
      \Biggr)
\\
a_{gq,Q}^{(3)}(10)&=&
T_FC_FC_A
      \Biggl( 
                -\frac{1453920909405842897}{130475834846280000}
                -\frac{1792}{1485}{\sf B_4}
                +\frac{1792}{165}\zeta_4
                -\frac{1016096}{147015}\zeta_3
\N \\ \N \\ &&
                +\frac{871711}{26952750}\zeta_2
      \Biggr)
+T_FC_F^2
      \Biggl( 
                -\frac{11703382372448370173}{667205973645750000}
                +\frac{3584}{1485}{\sf B_4}
                -\frac{1792}{165}\zeta_4
\N \\ \N \\ &&
                +\frac{62282416}{2695275}\zeta_3
                -\frac{6202346032}{2547034875}\zeta_2
      \Biggr)
+T_F^2C_F
      \Biggl( 
                -\frac{1346754066466}{756469357875}
                +\frac{28672}{4455}\zeta_3
\N \\ \N \\ &&
                -\frac{297472}{735075}\zeta_2
      \Biggr)
+n_fT_F^2C_F
      \Biggl( 
                 \frac{4251185859247}{756469357875}
                -\frac{12544}{4455}\zeta_3
                -\frac{148736}{735075}\zeta_2
      \Biggr)
\\
a_{gq,Q}^{(3)}(12)&=&
T_FC_FC_A
      \Biggl( 
                -\frac{1515875996003174876943331}{147976734123516602304000}
                -\frac{1264}{1287}{\sf B_4}
                +\frac{1264}{143}\zeta_4
                -\frac{999900989}{173918745}\zeta_3
\N \\ \N \\ &&
                -\frac{693594486209}{3798385390800}\zeta_2
      \Biggr)
+T_FC_F^2
      \Biggl( 
                -\frac{48679935129017185612582919}{4069360188396706563360000}
                +\frac{2528}{1287}{\sf B_4}
\N \\ \N \\ 
&&
                -\frac{1264}{143}\zeta_4
                +\frac{43693776149}{2260943685}\zeta_3
                -\frac{2486481253717}{1671289571952}\zeta_2
      \Biggr)
\nonumber
\end{eqnarray}\begin{eqnarray}
&&
+T_F^2C_F
      \Biggl( 
                -\frac{2105210836073143063}{1129248581528667600}
                +\frac{20224}{3861}\zeta_3
                -\frac{28514494}{57972915}\zeta_2
      \Biggr)
\N \\ \N \\ &&
+n_fT_F^2C_F
      \Biggl( 
                 \frac{9228836319135394697}{2258497163057335200}
                -\frac{8848}{3861}\zeta_3
                -\frac{14257247}{57972915}\zeta_2
      \Biggr)
\\
a_{gq,Q}^{(3)}(14)&=&
T_FC_FC_A
      \Biggl( 
                -\frac{1918253569538142572718209}{199199449781656964640000}
                -\frac{3392}{4095}{\sf B_4}
                +\frac{3392}{455}\zeta_4
\N \\ \N \\ &&
                -\frac{2735193382}{553377825}\zeta_3
                -\frac{1689839813797}{5113211103000}\zeta_2
      \Biggr)
\N \\ \N \\ &&
+T_FC_F^2
      \Biggl( 
                -\frac{143797180510035170802620917}{17429951855894984406000000}
                +\frac{6784}{4095}{\sf B_4}
                -\frac{3392}{455}\zeta_4
\N \\ \N \\ &&
                +\frac{12917466836}{774728955}\zeta_3
                -\frac{4139063104013}{4747981738500}\zeta_2
      \Biggr)
\N \\ \N \\ &&
+T_F^2C_F
      \Biggl( 
                -\frac{337392441268078561}{179653183425015300}
                +\frac{54272}{12285}\zeta_3
                -\frac{98112488}{184459275}\zeta_2
      \Biggr)
\N \\ \N \\ &&
+n_fT_F^2C_F
      \Biggl( 
                 \frac{222188365726202803}{71861273370006120}
                -\frac{3392}{1755}\zeta_3
                -\frac{49056244}{184459275}\zeta_2
      \Biggr)
~.
\end{eqnarray}

\vspace*{2mm}\noindent
\underline{$(vi)$~\large $a_{gg,Q}^{\rm (3)}(N)$} 
\begin{eqnarray}
a_{gg,Q}^{(3)}(2)&=&
T_FC_A^2
      \Biggl( 
                -\frac{170227}{4374}
                +\frac{88}{9}{\sf B_4}-72\zeta_4
                +\frac{31367}{324}\zeta_3
                -\frac{1076}{81}\zeta_2
      \Biggr)
+T_FC_FC_A
      \Biggl( 
                 \frac{154643}{729}
\N \\ \N \\ &&
                -\frac{208}{9}{\sf B_4}
                +104\zeta_4
                -\frac{7166}{27}\zeta_3+54\zeta_2
      \Biggr)
+T_FC_F^2
      \Biggl( 
                 \frac{15574}{243}
                +\frac{64}{9}{\sf B_4}-32\zeta_4
\N \\ \N \\ &&
                +\frac{3421}{81}\zeta_3
                -\frac{704}{27}\zeta_2
      \Biggr)
+T_F^2C_A
      \Biggl( 
                 \frac{20542}{2187}
                -\frac{4837}{162}\zeta_3
                +\frac{670}{81}\zeta_2
      \Biggr)
+T_F^2C_F
      \Biggl( 
                -\frac{11696}{729}
\N \\ \N \\ &&
                -\frac{569}{81}\zeta_3
                -\frac{256}{9}\zeta_2
      \Biggr)
                +\frac{64}{27}T_F^3\zeta_3
+n_fT_F^2C_A
      \Biggl( 
                -\frac{76498}{2187}
                +\frac{1232}{81}\zeta_3
                -\frac{40}{81}\zeta_2
      \Biggr)
\N \\ \N \\ &&
+n_fT_F^2C_F
      \Biggl( 
                 \frac{538}{27}
                -\frac{1792}{81}\zeta_3
                -\frac{28}{9}\zeta_2
      \Biggr)
\\
a_{gg,Q}^{(3)}(4)&=&
T_FC_A^2
      \Biggl( 
                 \frac{29043652079}{291600000}
                +\frac{533}{25}{\sf B_4}
                -\frac{4698}{25}\zeta_4
                +\frac{610035727}{2592000}\zeta_3
                +\frac{92341}{6750}\zeta_2
      \Biggr)
\N \\ \N \\ &&
+T_FC_FC_A
      \Biggl( 
                 \frac{272542528639}{874800000}
                -\frac{1088}{25}{\sf B_4}
                +\frac{4896}{25}\zeta_4
                -\frac{3642403}{17280}\zeta_3
                +\frac{73274237}{810000}\zeta_2
      \Biggr)
\nonumber
\end{eqnarray}\begin{eqnarray}
&&
+T_FC_F^2
      \Biggl( 
                 \frac{41753961371}{1749600000}
                +\frac{44}{25}{\sf B_4}
                -\frac{198}{25}\zeta_4
                +\frac{2676077}{64800}\zeta_3
                -\frac{4587077}{1620000}\zeta_2
      \Biggr)
\N \\ \N \\ &&
+T_F^2C_A
      \Biggl( 
                -\frac{1192238291}{14580000}
                -\frac{2134741}{25920}\zeta_3
                -\frac{16091}{675}\zeta_2
      \Biggr)
+T_F^2C_F
      \Biggl( 
                -\frac{785934527}{43740000}
\N \\ \N \\ &&
                -\frac{32071}{8100}\zeta_3
                -\frac{226583}{8100}\zeta_2
      \Biggr)
                +\frac{64}{27}T_F^3\zeta_3
+n_fT_F^2C_A
      \Biggl( 
                -\frac{271955197}{1822500}
                +\frac{13216}{405}\zeta_3
\N \\ \N \\ &&
                -\frac{6526}{675}\zeta_2
      \Biggr)
+n_fT_F^2C_F
      \Biggl( 
                -\frac{465904519}{27337500}
                -\frac{6776}{2025}\zeta_3
                -\frac{61352}{10125}\zeta_2
      \Biggr)
\\
a_{gg,Q}^{(3)}(6)&=&
T_FC_A^2
      \Biggl( 
                 \frac{37541473421359}{448084224000}
                +\frac{56816}{2205}{\sf B_4}
                -\frac{56376}{245}\zeta_4
                +\frac{926445489353}{2844979200}\zeta_3
                +\frac{11108521}{555660}\zeta_2
      \Biggr)
\N \\ \N \\ &&
+T_FC_FC_A
      \Biggl( 
                 \frac{18181142251969309}{54890317440000}
                -\frac{114512}{2205}{\sf B_4}
                +\frac{57256}{245}\zeta_4
                -\frac{12335744909}{67737600}\zeta_3
\N \\ \N \\ &&
                +\frac{94031857}{864360}\zeta_2
      \Biggr)
+T_FC_F^2
      \Biggl( 
                 \frac{16053159907363}{635304600000}
                +\frac{352}{441}{\sf B_4}
                -\frac{176}{49}\zeta_4
                +\frac{3378458681}{88905600}\zeta_3
\N \\ \N \\ &&
                -\frac{8325229}{10804500}\zeta_2
      \Biggr)
+T_F^2C_A
      \Biggl( 
                -\frac{670098465769}{6001128000}
                -\frac{25725061}{259200}\zeta_3
                -\frac{96697}{2835}\zeta_2
      \Biggr)
\N \\ \N \\ &&
+T_F^2C_F
      \Biggl( 
                -\frac{8892517283287}{490092120000}
                -\frac{12688649}{2540160}\zeta_3
                -\frac{2205188}{77175}\zeta_2
      \Biggr)
\N \\ \N \\ &&
                +\frac{64}{27}T_F^3\zeta_3
+n_fT_F^2C_A
      \Biggl( 
                -\frac{245918019913}{1312746750}
                +\frac{3224}{81}\zeta_3
                -\frac{250094}{19845}\zeta_2
      \Biggr)
\N \\ \N \\ &&
+n_fT_F^2C_F
      \Biggl( 
                -\frac{71886272797}{3403417500}
                -\frac{3872}{2835}\zeta_3
                -\frac{496022}{77175}\zeta_2
      \Biggr)
\\
a_{gg,Q}^{(3)}(8)&=&
T_FC_A^2
      \Biggl( 
                 \frac{512903304712347607}{18665908531200000}
                +\frac{108823}{3780}{\sf B_4}
                -\frac{162587}{630}\zeta_4
                +\frac{2735007975361}{6502809600}\zeta_3
\N \\ \N \\ &&
                +\frac{180224911}{7654500}\zeta_2
      \Biggr)
+T_FC_FC_A
      \Biggl( 
                 \frac{13489584043443319991}{43553786572800000}
                -\frac{163882}{2835}{\sf B_4}
                +\frac{81941}{315}\zeta_4
\N \\ \N \\ &&
                -\frac{3504113623243}{25082265600}\zeta_3
                +\frac{414844703639}{3429216000}\zeta_2
      \Biggr)
+T_FC_F^2
      \Biggl( 
                 \frac{5990127272073225467}{228657379507200000}
\N \\ \N \\ &&
                +\frac{37}{81}{\sf B_4}
                -\frac{37}{18}\zeta_4
                +\frac{3222019505879}{87787929600}\zeta_3
                -\frac{12144008761}{48009024000}\zeta_2
      \Biggr)
\N \\ \N \\ &&
+T_F^2C_A
      \Biggl( 
                -\frac{16278325750483243}{124439390208000}
                -\frac{871607413}{7962624}\zeta_3
                -\frac{591287}{14580}\zeta_2
      \Biggr)
\nonumber
\end{eqnarray}\begin{eqnarray}
&&
+T_F^2C_F
      \Biggl( 
                -\frac{7458367007740639}{408316749120000}
                -\frac{291343229}{52254720}\zeta_3
                -\frac{2473768763}{85730400}\zeta_2
      \Biggr)
\N \\ \N \\ &&
                +\frac{64}{27}T_F^3\zeta_3
+n_fT_F^2C_A
      \Biggl( 
                -\frac{102747532985051}{486091368000}
                +\frac{54208}{1215}\zeta_3
                -\frac{737087}{51030}\zeta_2
      \Biggr)
\N \\ \N \\ &&
+n_fT_F^2C_F
      \Biggl( 
                -\frac{1145917332616927}{51039593640000}
                -\frac{2738}{3645}\zeta_3
                -\frac{70128089}{10716300}\zeta_2
      \Biggr)
\end{eqnarray}
\begin{eqnarray}
a_{gg,Q}^{(3)}(10)&=&
T_FC_A^2
      \Biggl( 
                -\frac{15434483462331661005275759}{327337774462347264000000}
                +\frac{17788828}{571725}{\sf B_4}
                -\frac{17746492}{63525}\zeta_4
\N \\ \N \\ &&
                +\frac{269094476549521109}{519314374656000}\zeta_3
                +\frac{1444408720649}{55468759500}\zeta_2
      \Biggr)
\N \\ \N \\ &&
+T_FC_FC_A
      \Biggl( 
                 \frac{207095356146239371087405921}{771581896946961408000000}
                -\frac{35662328}{571725}{\sf B_4}
                +\frac{17831164}{63525}\zeta_4
\N \\ \N \\ &&
                -\frac{3288460968359099}{37093883904000}\zeta_3
                +\frac{6078270984602}{46695639375}\zeta_2
      \Biggr)
\N \\ \N \\ &&
+T_FC_F^2
      \Biggl( 
                 \frac{553777925867720521493231}{20667372239650752000000}
                +\frac{896}{3025}{\sf B_4}
                -\frac{4032}{3025}\zeta_4
\N \\ \N \\ &&
                +\frac{7140954579599}{198717235200}\zeta_3
                -\frac{282148432}{4002483375}\zeta_2
      \Biggr)
\N \\ \N \\ &&
+T_F^2C_A
      \Biggl( 
                -\frac{63059843481895502807}{433789788579840000}
                -\frac{85188238297}{729907200}\zeta_3
                -\frac{33330316}{735075}\zeta_2
      \Biggr)
\N \\ \N \\ &&
+T_F^2C_F
      \Biggl( 
                -\frac{655690580559958774157}{35787657557836800000}
                -\frac{71350574183}{12043468800}\zeta_3
                -\frac{3517889264}{121287375}\zeta_2
      \Biggr)
\N \\ \N \\ &&
                +\frac{64}{27}T_F^3\zeta_3
+n_fT_F^2C_A
      \Biggl( 
                -\frac{6069333056458984}{26476427525625}
                +\frac{215128}{4455}\zeta_3
                -\frac{81362132}{5145525}\zeta_2
      \Biggr)
\N \\ \N \\ &&
+n_fT_F^2C_F
      \Biggl( 
                -\frac{100698363899844296}{4368610541728125}
                -\frac{351232}{735075}\zeta_3
                -\frac{799867252}{121287375}\zeta_2
      \Biggr)
~.
\end{eqnarray}

\vspace*{2mm}\noindent
\underline{$(vii)$~\large $a_{qq,Q}^{(3), \sf NS}(N)$} 
\begin{eqnarray}
a_{qq,Q}^{(3), {\sf NS}}(1)&=& 0 \\
a_{qq,Q}^{(3), {\sf NS}}(2)&=&
T_FC_FC_A
      \Biggl( 
                 \frac{8744}{2187}
                +\frac{64}{9}{\sf B_4}
                -64\zeta_4
                +\frac{4808}{81}\zeta_3
                -\frac{64}{81}\zeta_2
      \Biggr)
+T_FC_F^2
      \Biggl( 
                 \frac{359456}{2187}
\nonumber
\end{eqnarray}\begin{eqnarray}
&&
                -\frac{128}{9}{\sf B_4}
                +64\zeta_4
                -\frac{848}{9}\zeta_3
                +\frac{2384}{81}\zeta_2
      \Biggr)
+T_F^2C_F
      \Biggl( 
                -\frac{28736}{2187}
                -\frac{2048}{81}\zeta_3
\N \\ \N \\ 
&&
                -\frac{512}{81}\zeta_2
      \Biggr)
+n_fT_F^2C_F
      \Biggl( 
                -\frac{100096}{2187}
                +\frac{896}{81}\zeta_3
                -\frac{256}{81}\zeta_2
      \Biggr)
\\
a_{qq,Q}^{(3), {\sf NS}}(3)&=&
T_FC_FC_A
      \Biggl( 
                 \frac{522443}{34992}
                +\frac{100}{9}{\sf B_4}-100\zeta_4
                +\frac{15637}{162}\zeta_3
                +\frac{175}{162}\zeta_2
      \Biggr)
\N \\ \N \\ &&
+T_FC_F^2
      \Biggl( 
                 \frac{35091701}{139968}
                -\frac{200}{9}{\sf B_4}+100\zeta_4
                -\frac{1315}{9}\zeta_3
                +\frac{29035}{648}\zeta_2
      \Biggr)
\N \\ \N \\ 
&&
+T_F^2C_F
      \Biggl( 
                -\frac{188747}{8748}
                -\frac{3200}{81}\zeta_3
                -\frac{830}{81}\zeta_2
      \Biggr)
\N \\ \N \\ &&
+n_fT_F^2C_F
      \Biggl( 
                -\frac{1271507}{17496}
                +\frac{1400}{81}\zeta_3
                -\frac{415}{81}\zeta_2
      \Biggr)
\\
a_{qq,Q}^{(3), {\sf NS}}(4)&=&
T_FC_FC_A
      \Biggl( 
                 \frac{419369407}{21870000}
                +\frac{628}{45}{\sf B_4}
                -\frac{628}{5}\zeta_4
                +\frac{515597}{4050}\zeta_3
                +\frac{10703}{4050}\zeta_2
      \Biggr)
\N \\ \N \\ &&
+T_FC_F^2
      \Biggl( 
                 \frac{137067007129}{437400000}
                -\frac{1256}{45}{\sf B_4}
                +\frac{628}{5}\zeta_4
                -\frac{41131}{225}\zeta_3
                +\frac{4526303}{81000}\zeta_2
      \Biggr)
\N \\ \N \\ &&
+T_F^2C_F
      \Biggl( 
                -\frac{151928299}{5467500}
                -\frac{20096}{405}\zeta_3
                -\frac{26542}{2025}\zeta_2
      \Biggr)
\N \\ \N \\ &&
+n_fT_F^2C_F
      \Biggl( 
                -\frac{1006358899}{10935000}
                +\frac{8792}{405}\zeta_3
                -\frac{13271}{2025}\zeta_2
      \Biggr)
\\
a_{qq,Q}^{(3), {\sf NS}}(5)&=&
T_FC_FC_A
      \Biggl( 
                 \frac{816716669}{43740000}
                +\frac{728}{45}{\sf B_4}
                -\frac{728}{5}\zeta_4
                +\frac{12569}{81}\zeta_3
                +\frac{16103}{4050}\zeta_2
      \Biggr)
\N \\ \N \\ &&
+T_FC_F^2
      \Biggl( 
                 \frac{13213297537}{36450000}
                -\frac{1456}{45}{\sf B_4}
                +\frac{728}{5}\zeta_4
                -\frac{142678}{675}\zeta_3
                +\frac{48391}{750}\zeta_2
      \Biggr)
\N \\ \N \\ &&
+T_F^2C_F
      \Biggl( 
                -\frac{9943403}{303750}
                -\frac{23296}{405}\zeta_3
                -\frac{31132}{2025}\zeta_2
      \Biggr)
\N \\ \N \\ &&
+n_fT_F^2C_F
      \Biggl( 
                -\frac{195474809}{1822500}
                +\frac{10192}{405}\zeta_3
                -\frac{15566}{2025}\zeta_2
      \Biggr)
\\
a_{qq,Q}^{(3), {\sf NS}}(6)&=&
T_FC_FC_A
      \Biggl( 
                 \frac{1541550898907}{105019740000}
                +\frac{5672}{315}{\sf B_4}
                -\frac{5672}{35}\zeta_4
                +\frac{720065}{3969}\zeta_3
                +\frac{1016543}{198450}\zeta_2
      \Biggr)
\N \\ \N \\ &&
+T_FC_F^2
      \Biggl( 
                 \frac{186569400917}{463050000}
                -\frac{11344}{315}{\sf B_4}
                +\frac{5672}{35}\zeta_4
                -\frac{7766854}{33075}\zeta_3
                +\frac{55284811}{771750}\zeta_2
      \Biggr)
\nonumber
\end{eqnarray}\begin{eqnarray}
&&
+T_F^2C_F
      \Biggl( 
                -\frac{26884517771}{729303750}
                -\frac{181504}{2835}\zeta_3
                -\frac{1712476}{99225}\zeta_2
      \Biggr)
\N \\ \N \\ 
&&
+n_fT_F^2C_F
      \Biggl( 
                -\frac{524427335513}{4375822500}
                +\frac{11344}{405}\zeta_3
                -\frac{856238}{99225}\zeta_2
      \Biggr)
\\
a_{qq,Q}^{(3), {\sf NS}}(7)&=&
T_FC_FC_A
      \Biggl( 
                 \frac{5307760084631}{672126336000}
                +\frac{2054}{105}{\sf B_4}
                -\frac{6162}{35}\zeta_4
                +\frac{781237}{3780}\zeta_3
                +\frac{19460531}{3175200}\zeta_2
      \Biggr)
\N \\ \N \\ &&
+T_FC_F^2
      \Biggl( 
                 \frac{4900454072126579}{11202105600000}
                -\frac{4108}{105}{\sf B_4}
                +\frac{6162}{35}\zeta_4
                -\frac{8425379}{33075}\zeta_3
\N \\ \N \\ 
&&
                +\frac{1918429937}{24696000}\zeta_2
      \Biggr)
+T_F^2C_F
      \Biggl( 
                -\frac{8488157192423}{210039480000}
                -\frac{65728}{945}\zeta_3
                -\frac{3745727}{198450}\zeta_2
      \Biggr)
\N \\ \N \\ &&
+n_fT_F^2C_F
      \Biggl( 
                -\frac{54861581223623}{420078960000}
                +\frac{4108}{135}\zeta_3
                -\frac{3745727}{396900}\zeta_2
      \Biggr)
\\
a_{qq,Q}^{(3), {\sf NS}}(8)&=&
T_FC_FC_A
      \Biggl( 
                -\frac{37259291367883}{38887309440000}
                +\frac{19766}{945}{\sf B_4}
                -\frac{19766}{105}\zeta_4
                +\frac{1573589}{6804}\zeta_3
\N \\ \N \\ &&
                +\frac{200739467}{28576800}\zeta_2
      \Biggr)
+T_FC_F^2
      \Biggl( 
                 \frac{3817101976847353531}{8166334982400000}
                -\frac{39532}{945}{\sf B_4}
                +\frac{19766}{105}\zeta_4
\N \\ \N \\ &&
                -\frac{80980811}{297675}\zeta_3
                +\frac{497748102211}{6001128000}\zeta_2
      \Biggr)
+T_F^2C_F
      \Biggl( 
                -\frac{740566685766263}{17013197880000}
                -\frac{632512}{8505}\zeta_3
\N \\ \N \\ &&
                -\frac{36241943}{1786050}\zeta_2
      \Biggr)
+n_fT_F^2C_F
      \Biggl( 
                -\frac{4763338626853463}{34026395760000}
                +\frac{39532}{1215}\zeta_3
\N \\ \N \\ &&
                -\frac{36241943}{3572100}\zeta_2
      \Biggr)
\\
a_{qq,Q}^{(3), {\sf NS}}(9)&=&
T_FC_FC_A
      \Biggl( 
                -\frac{3952556872585211}{340263957600000}
                +\frac{4180}{189}{\sf B_4}
                -\frac{4180}{21}\zeta_4
                +\frac{21723277}{85050}\zeta_3
\N \\ \N \\ &&
                +\frac{559512437}{71442000}\zeta_2
      \Biggr)
+T_FC_F^2
      \Biggl( 
                 \frac{1008729211999128667}{2041583745600000}
                -\frac{8360}{189}{\sf B_4}
                +\frac{4180}{21}\zeta_4
\N \\ \N \\ &&
                -\frac{85539428}{297675}\zeta_3
                +\frac{131421660271}{1500282000}\zeta_2
      \Biggr)
+T_F^2C_F
      \Biggl( 
                -\frac{393938732805271}{8506598940000}
                -\frac{133760}{1701}\zeta_3
\N \\ \N \\ &&
                -\frac{19247947}{893025}\zeta_2
      \Biggr)
+n_fT_F^2C_F
      \Biggl( 
                -\frac{2523586499054071}{17013197880000}
                +\frac{8360}{243}\zeta_3
\N \\ \N \\ &&
                -\frac{19247947}{1786050}\zeta_2
      \Biggr)
\end{eqnarray}\begin{eqnarray}
a_{qq,Q}^{(3), {\sf NS}}(10)&=&
T_FC_FC_A
      \Biggl( 
                -\frac{10710275715721975271}{452891327565600000}
                +\frac{48220}{2079}{\sf B_4}
                -\frac{48220}{231}\zeta_4
                +\frac{2873636069}{10291050}\zeta_3
\N \\ \N \\ 
&&
                +\frac{961673201}{112266000}\zeta_2
      \Biggr)
+T_FC_F^2
      \Biggl( 
                 \frac{170291990048723954490137}{328799103812625600000}
                -\frac{96440}{2079}{\sf B_4}
                +\frac{48220}{231}\zeta_4
\N \\ \N \\ 
&&
                -\frac{10844970868}{36018675}\zeta_3
                +\frac{183261101886701}{1996875342000}\zeta_2
      \Biggr)
+T_F^2C_F
      \Biggl( 
                -\frac{6080478350275977191}{124545115080540000}
\N \\ \N \\ &&
                -\frac{1543040}{18711}\zeta_3
                -\frac{2451995507}{108056025}\zeta_2
      \Biggr)
+n_fT_F^2C_F
      \Biggl( 
                -\frac{38817494524177585991}{249090230161080000}
\N \\ \N \\ 
&&
                +\frac{96440}{2673}\zeta_3
                -\frac{2451995507}{216112050}\zeta_2
      \Biggr)
\\
a_{qq,Q}^{(3), {\sf NS}}(11)&=&
T_FC_FC_A
      \Biggl( 
                -\frac{22309979286641292041}{603855103420800000}
                +\frac{251264}{10395}{\sf B_4}
                -\frac{251264}{1155}\zeta_4
\N \\ \N \\ &&
                +\frac{283300123}{935550}\zeta_3
                +\frac{1210188619}{130977000}\zeta_2
      \Biggr)
+T_FC_F^2
      \Biggl( 
                 \frac{177435748292579058982241}{328799103812625600000}
\N \\ \N \\ &&
                -\frac{502528}{10395}{\sf B_4}
                +\frac{251264}{1155}\zeta_4
                -\frac{451739191}{1440747}\zeta_3
                +\frac{47705202493793}{499218835500}\zeta_2
      \Biggr)
\N \\ \N \\ &&
+T_F^2C_F
      \Biggl( 
                -\frac{6365809346912279423}{124545115080540000}
                -\frac{8040448}{93555}\zeta_3
                -\frac{512808781}{21611205}\zeta_2
      \Biggr)
\N \\ \N \\ &&
+n_fT_F^2C_F
      \Biggl( 
                -\frac{40517373495580091423}{249090230161080000}
                +\frac{502528}{13365}\zeta_3
                -\frac{512808781}{43222410}\zeta_2
      \Biggr)
\\
a_{qq,Q}^{(3), {\sf NS}}(12)&=&
T_FC_FC_A
      \Biggl( 
                -\frac{126207343604156227942043}{2463815086971638400000}
                +\frac{3387392}{135135}{\sf B_4}
                -\frac{3387392}{15015}\zeta_4
\N \\ \N \\ &&
                +\frac{51577729507}{158107950}\zeta_3
                +\frac{2401246832561}{243486243000}\zeta_2
      \Biggr)
\N \\ \N \\ &&
+T_FC_F^2
      \Biggl( 
                 \frac{68296027149155250557867961293}{122080805651901196900800000}
                -\frac{6774784}{135135}{\sf B_4}
                +\frac{3387392}{15015}\zeta_4
\N \\ \N \\ &&
                -\frac{79117185295}{243486243}\zeta_3
                +\frac{108605787257580461}{1096783781593500}\zeta_2
      \Biggr)
\N \\ \N \\ &&
+T_F^2C_F
      \Biggl( 
                -\frac{189306988923316881320303}{3557133031815302940000}
                -\frac{108396544}{1216215}\zeta_3
\N \\ \N \\ &&
                -\frac{90143221429}{3652293645}\zeta_2
      \Biggr)
+n_fT_F^2C_F
      \Biggl( 
                -\frac{1201733391177720469772303}{7114266063630605880000}
\nonumber
\end{eqnarray}\begin{eqnarray}
&&
                +\frac{6774784}{173745}\zeta_3
                -\frac{90143221429}{7304587290}\zeta_2
      \Biggr)
\\
a_{qq,Q}^{(3), {\sf NS}}(13)&=&
T_FC_FC_A
      \Biggl( 
                -\frac{12032123246389873565503373}{181090408892415422400000}
                +\frac{3498932}{135135}{\sf B_4}
                -\frac{3498932}{15015}\zeta_4
\N \\ \N \\ &&
                +\frac{2288723461}{6548850}\zeta_3
                +\frac{106764723181157}{10226422206000}\zeta_2
      \Biggr)
\N \\ \N \\ 
&&
+T_FC_F^2
      \Biggl( 
                 \frac{10076195142551036234891679659}{17440115093128742414400000}
                -\frac{6997864}{135135}{\sf B_4}
                +\frac{3498932}{15015}\zeta_4
\N \\ \N \\ &&
                -\frac{81672622894}{243486243}\zeta_3
                +\frac{448416864235277759}{4387135126374000}\zeta_2
      \Biggr)
\N \\ \N \\ &&
+T_F^2C_F
      \Biggl( 
                -\frac{196243066652040382535303}{3557133031815302940000}
                -\frac{111965824}{1216215}\zeta_3
                -\frac{93360116539}{3652293645}\zeta_2
      \Biggr)
\N \\ \N \\ &&
+n_fT_F^2C_F
      \Biggl( 
                -\frac{1242840812874342588467303}{7114266063630605880000}
                +\frac{6997864}{173745}\zeta_3
\N \\ \N \\ &&
                -\frac{93360116539}{7304587290}\zeta_2
      \Biggr)
\\
a_{qq,Q}^{(3), {\sf NS}}(14)&=&
T_FC_FC_A
      \Biggl( 
                -\frac{994774587614536873023863}{12072693926161028160000}
                +\frac{720484}{27027}{\sf B_4}
                -\frac{720484}{3003}\zeta_4
\N \\ \N \\ 
&&
                +\frac{6345068237}{17027010}\zeta_3
                +\frac{37428569944327}{3408807402000}\zeta_2
      \Biggr)
\N \\ \N \\ 
&&
+T_FC_F^2
      \Biggl( 
                 \frac{72598193631729215117875463981}{122080805651901196900800000}
                -\frac{1440968}{27027}{\sf B_4}
                +\frac{720484}{3003}\zeta_4
\N \\ \N \\ &&
                -\frac{2101051892878}{6087156075}\zeta_3
                +\frac{461388998135343407}{4387135126374000}\zeta_2
      \Biggr)
\N \\ \N \\ &&
+T_F^2C_F
      \Biggl( 
                -\frac{40540032063650894708251}{711426606363060588000}
                -\frac{23055488}{243243}\zeta_3
                -\frac{481761665447}{18261468225}\zeta_2
      \Biggr)
\N \\ \N \\ &&
+n_fT_F^2C_F
      \Biggl( 
                -\frac{256205552272074402170491}{1422853212726121176000}
                +\frac{1440968}{34749}\zeta_3
\N \\ \N \\ &&
                -\frac{481761665447}{36522936450}\zeta_2
      \Biggr)
~.
\end{eqnarray}


\newpage
\begin{thebibliography}{99}
%
\bibitem{UNIV}
  H.~Georgi and S.~L.~Glashow,
  Phys.\ Rev.\ Lett.\  {\bf 32} (1974) 438;\\
  H.~Fritzsch and P.~Minkowski,
  Annals Phys.\  {\bf 93} (1975) 193.
%
\bibitem{HERALHC}
  H.~Jung {\it et al.},
  {\sf Proceedings of the workshop: HERA and the LHC workshop series on the
       implications of HERA for LHC physics},
       arXiv:0903.3861 [hep-ph];\\
  S.~Alekhin {\it et al.},
  {\sf HERA and the LHC - A workshop on the implications of HERA for LHC  physics}
  arXiv:hep-ph/0601012;
  arXiv:hep-ph/0601013.
%
\bibitem{HIGGS}
See e.g.: 
  A.~Djouadi,
  Phys.\ Rept.\  {\bf 457} (2008) 1
  [arXiv:hep-ph/0503172];
  Phys.\ Rept.\  {\bf 459} (2008) 1
  [arXiv:hep-ph/0503173].
%
\bibitem{DRELL}
  S.~D.~Drell and T.~M.~Yan,
  Annals Phys.\  {\bf 66} (1971) 578
  [Annals Phys.\  {\bf 281} (2000) 450].
%
\bibitem{HEXP}
  P.~D.~Thompson,
  J.\ Phys.\ G {\bf 34} (2007) N177
  [arXiv:hep-ph/0703103];\\
  A.~Aktas {\it et al.}  [H1 Collaboration],
  Eur.\ Phys.\ J.\  C {\bf 40} (2005) 349
  [arXiv:hep-ex/0411046];\\
  S.~Chekanov {\it et al.}  [ZEUS Collaboration],
  Phys.\ Rev.\  D {\bf 69} (2004) 012004
  [arXiv:hep-ex/0308068].
%
\bibitem{BR}
For early studies see:~
  E.~Eichten, I.~Hinchliffe, K.~D.~Lane and C.~Quigg,
  Rev.\ Mod.\ Phys.\  {\bf 56} (1984) 579
  [Addendum-ibid.\  {\bf 58} (1986) 1065];\\
  M.~Gl\"uck, E.~Reya and M.~Stratmann,
  Nucl.\ Phys.\  B {\bf 422} (1994) 37;\\
  J.~Bl\"umlein and S.~Riemersma,
  arXiv:hep-ph/9609394.
%
\bibitem{LO}
  E.~Witten,
  Nucl.\ Phys.\  B {\bf 104} (1976) 445;\\
  J.~Babcock, D.~W.~Sivers and S.~Wolfram,
  Phys.\ Rev.\  D {\bf 18} (1978) 162;\\
  M.~A.~Shifman, A.~I.~Vainshtein and V.~I.~Zakharov,
  Nucl.\ Phys.\  B {\bf 136} (1978) 157
  [Yad.\ Fiz.\  {\bf 27} (1978) 455];\\
  J.~P.~Leveille and T.~J.~Weiler,
  Nucl.\ Phys.\  B {\bf 147} (1979) 147;\\
  M.~Gl\"uck, E.~Hoffmann and E.~Reya,
  Z.\ Phys.\  C {\bf 13} (1982) 119.
%
\bibitem{NLO}
  E.~Laenen, S.~Riemersma, J.~Smith and W.~L.~van Neerven,
  Nucl.\ Phys.\  B {\bf 392} (1993) 162;
  229;\\
  S.~Riemersma, J.~Smith and W.~L.~van Neerven,
  Phys.\ Lett.\  B {\bf 347} (1995) 143
  [arXiv:hep-ph/9411431].
%
\bibitem{LAMB}
  S.~Bethke,
  Nucl.\ Phys.\ Proc.\ Suppl.\  {\bf 135} (2004) 345
  [arXiv:hep-ex/0407021];
  J.\ Phys.\ G {\bf 26} (2000) R27
  [arXiv:hep-ex/0004021];\\
  J.~Bl\"umlein, H.~B\"ottcher and A.~Guffanti,
  Nucl.\ Phys.\  B {\bf 774} (2007) 182
  [arXiv:hep-ph/0607200];\\
  J.~Bl\"umlein,
  arXiv:0706.2430 [hep-ph].
%
\bibitem{Floratos:1977au}
  E.~G.~Floratos, D.~A.~Ross and C.~T.~Sachrajda,
  Nucl.\ Phys.\  B {\bf 129} (1977) 66
  [Erratum-ibid.\  B {\bf 139} (1978) 545].
%
\bibitem{Floratos:1978ny}
  E.~G.~Floratos, D.~A.~Ross and C.~T.~Sachrajda,
  Nucl.\ Phys.\  B {\bf 152} (1979) 493.
%
\bibitem{GonzalezArroyo:1979df}
  A.~Gonzalez-Arroyo, C.~Lopez and F.~J.~Yndurain,
  Nucl.\ Phys.\  B {\bf 153} (1979) 161.
%
\bibitem{GonzalezArroyo:1979he}
  A.~Gonzalez-Arroyo and C.~Lopez,
  Nucl.\ Phys.\  B {\bf 166} (1980) 429.
%
\bibitem{Curci:1980uw}
  G.~Curci, W.~Furmanski and R.~Petronzio,
  Nucl.\ Phys.\  B {\bf 175} (1980) 27.
%
\bibitem{Furmanski:1980cm}
  W.~Furmanski and R.~Petronzio,
  Phys.\ Lett.\  B {\bf 97} (1980) 437.
%
\bibitem{Hamberg:1991qt}
  R.~Hamberg and W.~L.~van Neerven,
  Nucl.\ Phys.\  B {\bf 379} (1992) 143.
%
\bibitem{Ellis:1996nn}
  R.~K.~Ellis and W.~Vogelsang,
  arXiv:hep-ph/9602356.
%
\bibitem{Moch:2004pa}
  S.~Moch, J.~A.~M.~Vermaseren and A.~Vogt,
  Nucl.\ Phys.\  B {\bf 688} (2004) 101
  [arXiv:hep-ph/0403192].
%
\bibitem{Vogt:2004mw}
  A.~Vogt, S.~Moch and J.~A.~M.~Vermaseren,
  Nucl.\ Phys.\  B {\bf 691} (2004) 129
  [arXiv:hep-ph/0404111].
%
\bibitem{WIL1}
  W.~Furmanski and R.~Petronzio,
  Z.\ Phys.\  C {\bf 11} (1982) 293 and references therein.
%
\bibitem{WIL2}
  W.~L.~van Neerven and E.~B.~Zijlstra,
  Phys.\ Lett.\  B {\bf 272} (1991) 127;\\
  E.~B.~Zijlstra and W.~L.~van Neerven,
  Phys.\ Lett.\  B {\bf 273} (1991) 476;
  Nucl.\ Phys.\  B {\bf 383} (1992) 525;\\
  S.~A.~Larin and J.~A.~M.~Vermaseren,
  Z.\ Phys.\  C {\bf 57} (1993) 93;\\
%
  S.~Moch and J.~A.~M.~Vermaseren,
  Nucl.\ Phys.\  B {\bf 573} (2000) 853
  [arXiv:hep-ph/9912355].
%
\bibitem{Larin:1993vu}
  S.~A.~Larin, T.~van Ritbergen and J.~A.~M.~Vermaseren,
  Nucl.\ Phys.\  B {\bf 427} (1994) 41.
%
\bibitem{Larin:1996wd}
  S.~A.~Larin, P.~Nogueira, T.~van Ritbergen and J.~A.~M.~Vermaseren,
  Nucl.\ Phys.\  B {\bf 492} (1997) 338
  [arXiv:hep-ph/9605317].
%
\bibitem{Retey:2000nq}
  A.~Retey and J.~A.~M.~Vermaseren,
  Nucl.\ Phys.\  B {\bf 604} (2001) 281
  [arXiv:hep-ph/0007294]l.
%
\bibitem{Blumlein:2004xt}
  J.~Bl\"umlein and J.~A.~M.~Vermaseren,
  Phys.\ Lett.\  B {\bf 606} (2005) 130
  [arXiv:hep-ph/0411111].
%
\bibitem{Vermaseren:2005qc}
  J.~A.~M.~Vermaseren, A.~Vogt and S.~Moch,
  Nucl.\ Phys.\  B {\bf 724} (2005) 3
  [arXiv:hep-ph/0504242].
%
\bibitem{AB}
  S.~I.~Alekhin and J.~Bl\"umlein,
  Phys.\ Lett.\  B {\bf 594} (2004) 299
  [arXiv:hep-ph/0404034].
%
\bibitem{Blumlein:1998sh}
  J.~Bl\"umlein and W.~L.~van Neerven,
  Phys.\ Lett.\  B {\bf 450} (1999) 417
  [arXiv:hep-ph/9811351].
%
\bibitem{Buza:1995ie}
  M.~Buza, Y.~Matiounine, J.~Smith, R.~Migneron and W.~L.~van Neerven,
  Nucl.\ Phys.\  B {\bf 472} (1996) 611
  [arXiv:hep-ph/9601302].
%
\bibitem{BFNK}
  J.~Bl\"umlein, A.~De Freitas, W.~L.~van Neerven and S.~Klein,
  Nucl.\ Phys.\  B {\bf 755} (2006) 272
  [arXiv:hep-ph/0608024].
%
\bibitem{Buza:1996wv}
  M.~Buza, Y.~Matiounine, J.~Smith and W.~L.~van Neerven,
  Eur.\ Phys.\ J.\  C {\bf 1} (1998) 301
  [arXiv:hep-ph/9612398].
%
\bibitem{Bierenbaum:2007dm}
  I.~Bierenbaum, J.~Bl\"umlein and S.~Klein,
  Phys.\ Lett.\  B {\bf 648} (2007) 195
  [arXiv:hep-ph/0702265].
%
\bibitem{Bierenbaum:2007qe}
  I.~Bierenbaum, J.~Bl\"umlein and S.~Klein,
  Nucl.\ Phys.\  B {\bf 780} (2007) 40
  [arXiv:hep-ph/0703285].
%
\bibitem{Bierenbaum:2008yu}
  I.~Bierenbaum, J.~Bl\"umlein, S.~Klein and C.~Schneider,
  Nucl.\ Phys.\  B {\bf 803} (2008) 1
  [arXiv:0803.0273 [hep-ph]].
%
\bibitem{Bierenbaum:2009zt}
  I.~Bierenbaum, J.~Bl\"umlein and S.~Klein,
  Phys.\ Lett.\  B {\bf 672} (2009) 401
  [arXiv:0901.0669 [hep-ph]].
%
\bibitem{ANCONT}
  J.~Bl\"umlein,
  Comput.\ Phys.\ Commun.\  {\bf 133} (2000) 76
  [arXiv:hep-ph/0003100];\\
  J.~Bl\"umlein and S.~O.~Moch,
  Phys.\ Lett.\  B {\bf 614} (2005) 53
  [arXiv:hep-ph/0503188];\\
  J.~Bl\"umlein,
  arXiv:0901.3106 [hep-ph];
  arXiv:0901.0837 [math-ph].
%
\bibitem{Geyer:1977gv}
  B.~Geyer, D.~Robaschik and E.~Wieczorek,
  Fortsch.\ Phys.\  {\bf 27} (1979) 75.
%
\bibitem{Harris:1994tp}
  B.~W.~Harris and J.~Smith,
  Phys.\ Rev.\  D {\bf 51} (1995) 4550
  [arXiv:hep-ph/9409405].
%
\bibitem{Collins:1994ee}
  J.~C.~Collins and R.~J.~Scalise,
  Phys.\ Rev.\  D {\bf 50} (1994) 4117
  [arXiv:hep-ph/9403231].
%
\bibitem{Chetyrkin:2008jk}
  K.~G.~Chetyrkin, B.~A.~Kniehl and M.~Steinhauser,
  Nucl.\ Phys.\  B {\bf 814} (2009) 231
  [arXiv:0812.1337 [hep-ph]].
%
\bibitem{Tarrach:1980up}
  R.~Tarrach,
  Nucl.\ Phys.\  B {\bf 183} (1981) 384.
%
\bibitem{Nachtmann:1981zg}
  O.~Nachtmann and W.~Wetzel,
  Nucl.\ Phys.\  B {\bf 187} (1981) 333.
%
\bibitem{Gray:1990yh}
  N.~Gray, D.~J.~Broadhurst, W.~Grafe and K.~Schilcher,
  Z.\ Phys.\  C {\bf 48},  (1990) 673.
%
\bibitem{Broadhurst:1991fy}
  D.~J.~Broadhurst, N.~Gray and K.~Schilcher,
  Z.\ Phys.\  C {\bf 52} (1991) 111.
%
\bibitem{Fleischer:1998dw}
  J.~Fleischer, F.~Jegerlehner, O.~V.~Tarasov and O.~L.~Veretin,
  Nucl.\ Phys.\  B {\bf 539} (1999) 671
  [Erratum-ibid.\  B {\bf 571} (2000) 511]
  [arXiv:hep-ph/9803493].
%
\bibitem{Khriplovich:1969aa}
  I.~B.~Khriplovich,
  Yad.\ Fiz.\  {\bf 10} (1969) 409.
%
\bibitem{tHooft:unpub}
G. t'Hooft, unpublished.
%
\bibitem{Politzer:1973fx}
  H.~D.~Politzer,
  Phys.\ Rev.\ Lett.\  {\bf 30} (1973) 1346.
%
\bibitem{Gross:1973id}
  D.~J.~Gross and F.~Wilczek,
  Phys.\ Rev.\ Lett.\  {\bf 30} (1973) 1343.
%
\bibitem{Caswell:1974gg}
  W.~E.~Caswell,
  Phys.\ Rev.\ Lett.\  {\bf 33} (1974) 244.
%
\bibitem{Jones:1974mm}
  D.~R.~T.~Jones,
  Nucl.\ Phys.\  B {\bf 75}  (1974) 531.
%
\bibitem{Abbott:1980hw}
  L.~F.~Abbott,
  Nucl.\ Phys.\  B {\bf 185} (1981) 189.
%
\bibitem{Rebhan:1985yf}
  A.~Rebhan,
  Z.\ Phys.\  C {\bf 30} (1986) 309.
%
\bibitem{Jegerlehner:1998zg}
  F.~Jegerlehner and O.~V.~Tarasov,
  Nucl.\ Phys.\  B {\bf 549} (1999) 481
  [arXiv:hep-ph/9809485].
%
\bibitem{YND}
F.J. Yndurain, {\sf The Theory of Quark and Gluon Interactions}, (Springer, 
Berlin, 2006, 4th Edition).
%
\bibitem{Matiounine:1998ky}
  Y.~Matiounine, J.~Smith and W.~L.~van Neerven,
  Phys.\ Rev.\  D {\bf 57} (1998) 6701
  [arXiv:hep-ph/9801224].
%
\bibitem{Gross:1973ju}
  D.~J.~Gross and F.~Wilczek,
  Phys.\ Rev.\  D {\bf 8} (1973) 3633;
  Phys.\ Rev.\  D {\bf 9} (1974) 980.
%
\bibitem{Georgi:1951sr}
  H.~Georgi and H.~D.~Politzer,
  Phys.\ Rev.\  D {\bf 9} (1974) 416.
%
\bibitem{Altarelli:1977zs}
  G.~Altarelli and G.~Parisi,
  Nucl.\ Phys.\  B {\bf 126} (1977) 298.
%
\bibitem{Steinhauser:2000ry}
    M.~Steinhauser,
    Comput.\ Phys.\ Commun.\  {\bf 134} (2001) 335,
    [arXiv:hep-ph/0009029]; code~{\sf MATAD 3.0}.
%
\bibitem{CS1}
  K.~G.~Chetyrkin and M.~Steinhauser,
  Nucl.\ Phys.\  B {\bf 573}, 617 (2000)
  [arXiv:hep-ph/9911434];
  Phys.\ Rev.\ Lett.\  {\bf 83}, 4001 (1999)
  [arXiv:hep-ph/9907509].
%
\bibitem{TRANS}
  J. Bl\"umlein, S. Klein, and B. T\"odtli, DESY 09-060. 
%
\bibitem{Nogueira:1991ex}
  P.~Nogueira,
  J.\ Comput.\ Phys.\  {\bf 105} (1993) 279.
%
\bibitem{Gorishnii:1989gt}
  S.~G.~Gorishnii, S.~A.~Larin, L.~R.~Surguladze and F.~V.~Tkachov,
  Comput.\ Phys.\ Commun.\  {\bf 55} (1989) 381.
%
\bibitem{Larin:1991fz}
  S.~A.~Larin, F.~V.~Tkachov and J.~A.~M.~Vermaseren,
  {\sf The Form Version Of Mincer}, Preprint NIKHEF-H-91-18.
%
\bibitem{Vermaseren:2000nd}
  J.~A.~M.~Vermaseren,
  arXiv:math-ph/0010025.
%
\bibitem{LCE}
  K.~G.~Wilson,
  Phys.\ Rev.\  {\bf 179} (1969) 1499;\\
R.A.~Brandt and G.~Preparata, Fortschr. Phys. {\bf 18} (1970)
249;\\
W.~Zimmermann, {\sf Lect. on Elementary Particle Physics and
Quantum
Field Theory}, Brandeis Summer Inst., Vol.~1,
(MIT Press, Cambridge, 1970),~p. 395;\\
  Y.~Frishman,
  Annals Phys.\  {\bf 66} (1971) 373.
%
\bibitem{vanRitbergen:1998pn}
  T.~van Ritbergen, A.~N.~Schellekens and J.~A.~M.~Vermaseren,
  Int.\ J.\ Mod.\ Phys.\  A {\bf 14}  (1999) 41
  [arXiv:hep-ph/9802376].
%
\bibitem{Blumlein:1998if}
  J.~Bl\"umlein and S.~Kurth,
  Phys.\ Rev.\  D {\bf 60} (1999) 014018
  [arXiv:hep-ph/9810241].
%
\bibitem{Vermaseren:1998uu}
  J.~A.~M.~Vermaseren,
  Int.\ J.\ Mod.\ Phys.\  A {\bf 14} (1999) 2037
  [arXiv:hep-ph/9806280].
%
\bibitem{Borwein:1999js}
  J.~M.~Borwein, D.~M.~Bradley, D.~J.~Broadhurst and P.~Lisonek,
  Trans.\ Am.\ Math.\ Soc.\  {\bf 353} (2001) 907
  [arXiv:math/9910045].
%
\bibitem{Remiddi:1999ew}
  E.~Remiddi and J.~A.~M.~Vermaseren,
  Int.\ J.\ Mod.\ Phys.\  A {\bf 15}  (2000) 725
  [arXiv:hep-ph/9905237].
%
\bibitem{Moch:2001zr}
  S.~Moch, P.~Uwer and S.~Weinzierl,
  J.\ Math.\ Phys.\  {\bf 43} (2002) 3363
  [arXiv:hep-ph/0110083].
%
\bibitem{Blumlein:2003gb}
  J.~Bl\"umlein,
  Comput.\ Phys.\ Commun.\  {\bf 159} (2004) 19
  [arXiv:hep-ph/0311046].
%
\bibitem{Vermaseren:mincer}
J.A.M. Vermaseren, {\sf The Form version of MINCER}, unpublished.
%
\bibitem{Tentyukov:2007mu}
  M.~Tentyukov and J.~A.~M.~Vermaseren,
  arXiv:hep-ph/0702279.
%
\bibitem{Bierenbaum:2008tt}
  I.~Bierenbaum, J.~Bl\"umlein and S.~Klein,
  arXiv:0812.2427 [hep-ph].
%
\bibitem{Bierenbaum:2008dk}
  I.~Bierenbaum, J.~Bl\"umlein and S.~Klein,
  Nucl.\ Phys.\ Proc.\ Suppl.\  {\bf 183} (2008) 162
  [arXiv:0806.4613 [hep-ph]].
%
\bibitem{Gracey:1993nn}
  J.~A.~Gracey,
  Phys.\ Lett.\  B {\bf 322} (1994) 141
  [arXiv:hep-ph/9401214].
%
\bibitem{Moch:2002sn}
  S.~Moch, J.~A.~M.~Vermaseren and A.~Vogt,
  Nucl.\ Phys.\  B {\bf 646} (2002) 181
  [arXiv:hep-ph/0209100].
%
\bibitem{B4cite}
  D.~J.~Broadhurst,
  Z.\ Phys.\  C {\bf 54} (1992) 599;\\
  L.~Avdeev, J.~Fleischer, S.~Mikhailov and O.~Tarasov,
  Phys.\ Lett.\  B {\bf 336} (1994) 560
  [Erratum-ibid.\  B {\bf 349} (1995) 597]
  [arXiv:hep-ph/9406363];\\
  S.~Laporta and E.~Remiddi,
  Phys.\ Lett.\  B {\bf 379} (1996) 283
  [arXiv:hep-ph/9602417].
  D.~J.~Broadhurst,
  Eur.\ Phys.\ J.\  C {\bf 8} (1999) 311
  [arXiv:hep-th/9803091];\\
  R.~Boughezal, J.~B.~Tausk and J.~J.~van der Bij,
  Nucl.\ Phys.\  B {\bf 713} (2005) 278
  [arXiv:hep-ph/0410216].
%
\bibitem{BROAD}
D. Broadhurst, private communication.
%
\bibitem{BK9}
J. Bl\"umlein and S. Klein, in preparation.
%
\bibitem{BKKS}
  J.~Bl\"umlein, M.~Kauers, S.~Klein and C.~Schneider,
  arXiv:0902.4091 [hep-ph];
  arXiv:0902.4095 [hep-ph].
%
\bibitem{Mertig:1995ny}
  R.~Mertig and W.~L.~van Neerven,
  Z.\ Phys.\  C {\bf 70} (1996) 637
  [arXiv:hep-ph/9506451].
%
\bibitem{Buza:1996xr}
  M.~Buza, Y.~Matiounine, J.~Smith and W.~L.~van Neerven,
  Nucl.\ Phys.\  B {\bf 485} (1997) 420
  [arXiv:hep-ph/9608342].
%
\bibitem{Matiounine:1998re}
  Y.~Matiounine, J.~Smith and W.~L.~van Neerven,
  Phys.\ Rev.\  D {\bf 58} (1998) 076002
  [arXiv:hep-ph/9803439].
\end{thebibliography}
\end{document}